\documentclass[%
 reprint,
 12pt,
 amsmath,amssymb,
 aps,
 onecolumn,
 showkeys,
]{revtex4-2}

\usepackage{graphicx}      
\usepackage{dcolumn}       
\usepackage{bm}            
\usepackage{amsmath}
\usepackage{amssymb}
\usepackage{cancel}
\usepackage{esint}
\usepackage{fixmath}
\usepackage{algorithm}
\usepackage{algorithmic}
\usepackage{theorem}
\usepackage{comment}

\usepackage{geometry}
\usepackage{setspace}
\usepackage{float}

\usepackage{booktabs}
\usepackage{svg}
\usepackage{xcolor}
\usepackage[export]{adjustbox}
\usepackage{pifont}

\usepackage{tikz}
\usetikzlibrary{arrows.meta}
\definecolor{shockblue}{RGB}{0,142,184}
\definecolor{bodyfill}{RGB}{243,245,246}
\definecolor{bodyline}{RGB}{30,30,30}

\usepackage{caption}
\usepackage{subcaption}
\usepackage{ragged2e}
\DeclareCaptionJustification{justified}{\justifying}

\usepackage{hyperref}

\begin{document}

\title{Bow-shock instability in entry, descent, and
  landing vehicles under high-enthalpy conditions}

\author{Adri\'an Ant\'on-\'Alvarez\textsuperscript{1,\dag}}
\author{Adri\'an Lozano-Dur\'an\textsuperscript{1,2}}

\affiliation{\textsuperscript{1}Graduate Aerospace Laboratories, California Institute of Technology, Pasadena, CA 91125, USA}
\affiliation{\textsuperscript{2}Department of Aeronautics and Astronautics, Massachusetts Institute of Technology, Cambridge, MA 02139, USA}
\affiliation{\vspace{0.2em}\rm{\textsuperscript{\dag}Corresponding author: \href{mailto:aantonal@caltech.edu}{aantonal@caltech.edu}}}





\begin{abstract}
Laminar--turbulent transition remains a major uncertainty in the
aerothermal design of entry, descent, and landing (EDL) vehicles. We
show that, under high-enthalpy Mars-entry conditions, the detached bow
shock and the shock-generated shear--entropy layer can become unstable
under freestream disturbances, leading to nonlinear breakdown and
enhanced wall heating.  The analysis is carried out at high altitude
with freestream Mach numbers ($M_\infty$) up to 30 for both Earth and
Mars, with the latter being more susceptible to this instability.
The freestream receptivity analysis shows that disturbance
amplification occurs through a three-step mechanism: (i) transmission
and amplification of acoustic and entropic freestream components
across the bow shock; (ii) further convective amplification within the
post-shock shear--entropy layer; and (iii) bow-shock corrugation
driven by the downstream pressure field, which reinforces the
instability. The dominant response is localized in the shock layer,
and no classical boundary-layer mode is required to explain the
disturbance amplification.
The total optimal energy gain scales as $\overline{G}_T^{\rm opt}\sim
\gamma_2^*M_\infty^2 \exp[(\rho_2/\rho_1)/C-B/\sqrt{Re_\infty}]$,
where $\gamma_2^*$ is an effective specific-heat ratio, $\rho_1$ and
$\rho_2$ are the pre- and post-shock densities, $Re_\infty$ is the
freestream Reynolds number, and $B$ and $C$ are geometry-dependent
constants. For a representative EDL vehicle during Mars entry, the
resulting amplification factors are of order $10^6$.
We further show that flight measurements from the Mars Science
Laboratory (MSL) and Mars 2020/Perseverance
capsules~\citep{bose2014reconstruction, edquist2022mars} are
consistent with these results. The findings are also supported by
wall-modeled large-eddy simulations of MSL under representative
Mars-entry conditions. These results suggest that bow-shock
instabilities may constitute a transition mechanism for blunt
hypersonic entry vehicles, either as a standalone process or in
combination with others.
\end{abstract}

\maketitle


\section{Introduction}
\label{sec::Introduction}

Accurate prediction of laminar-to-turbulent boundary-layer transition
remains a fundamental challenge in the design of hypersonic
vehicles~\citep{schneider2004hypersonic}. The state of the boundary
layer directly governs key aerodynamic forces and thermal loads, which
are critical for assessing the operational and survival margins of the
vehicle. In the turbulent regime, boundary layers produce
substantially higher skin friction and, more importantly,
significantly increased aerodynamic heating compared to their laminar
counterparts~\citep{zoby1981approximate, hollis2010blunt}. These
intensified thermal loads impose stringent requirements on the thermal
protection system (TPS) design, affecting vehicle mass budgets,
mission payload fractions, and structural durability. Inaccurate
prediction of these forces and thermal loads can lead either to overly
conservative designs with excessive weight penalties or to optimistic
designs that are vulnerable to thermal failure. The Mars Science
Laboratory (MSL) and Mars 2020/Perseverance entry, descent, and
landing (EDL) missions exemplify these challenges, with reconstructed
flight aeroheating data indicating transition and associated elevated
heating during entry~\citep{bose2014reconstruction, edquist2022mars,
  alpert2022inverse}.

The transition characteristics of MSL were experimentally studied
through campaigns spanning multiple hypersonic test
facilities~\citep{hollis2005transition,hollis2007turbulent,
  hollis2010blunt,liechty2006mars}. Those ground-test data showed
transition and turbulent-heating augmentation preferentially on the
leeside of the capsule rather than on the windside.  The associated
leeside heat-flux augmentation became a critical consideration in
sizing the TPS. During the actual MSL mission, post-flight heat-flux
data were reconstructed, indicating elevated heating on the leeside
consistent with transition~\citep{bose2014reconstruction}. Similar
leeside heating behavior was subsequently documented for the
Perseverance mission~\citep{edquist2022mars,alpert2022inverse},
suggesting a recurring trend for MSL-class Mars entry capsules and
underscoring the importance of understanding the underlying mechanisms
for future mission design.

Drawing on the database of experimental measurements,
\citet{hollis2010blunt} proposed an empirical criterion for smooth
blunt-body transition onset based on the momentum-thickness Reynolds
number, approximately $Re_\theta \approx 200$. However, the numerical
simulations used during the MSL design phase with the steady-state
LAURA solver were generally restricted to operating in either fully
laminar or fully turbulent
mode~\citep{edquist2007aerothermodynamic,wright2006modeling}. While
these simulations successfully estimated the expected range of wall
heating and reproduced observed transition locations when
appropriately tuned, they lack inherent predictive capability
regarding the physical mechanisms that trigger transition or the
conditions governing its onset and location. This limitation
represents a significant gap in the theory required for extrapolation
to realistic flight conditions, which cannot be fully reproduced
experimentally on Earth.

The present work addresses the following question: in high-enthalpy
hypersonic flow over blunt bodies, can freestream disturbances be
amplified through interactions between the bow shock and the
shear--entropy layer before conventional boundary-layer instabilities
become dominant?

\subsection{Freestream receptivity}
\label{sec::freestream_receptivity}

Experimental evidence has long indicated that transition in high-speed
regimes is strongly influenced by both the structure and intensity of
freestream disturbances~\citep{laufer1954factors,kendall1975wind},
making receptivity theory a central link between environmental
disturbances and the onset of transition over hypersonic vehicles.

Early theoretical investigations employed linear interaction analysis
(LIA) to derive analytical scaling laws for infinitesimal disturbance
amplification across normal and oblique shock waves. In the
strong-shock limit, \citet{mckenzie1968interaction} showed that the
pressure amplitude and the shock-normal wave-energy flux associated
with transmitted acoustic/entropic perturbations scale as
$O(M_\infty^2)$. For incident vortical disturbances, classical LIA
predicts more modest amplification of the post-shock vortical
(kinetic-energy-containing) fluctuations, with gains that depend on
shock strength and incidence~\citep{ribner1954convection}. This
qualitative picture is consistent with later numerical studies of
shock/turbulence interaction~\citep{larsson2009direct}. A key
implication of LIA is that, because the shock jump conditions couple
thermodynamic and velocity perturbations, a single upstream Kov\'asznay
mode (acoustic, vortical, or entropic) typically generates all three
modes downstream, providing multiple pathways for post-shock base-flow
excitation.

Once disturbances are transmitted and amplified across the bow shock,
receptivity studies have shown that coupling between Mack modes and
slow acoustic perturbations often dominates the receptivity process in
canonical high-speed boundary layers on flat plates, wedges, and
slender bodies. \citet{fedorov2001prehistory} and
\citet{fedorov2003receptivity} applied the Kov\'asznay decomposition to
analyze post-shock disturbance fields and demonstrated that slow
acoustic modes can synchronize with Mack's second mode, i.e., a
resonance-like matching of phase speeds that enables sustained forcing
of the instability by the slow acoustic wave. This synchronization
facilitates efficient energy transfer from freestream disturbances to
modal instabilities.
\citet{ma2003receptivity1,ma2003receptivity2} extended receptivity
theory by identifying intermediate modes that bridge freestream
acoustic forcing and boundary-layer instabilities. Although these
modes are stable, they are excited preferentially and subsequently
transfer energy to the unstable second mode via slow-acoustic
synchronization, enabled by phase-speed matching and spatial
overlap. In this view, the shock does more than amplify incoming
disturbances: it organizes the post-shock disturbance field in a way
that promotes efficient coupling into the dominant instability waves.
Building on this perspective, \citet{qin2016response} uncovered a
continuous resonance process in which slow acoustic modes repeatedly
reflect between the shock and the wall, leading to a persistent
excitation mechanism that is intrinsic to shock-bounded flows and has
no analogue in shock-free configurations. Motivated by these
observations, more recent work has sought to capture the global
interaction of shock transmission, reflection, and instability
development within unified receptivity frameworks. For example,
\citet{kamal2023global} developed a global receptivity approach with
realizable excitations of acoustic, entropic, and vortical
disturbances, offering a detailed description of how freestream
fluctuations penetrate and couple into the boundary layer across the
entire flow field.

\subsection{Current challenges and limitations}
\label{sec::challenges}

In this work, we address challenges and limitations that arise when
the studies surveyed above are viewed in the context of hypersonic EDL
vehicles under flight-relevant conditions.
\begin{itemize}

\item \emph{Stability and receptivity mechanisms specific to EDL
vehicles.} Many hypersonic transition studies have focused on slender
  configurations (e.g., HiFiRE-type cones), where Mack-mode dynamics
  are a central mechanism~\citep{schneider2004hypersonic}. Blunt entry
  capsules, however, exhibit a fundamentally different flow topology:
  a strong, detached bow shock produces substantial deceleration and
  thermodynamic changes, yielding post-shock edge Mach numbers that
  are often subsonic or only weakly supersonic over much of the
  forebody. Moreover, bow-shock curvature and shock-layer
  non-parallelism can introduce global coupling effects that are
  absent in canonical slender-body settings. As a result, the
  instability mechanisms governing EDL vehicles may differ from those
  of slender bodies, and their characterization remains less well
  established. These differences extend to receptivity, where prior work has largely emphasized the excitation of Mack-mode instabilities by freestream disturbances on cone-like bodies. Here,
  we perform receptivity analyses tailored to
  EDL-vehicle-like flows, explicitly accounting for bow-shock
  curvature and shock-layer non-parallelism.

\item \emph{High-enthalpy, realistic flight conditions.} A large
  fraction of the available computational and experimental literature
  focuses on low-enthalpy conditions under calorically perfect-gas
  assumptions. Realistic EDL trajectories, however, involve high
  enthalpy, thermochemical effects, and freestream Mach numbers
  approaching $M_\infty \approx 30$. These effects modify the base
  flow (shock-layer structure, temperature, and composition profiles)
  and the disturbance dynamics (acoustic propagation, entropic
  production, and mode coupling), limiting the direct applicability of
  low-enthalpy insights. In our analysis, we account for
  thermochemical effects at flight-relevant Mach numbers using an
  equilibrium model to better quantify these sensitivities.

\item \emph{Experimental limitations.} The disturbance levels in
  ground-based experiments can differ markedly from those in
  atmospheric flight. In conventional hypersonic tunnels, acoustic
  radiation from turbulent boundary layers on the tunnel walls often
  constitutes a dominant source of freestream forcing, with levels
  that can exceed those in flight by orders of
  magnitude~\citep{schneider2001effects}. Many of the experiments used
  to support the MSL design were conducted in such non-quiet
  facilities~\citep{hollis2005transition,
    hollis2007turbulent,hollis2010blunt,liechty2006mars,
    wright2006modeling}. This facility-specific disturbance
  environment raises questions about how directly tunnel-based
  transition trends and criteria translate to free flight, where
  freestream acoustic levels are expected to be substantially
  lower. Complicating this comparison further, entry trajectories
  typically involve freestream Mach numbers well above those
  achievable in many tunnels, while the tunnel acoustic amplitudes are
  often higher. Hence, we consider flight-relevant conditions to
  better assess the conclusions from tunnel-to-flight extrapolations.

\end{itemize}

The remainder of the paper is organized as follows.
Section~\ref{sec:methods} presents the theoretical and computational
methodology. Section~\ref{sec::Optimal_disturbance} examines the
receptivity of the shock layer to freestream disturbances, and
\S\ref{sec::budget} analyses the energy budgets of the resulting
optimal disturbance.  Section~\ref{sec::scaling} develops scaling laws
for the identified optimal response,
\S\ref{sec::corrugation_feedback_effect} examines the feedback
associated with shock corrugation, and \S\ref{sec::path} maps the
resulting gain over velocity--altitude space for Mars and Earth entry
conditions. Finally, \S\ref{sec::non_linear_effects} investigates the
nonlinear instability dynamics using wall-modeled large-eddy
simulation.

\section{Methods}
\label{sec:methods}

\subsection{Problem setup}
\label{sec:problem-setup}

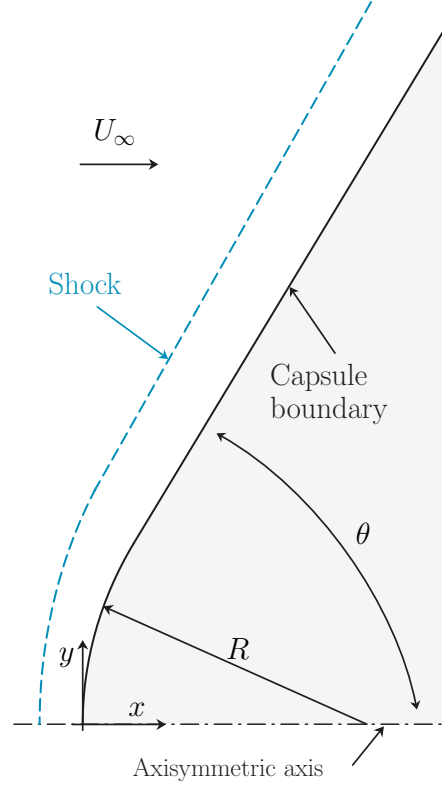
\begin{figure}[t]
\centering
\resizebox{0.34\linewidth}{!}{%
\begin{tikzpicture}[
  x=1pt,y=-1pt,
  line cap=round,
  line join=round,
  >=Stealth,
  body/.style={draw=bodyline,line width=1.35pt},
  shock/.style={draw=shockblue,line width=1.35pt,dash pattern=on 8pt off 4.4pt},
  axisline/.style={draw=bodyline,line width=1.0pt,dash pattern=on 12pt off 4.5pt on 1.5pt off 4.5pt},
  guide/.style={draw=bodyline,line width=1.15pt},
  guideblue/.style={draw=shockblue,line width=1.15pt},
  ann/.style={font=\fontsize{20}{22}\selectfont},
  annsmall/.style={font=\fontsize{19}{21}\selectfont},
  mathann/.style={font=\fontsize{21}{21}\selectfont},
  every node/.style={inner sep=0pt}
]

\path[use as bounding box] (0,0) rectangle (288,540);
\clip (0,0) rectangle (288,540);

\fill[bodyfill]
  (80.722656,367.570312)
  -- (291.089844,18.660156)
  -- (291.089844,489.773438)
  -- (46.375000,489.773438)
  .. controls (46.375000,447.121094) and (58.175781,405.128906) ..
  (80.722656,367.570312)
  -- cycle;

\draw[body]
  (80.722656,367.570312)
  .. controls (58.175781,405.128906) and (46.375000,447.121094) ..
  (46.375000,489.773438);

\draw[body]
  (80.722656,367.570312) -- (291.089844,18.660156);

\draw[axisline]
  (0,489.773438) -- (295,489.773438);

\draw[shock]
  (55.406250,329.949219)
  .. controls (30.609375,376.839844) and (17.316406,432.621094) ..
  (17.316406,489.773438);

\draw[shock]
  (54.035156,332.441406) -- (241.621094,0);

\draw[guide,-{Stealth[length=6.5pt,width=7.5pt]}]
  (238.0,489.8) -- (60.144531,410.050781);

\node[mathann] at (151.0,438.5) {$R$};

\draw[guide,{Stealth[length=5.8pt,width=7pt]}-{Stealth[length=5.8pt,width=7pt]}]
  (136.6,293.1)
  .. controls (205,332) and (258,410) ..
  (271.7,479.9);

\node[mathann] at (235.5,360.5) {$\theta$};

\draw[guide,-{Stealth[length=6.5pt,width=7.5pt]}]
  (39.589844,489.941406) -- (102.558594,489.941406);

\draw[guide,-{Stealth[length=6.5pt,width=7.5pt]}]
  (46.375000,493.910156) -- (46.375000,432.898438);

\node[mathann] at (83.0,479.5) {$x$};
\node[mathann] at (35.5,444.4) {$y$};

\node[mathann] at (68.0,89.3) {$U_\infty$};

\draw[guide,-{Stealth[length=6.5pt,width=7.5pt]}]
  (44.769531,111.269531) -- (97.382812,111.269531);

\node[shockblue,ann,anchor=west] at (23.2,192.0) {Shock};

\draw[guideblue,-{Stealth[length=6pt,width=7pt]}]
  (56.2,207.6) -- (104.0,242.85);

\node[bodyline,ann,anchor=west,align=left] at (172.6,266.5) {Capsule\\boundary};

\draw[guide,-{Stealth[length=6pt,width=7pt]}]
  (216.5,238.9) -- (185.9,194.15);

\node[bodyline,annsmall,anchor=west] at (80,520.3) {Axisymmetric axis};

\draw[guide,-{Stealth[length=6.5pt,width=7.5pt]}]
  (228.4,517.4) -- (249.5,491.2);

\end{tikzpicture}%
}
\caption{Reference coordinate system and geometric notation for the simplified capsule.}
\label{fig:setup}
\end{figure}

We consider flow over a simplified axisymmetric capsule intended to
isolate the forebody shock-layer dynamics relevant to blunt EDL
vehicles. The geometry consists of a spherical nose cap of radius $R$
that blends smoothly into a conical frustum of half-angle
$\theta=52.7^\circ$, as shown in figure~\ref{fig:setup}. The spherical
and conical portions meet tangentially. The value $\theta=52.7^\circ$
is chosen as an effective half-angle, $\theta=70^\circ-\alpha$, so
that the axisymmetric configuration approximates the leeside
inclination of the MSL $70^\circ$ sphere-cone at the representative
flight angle of attack $\alpha=17.3^\circ$.

A Cartesian reference frame is adopted in the meridional plane. The
$x$-axis is aligned with both the freestream direction and the axis of
symmetry of the body. The origin is located at the nose apex. For the
axisymmetric zero-angle-of-attack base flow considered here, this
point is also the wall stagnation point. The coordinate $y$ denotes
the radial direction in the meridional plane.

The freestream state is specified by the density $\rho_\infty$,
velocity $U_\infty$, temperature $T_\infty$, and species mole
fractions $X_i$, resulting in the freestream composition
$\boldsymbol{X}_\infty$. Table~\ref{tab:freestream} lists the
representative Mars-entry baseline used to define the reference
configuration. This state is selected because it lies near the
velocity--altitude region of the MSL and Mars 2020/Perseverance
entries where reconstructed aeroheating data indicate enhanced leeside
heating associated with
transition~\citep{bose2014reconstruction,edquist2022mars,
  alpert2022inverse}. Additional freestream states used in the
Reynolds-number, Mach-number and trajectory sweeps are introduced in
\S\ref{sec::scaling} and \S\ref{sec::path}. For the class of
conditions considered here, a detached bow shock forms ahead of the
nose. The shock standoff distance, post-shock thermochemical state and
shock-layer gradients determine the stability and receptivity
properties analyzed below.
%
%
\begin{table}
\centering
\small
\begin{tabular}{@{}ccccc@{}}
$U_\infty$ [m/s] & $\rho_\infty$ [kg/m$^{3}$] & $T_\infty$ [K] & Freestream composition & $M_\infty$ \\[3pt] \midrule
5\,690 & 0.000351 & 158 &
$X_{\text{CO}_2}:0.9556$, $X_{\text{N}_2}:0.0270$,
$X_{\text{Ar}}:0.0160$, $X_{\text{O}_2}:0.0014$ & 28.7
\end{tabular}
\caption{Representative Mars-entry baseline used to define the
  reference configuration. The species mole fractions $X_i$
  correspond to the Martian atmosphere.}
\label{tab:freestream}
\end{table}

\subsection{Modeling equations}
\label{sec:modeling-equations}

\subsubsection{Compressible Navier--Stokes equations}
\label{sec:ns}

The flow is modeled as a continuum, viscous, heat-conducting,
single-temperature equilibrium gas mixture. Unless otherwise stated,
length, velocity, density, pressure, and time are non-dimensionalized by
$R$, $U_\infty$, $\rho_\infty$, $\rho_\infty U_\infty^2$ and
$R/U_\infty$, respectively. The specific internal energy is
non-dimensionalized by $U_\infty^2$, the temperature by
$U_\infty^2/c_{v\infty}$, and the entropy by $c_{v\infty}$. With these
conventions, the governing equations are
\begin{subequations}
\label{eq:ns}
\begin{align}
\frac{\partial \rho}{\partial t}
+\frac{\partial(\rho u_i)}{\partial x_i}
&=0,
\label{eq:ns-a}\\[3pt]
\frac{\partial(\rho u_i)}{\partial t}
+\frac{\partial(\rho u_i u_j)}{\partial x_j}
&=
-\frac{\partial\bigl[(\gamma^* - 1)\rho e\bigr]}{\partial x_i}
+\frac{\partial\tau_{ij}}{\partial x_j},
\label{eq:ns-b}\\[3pt]
\frac{\partial \left(\frac{1}{2}\rho u_i u_i + \rho e\right)}{\partial t}
+ \frac{\partial \left[\left(\frac{1}{2}\rho u_i u_i + \gamma^*\rho e\right)u_j\right]}{\partial x_j}
&= \frac{\partial (\tau_{ij}u_i)}{\partial x_j} - \frac{\partial q_i}{\partial x_i},
\label{eq:ns-c}\\[3pt]
\tau_{ij}
&=
\frac{\mu^*}{Re_\infty}
\left(
\frac{\partial u_j}{\partial x_i}
+\frac{\partial u_i}{\partial x_j}
-\frac{2}{3}
\frac{\partial u_k}{\partial x_k}\delta_{ij}
\right),
\label{eq:ns-d}\\[3pt]
q_i
&=
-\frac{\gamma_\infty k^*}{Re_\infty Pr_\infty}
\frac{\partial}{\partial x_i}
\left(\frac{e}{c_v^*}\right).
\label{eq:ns-e}
\end{align}
\end{subequations}
Here $u_i$ are the velocity components, $\rho$ is the density,
$e$ is the specific internal energy, $p$ is the pressure, $T$ is
the temperature, $\tau_{ij}$ is the viscous stress and $q_i$ is
the heat flux. The effective thermodynamic quantities are
$\gamma^* = 1 + p/(\rho e)$ and
$c_v^* = e/T$, while the non-dimensional transport
coefficients are $\mu^*=\mu/\mu_\infty$ and
$k^*=k/k_\infty$. The corresponding non-dimensional groups are
\begin{equation}
Re_\infty
=
\frac{\rho_\infty U_\infty R}{\mu_\infty},
\qquad
\gamma_\infty
=
\frac{c_{p\infty}}{c_{v\infty}},
\qquad
Pr_\infty
=
\frac{\mu_\infty c_{p\infty}}{k_\infty},
\label{eq:nondim_groups}
\end{equation}
where $c_{p\infty}$ is the freestream specific heat at constant
pressure. The thermochemical models for $\gamma^*$ and $c_v^*$ along
with the transport models for $k^*$ and $\mu^*$ are discussed in the
next section. The governing equations are closed with the
ideal-mixture equation of state
\[
p=\rho R_g T,
\]
where the local mixture gas constant $R_g$ is non-dimensionalized by
$c_{v\infty}$. 

The use of the continuum equations is justified by the small
body-scale Knudsen number:
\[
Kn_\infty
\simeq
\sqrt{\frac{\pi\gamma_\infty}{2}}\,
\frac{M_\infty}{Re_\infty}.
\]
At the freestream state in table~\ref{tab:freestream}, this estimate
gives $Kn_\infty\simeq 4\times10^{-4}$ for
$Re_\infty=10^5$. Additionally, the local Knudsen number is even
smaller in the compressed post-shock region. Therefore, the flow is
well within the continuum regime at the body scale relevant to the
present analyses.

\subsubsection{Thermochemical and transport models}
\label{sec:thermochem}

High-enthalpy Mars-entry conditions require a thermochemical model
that accounts for the redistribution of internal energy among
translational, rotational, vibrational, and chemical
modes. Vibrational excitation and finite-rate chemistry, particularly
dissociation at sufficiently high temperature, can alter the shock
standoff distance, density ratio, temperature field and entropy
gradients that control the stability of the shock layer.

We assume a chemically and thermally equilibrated ideal-gas mixture
for both the base-flow calculations and the receptivity
analysis. Under this assumption, separate species-transport and
vibrational-energy equations are not solved. Instead, the local
composition is obtained algebraically from the equilibrium
relations. The constitutive models for
\[
\gamma^*=\gamma^*(\rho,e;\boldsymbol{X}_\infty), \
c_v^*=c_v^*(\rho,e;\boldsymbol{X}_\infty), \
\mu^*=\mu^*(\rho,e;\boldsymbol{X}_\infty), \
k^*=k^*(\rho,e;\boldsymbol{X}_\infty)
\]
are described in~\cite{AntonAlvarez2026HYMOR}. The equilibrium thermochemical model
also provides $R_g$, the local equilibrium composition, and the local
speed of sound.

The validity of the equilibrium assumption is assessed in
Appendix~\ref{sec::validity_equilibrium}. For reference, we define the
chemical and vibrational Damk\"{o}hler numbers using the body-scale
post-shock flow time $ \tau_{\mathrm{flow}} = R/U_2^{\mathrm{RH}}$,
where $U_2^{\mathrm{RH}}$ is the post-shock normal velocity from the
corresponding normal-shock Rankine--Hugoniot estimate. Thus
\begin{equation}
Da_{\mathrm{chem}}
=
\frac{\tau_{\mathrm{flow}}}{\tau_{\mathrm{chem}}},
\qquad
Da_{\mathrm{vib}}
=
\frac{\tau_{\mathrm{flow}}}{\tau_{\mathrm{vib}}}.
\label{eq:damkohler_methods}
\end{equation}
For the baseline Mars-entry state,
Appendix~\ref{sec::validity_equilibrium} gives
$Da_{\mathrm{chem}}=O(10^2)$--$O(10^3)$ and
$Da_{\mathrm{vib}}=O(10^4)$--$O(10^5)$ in the critical
velocity--altitude region. The relaxation times are therefore short
relative to the residence time of the gas in the forebody shock layer,
supporting the use of an equilibrium model for the stability
calculations reported here. The equilibrium calculations and the
relaxation-time estimates are obtained with Cantera~\citep{cantera},
as described in Appendix~\ref{sec::validity_equilibrium}.

The mixture transport coefficients,
$\mu^*=\mu^*(\rho,e;\boldsymbol{X}_\infty)$ and
$k^*=k^*(\rho,e;\boldsymbol{X}_\infty)$, are evaluated using standard
mixture rules: Wilke's formula for
viscosity~\citep{wilke1950viscosity} and the Mason--Saxena relation
for thermal conductivity~\citep{mason1958approximate}. Species
diffusion and diffusive enthalpy transport are neglected in the
present equilibrium-mixture formulation.

\subsubsection{Boundary conditions}
\label{sec:boundary-conditions}

Boundary conditions are imposed at the fitted bow shock, the capsule
surface, the symmetry axis and the downstream outflow plane.
\begin{itemize}

\item \emph{Bow shock.} Only the post-shock region is discretized. For
  the steady base flow, the state immediately downstream of the fitted
  shock satisfies the Rankine--Hugoniot jump relations with the
  uniform freestream, closed by the equilibrium-mixture thermodynamic
  model. In the linearized calculations, the jump relations and the
  shock kinematic condition are linearized about the fitted-shock base
  state, so that shock displacement and post-shock perturbations are
  coupled consistently.

\item \emph{Wall.} No-slip and adiabatic-wall conditions are imposed
  at the capsule surface,
  \[
  \bm{u}=\bm{0},
  \qquad
  q_j n_j=0 ,
  \]
  where $\bm{n}$ is the unit normal to the wall. This is a deliberate
  simplification relative to a flight heat shield, for which wall
  cooling, catalysis, ablation, pyrolysis and radiation may affect the
  near-wall thermal state. The dominant linear responses identified
  below are localized primarily in the outer shock layer.  Additional
  calculations with cooled-wall conditions produced the same
  shock-layer amplification mechanism.

\item \emph{Axis of symmetry.} Symmetry conditions are
  imposed along the geometric symmetry axis.

\item \emph{Outflow.} Non-reflecting characteristic conditions are
  applied at the downstream boundary following
  \citet{poinsot1992boundary}.

\end{itemize}

\subsection{Numerical solver}
\label{sec:discretization}

The computations are performed with the Hypersonic Linear Stability
Toolkit (HYMOR), an open-source solver for global modal, non-modal and
receptivity analyses of hypersonic shock
layers~\citep{AntonAlvarez2026HYMOR}. Only the key information needed
for the present study is summarized here.

The receptivity calculations reported in
\S\ref{sec::Results} are axisymmetric: the base flow and perturbations are
represented in the meridional plane and no azimuthal modes are
included. HYMOR solves the axisymmetric compressible Navier--Stokes
equations with a second-order cell-centered finite-volume method on a
body-fitted curvilinear grid. Cell-centered states are linearly
interpolated to face midpoints, face integrals are evaluated with
midpoint quadrature and time integration of the nonlinear base
flow uses the classical four-stage fourth-order Runge--Kutta scheme.

The bow shock is treated by shock fitting rather than shock capturing.
The fitted shock is represented by a cubic spline whose degrees of
freedom evolve with the shock kinematic condition. The discontinuity
is therefore imposed as a boundary condition of the computational
domain. This avoids shock-capturing pathologies such as the carbuncle
phenomenon~\citep{pandolfi2001numerical} and, more importantly for the
present study, retains the linear coupling between infinitesimal shock
motion and the post-shock perturbation field. In the linearized
solver, the Rankine--Hugoniot relations, shock kinematics and interior
finite-volume residual are differentiated consistently about the
shock-fitted base state.

Computations were performed on three progressively refined meshes,
denoted coarse, base and fine. At the representative condition
$Re_\infty=285\,000$, these meshes contain $1000\times160$,
$2000\times320$ and $4000\times640$ control volumes in the streamwise
and wall-normal directions, respectively.  The grid is clustered near
the wall, across the post-shock shear--entropy layer and near the
fitted shock. A systematic refinement study confirmed that the leading
long-time optimal gain differs by less than $3\,\%$ between the base
and fine meshes. The base mesh is therefore used for the results
reported below.

\subsubsection{Freestream receptivity}
\label{sec::freestream_receptivity_implementation}

We use non-bold symbols for continuous fields and bold symbols for
discrete vectors and matrices. The continuous conservative flow state
is
\[
q=[\rho,\rho u,\rho v,\rho E]^T,
\qquad
E=e+\frac{1}{2}u_i u_i ,
\]
where $u$ and $v$ are the meridional velocity components. The
freestream state is denoted by $q_\infty$, while the flow conditions after the shock are given by  $q$. The total upstream and
post-shock fields are decomposed as
\begin{equation}
  \begin{bmatrix}
    q_\infty\\ q
  \end{bmatrix}
  =
  \begin{bmatrix}
    q_{\infty,0}+q_\infty'\\ q_0+q'
  \end{bmatrix},
\label{eq:field_decomposition_methods}
\end{equation}
where the subscript $0$ denotes the steady base state and primes
denote infinitesimal perturbations.

Because the bow shock is fitted, the discrete perturbation state
contains both the post-shock flow perturbations and the displacement
of the fitted shock. We write the extended post-shock disturbance as
\begin{equation}
\bm{q}'_e
=
\begin{bmatrix}
\bm{q}'\\
\bm{\eta}'
\end{bmatrix},
\label{eq:augmented_state}
\end{equation}
where $\bm{q}'$ contains the cell-centered conservative perturbation
variables and $\bm{\eta}'$ contains the perturbations of the
shock-spline degrees of freedom. The freestream receptivity problem is
posed as an input--output problem: a prescribed perturbation
immediately upstream of the fitted shock interacts with the
Rankine--Hugoniot boundary conditions and forces the post-shock
response. The post-shock perturbation is initially zero.

Linearization about the steady shock-fitted base state
$\bm{q}_{e,0}$ gives the semi-discrete forced system
\begin{equation}
  \frac{\mathrm{d}\bm{q}'_e}{\mathrm{d} t}
  =
  \bm{L}(\bm{q}_{e,0})\bm{q}'_e
  +
  \bm{B}\bm{q}_\infty'(t),
  \qquad
  \bm{L}(\bm{q}_{e,0})
  \equiv
  \left.
  \frac{\delta \bm{N}}{\delta \bm{q}_e}
  \right|_{\bm{q}_{e,0}} .
  \label{eq:linearized_system}
\end{equation}
Here $\bm{N}$ is the nonlinear shock-fitted residual, including the
interior finite-volume equations, the shock kinematic condition and
the Rankine--Hugoniot compatibility conditions. The matrix $\bm{L}$ is
evaluated numerically by finite differences and is written simply as
$\bm{L}$ hereafter. The vector $\bm{q}_\infty'(t)$ denotes the
discrete freestream perturbation of the conservative variables
immediately upstream of the shock points,
\begin{equation}
  \bm{q}_\infty'
  =
  \left[
  \bm{\rho}_\infty',
  (\bm{\rho u})_\infty',
  (\bm{\rho v})_\infty',
  (\bm{\rho E})_\infty'
  \right]^T .
\end{equation}
The coupling matrix $\bm{B}$ is the linearized map from this upstream
shock trace to the post-shock residual.  This finite-difference
construction automatically includes the linearized response of the
shock-fitting boundary treatment and the Rankine--Hugoniot jump
conditions.

The freestream input is represented only through its trace on the
upstream side of the fitted shock. The imposed upstream disturbance is
expanded as a finite sum of harmonic components,
\begin{equation}
  q_\infty'(s,t)
  =
  \Re\left\{
  \sum_{l=0}^{N_\omega}
  \widehat{q}'_{\infty,l}(s)
  \exp(-i\omega_l t)
  \right\},
  \label{eq:freestream_trace_expansion}
\end{equation}
where $s$ denotes the arc length along the shock surface and the
temporal frequencies $\omega_l$ are specified by the user.  The actual
optimization is performed for the disturbance over the bow shock
through the functions $\widehat{q}'_{\infty,l}(s)$ in
\eqref{eq:freestream_trace_expansion}. After discretization of the
shock surface, each complex amplitude $\widehat{\bm q}_{\infty,l}'$
contains the conservative-variable perturbations at the shock points
for the $l$th frequency. The frequency blocks satisfy
\begin{equation}
  \frac{\mathrm{d}\widehat{\bm q}_{\infty,l}'}{\mathrm{d} t}
  =
  -i\omega_l
  \widehat{\bm q}_{\infty,l}',
  \qquad
  \bm{q}_\infty'(t)
  =
  \sum_{l=0}^{N_\omega}
  \widehat{\bm q}_{\infty,l}'(t).
  \label{eq:freestream_frequency_blocks}
\end{equation}
Thus, after discretization, the unknowns are the complex amplitude
values of $\widehat{q}'_{\infty,l}(s)$ at the shock points for the
selected set of temporal frequencies. No upstream flow field is solved
for, and the computation does not require any prescribed relation
between temporal frequency and wavenumber. The relevant dispersion
relations are introduced only afterward, as an external interpretation
of the shock trace, in which the projection of the incident
disturbance onto Kov\'asznay modes is evaluated \textit{a posteriori},
see Appendix~\ref{app:Kovasznay}.  This makes it possible to determine
the admissible freestream modes---vortical, entropic, and
acoustic---excited by the optimal disturbance~\citep{kamal2023global}.

Combining \eqref{eq:linearized_system} and
\eqref{eq:freestream_frequency_blocks} gives a time-dependent
input--output problem for the post-shock disturbance. We first collect
the initial complex amplitudes of all retained freestream frequencies
into the vector
\begin{equation}
  \widehat{\bm q}_\infty'(0)
  =
  \begin{bmatrix}
    \widehat{\bm q}_{\infty,0}'(0)\\
    \widehat{\bm q}_{\infty,1}'(0)\\
    \vdots\\
    \widehat{\bm q}_{\infty,N_\omega}'(0)
  \end{bmatrix}.
  \label{eq:freestream_initial_vector}
\end{equation}
Let $\bm{\Omega}$ and $\bm{\Sigma}$ be the operators defined by
\begin{equation}
  \left[\bm{\Omega}\widehat{\bm q}_\infty'\right]_l
  =
  \omega_l\widehat{\bm q}_{\infty,l}',
  \qquad
  \bm{\Sigma}\widehat{\bm q}_\infty'
  =
  \sum_{l=0}^{N_\omega}
  \widehat{\bm q}_{\infty,l}' ,
  \label{eq:omega_sigma_operators}
\end{equation}
where $[\cdot]_l$ denotes the block associated with the frequency
$\omega_l$. Thus, $\bm{\Omega}$ applies the prescribed temporal
frequency to each harmonic block, and $\bm{\Sigma}$ sums the frequency
blocks to recover the instantaneous upstream perturbation at the shock.
Consequently,
\begin{equation}
  \widehat{\bm q}_\infty'(t)
  =
  \exp(-i\bm{\Omega}t)
  \widehat{\bm q}_\infty'(0),
  \qquad
  \bm{q}_\infty'(t)
  =
  \bm{\Sigma}\exp(-i\bm{\Omega}t)
  \widehat{\bm q}_\infty'(0).
  \label{eq:freestream_reconstruction_operator}
\end{equation}
Substitution into \eqref{eq:linearized_system} yields the forced
post-shock system
\begin{equation}
  \frac{\mathrm{d}\bm{q}'_e}{\mathrm{d} t}
  =
  \bm{L}\bm{q}'_e
  +
  \bm{F}(t)\widehat{\bm q}_\infty'(0),
  \qquad
  \bm{F}(t)
  =
  \bm{B}\bm{\Sigma}\exp(-i\bm{\Omega}t),
  \qquad
  \bm{q}'_e(0)=\bm{0}.
  \label{eq:forced_system}
\end{equation}
The corresponding response operator is
\begin{equation}
  \bm{q}'_e(t)
  =
  \int_0^t
  \exp[\bm{L}(t-\tau)]
  \bm{F}(\tau)\,\mathrm{d}\tau\,
  \widehat{\bm q}_\infty'(0)
  \equiv
  \bm{R}(t)\widehat{\bm q}_\infty'(0).
  \label{eq:response_operator}
\end{equation}
Therefore, the downstream response is obtained through the linear
input--output map $\bm{R}(t)$. 

The downstream response is measured with Chu's disturbance-energy
norm,
\begin{equation}
E(t) = \int_{V_D}\left[
\underbrace{\frac{\rho_0 a_0^2}{2(\gamma_0^*p_0)^2}p'^2}_{\text{pressure}}
+ \underbrace{\frac{\rho_0}{2}u_i'u_i'}_{\text{kinetic}}
+ \underbrace{\frac{(\gamma_0^*-1)p_0}{2\gamma_0^*}
\left(\frac{s'}{R_{g,0}}\right)^2}_{\text{entropic}}
\right]\mathrm{d} V,
\label{eq:chu}
\end{equation}
where $V_D$ is the post-shock domain, $a_0$ is the base-flow speed of
sound, $R_{g,0}$ is the base-flow gas constant and $s'$ is the entropy
perturbation.  In discrete form, we obtain
\begin{equation}
  E(t)
  =
  \bm{q}'_e(t)^\dagger
  \bm{P}^\dagger
  \bm{Q}^\dagger
  \bm{M}
  \bm{Q}
  \bm{P}
  \bm{q}'_e(t),
  \label{eq:discrete_chu}
\end{equation}
where $\bm{P}$ extracts the flow perturbation components from
$\bm{q}'_e$ excluding any shock-displacement degrees of freedom,
$\bm{Q}$ maps conservative perturbations to Chu variables and $\bm{M}$
contains the quadrature and thermodynamic weights over the post-shock
control volumes. The dagger denotes the Hermitian transpose.

The incident input energy is defined from the Chu-energy flux of the
prescribed freestream disturbance through the fitted shock. Let
$\mathcal{E}_\infty$ denote the freestream version of the energy
density in \eqref{eq:chu}. The instantaneous incident energy flux is
\begin{equation}
  \dot E_\infty(t)
  =
  \int_{S_s}
  \mathcal{E}_\infty(t)\,
  U_\infty
  \left|\bm{e}_x\cdot\bm{n}_s\right|
  \,\mathrm{d} S ,
  \label{eq:incident_flux}
\end{equation}
where $S_s$ is the fitted-shock surface, $\bm{n}_s$ is the shock
normal and $\bm{e}_x$ is the unit vector in the freestream direction.
The reference input energy is
\begin{equation}
  E_\infty^{\mathrm{ref}}
  =
  \overline{\dot E_\infty}\,T^{\mathrm{ref}},
  \qquad
  T^{\mathrm{ref}}
  =
  \frac{m_D}{\dot m_\infty},
  \label{eq:reference_energy}
\end{equation}
where the overbar denotes averaging over the harmonic input, $m_D =
\int_{V_D}\rho_0\,\mathrm{d} V$ is the base-flow mass contained in the
post-shock domain, and
\begin{equation}
  \dot m_\infty
  =
  \int_{S_s}
  \rho_\infty U_\infty
  \left|\bm{e}_x\cdot\bm{n}_s\right|
  \,\mathrm{d} S
  \label{eq:reference_mass_flux}
\end{equation}
is the freestream mass flux through the fitted shock. Thus
$T^{\mathrm{ref}}$ is the time required for the base-flow mass flux
entering through the shock to replenish the mass contained in
$V_D$. Then
\begin{equation}
  E_\infty^{\mathrm{ref}}
  =
  \widehat{\bm q}_\infty'(0)^\dagger
  \bm{D}_\infty
  \widehat{\bm q}_\infty'(0),
  \label{eq:discrete_input_energy}
\end{equation}
where the Hermitian matrix $\bm{D}_\infty$ is described
in~\citet{AntonAlvarez2026HYMOR}.

The total gain produced by a specified freestream input at time $t$ is
\begin{equation}
  G_T(t)
  =
  \frac{E(t)}{E_\infty^{\mathrm{ref}}}.
  \label{eq::gain_ref}
\end{equation}
Using \eqref{eq:response_operator}, this gain can be written as the
generalized Rayleigh quotient
\begin{equation}
  G_T(t)
  =
  \frac{
  \widehat{\bm q}_\infty'(0)^\dagger
  \bm{C}_\infty(t)
  \widehat{\bm q}_\infty'(0)}
  {
  \widehat{\bm q}_\infty'(0)^\dagger
  \bm{D}_\infty
  \widehat{\bm q}_\infty'(0)},
  \label{eq:receptivity_rayleigh}
\end{equation}
where
\begin{equation}
  \bm{C}_\infty(t)
  =
  \bm{R}(t)^\dagger
  \bm{P}^\dagger
  \bm{Q}^\dagger
  \bm{M}
  \bm{Q}
  \bm{P}
  \bm{R}(t).
  \label{eq:C_operator}
\end{equation}
The optimal freestream disturbance for the prescribed horizon $t$ is
the leading generalized eigenvector of
\begin{equation}
  \bm{C}_\infty(t)\widehat{\bm q}_\infty'(0)
  =
  G_T(t)\,\bm{D}_\infty\widehat{\bm q}_\infty'(0).
  \label{eq:freestream_generalized_eigenproblem}
\end{equation}
The matrix $\bm{C}_\infty(t)$ is not formed explicitly. Its action is
evaluated in matrix-free form by applying the forward and adjoint
input--output maps associated with \eqref{eq:forced_system} and
\eqref{eq:response_operator}. In the implementation, this action is
computed using the equivalent autonomous harmonic-forcing operator and
the same Taylor-series approximation of the matrix exponential
employed in the transient-growth calculation. The Hermitian
generalized eigenvalue problem is solved with a Lanczos iteration.

For the frequency sets used below, the forcing produces a periodic
response after the initial transient has decayed, provided the
unforced post-shock linearized operator is stable. Once an optimal
input $\widehat{\bm q}_{\infty}^{\prime\,\mathrm{opt}}(0)$ has been
obtained from \eqref{eq:freestream_generalized_eigenproblem}, the
instantaneous gain is evaluated from \eqref{eq::gain_ref}. The
cycle-averaged gain reported below is
\begin{equation}
  \overline{G}_T^{\mathrm{opt}}
  =
  \frac{1}{T_p}
  \int_{t_0}^{t_0+T_p}
  G_T(t;\widehat{\bm q}_{\infty}^{\prime\,\mathrm{opt}}(0))\,\mathrm{d} t ,
  \label{eq:cycle_average_gain}
\end{equation}
where $T_p$ is the forcing period and $t_0$ is chosen after transients
have decayed. The maximum-over-phase gain is
\begin{equation}
  G_{T,\max}^{\mathrm{opt}}
  =
  \max_{t\in[t_0,t_0+T_p]}
  G_T(t;\widehat{\bm q}_{\infty}^{\prime\,\mathrm{opt}}(0)).
  \label{eq:peak_gain}
\end{equation}
These quantities are postprocessed from the optimal freestream forcing
mode obtained from the fixed-time eigenproblem
\eqref{eq:freestream_generalized_eigenproblem}.

\section{Results}
\label{sec::Results}

The results are organized as
follows. Section~\ref{sec::Optimal_disturbance} identifies the leading
freestream-receptivity response and decomposes its amplification into
shock-transmission and post-shock contributions.
Section~\ref{sec::budget} examines the Chu-energy budgets of that
response.  Section~\ref{sec::scaling} develops scaling laws for the
shock and downstream gains, and
\S\ref{sec::corrugation_feedback_effect} estimates the
finite-amplitude feedback associated with bow-shock corrugation.
Section~\ref{sec::path} maps the resulting gain over
velocity--altitude space for Mars and Earth entry conditions. Finally,
\S\ref{sec::non_linear_effects} uses wall-modeled large-eddy
simulation of the MSL configuration to assess whether the mechanism
identified by the linear analysis can lead to nonlinear breakdown.

\subsection{Optimal disturbance}
\label{sec::Optimal_disturbance}

\begin{figure}[]
  \centering
  \begin{subfigure}[t]{0.42\textwidth}
    \includegraphics[width=\textwidth]{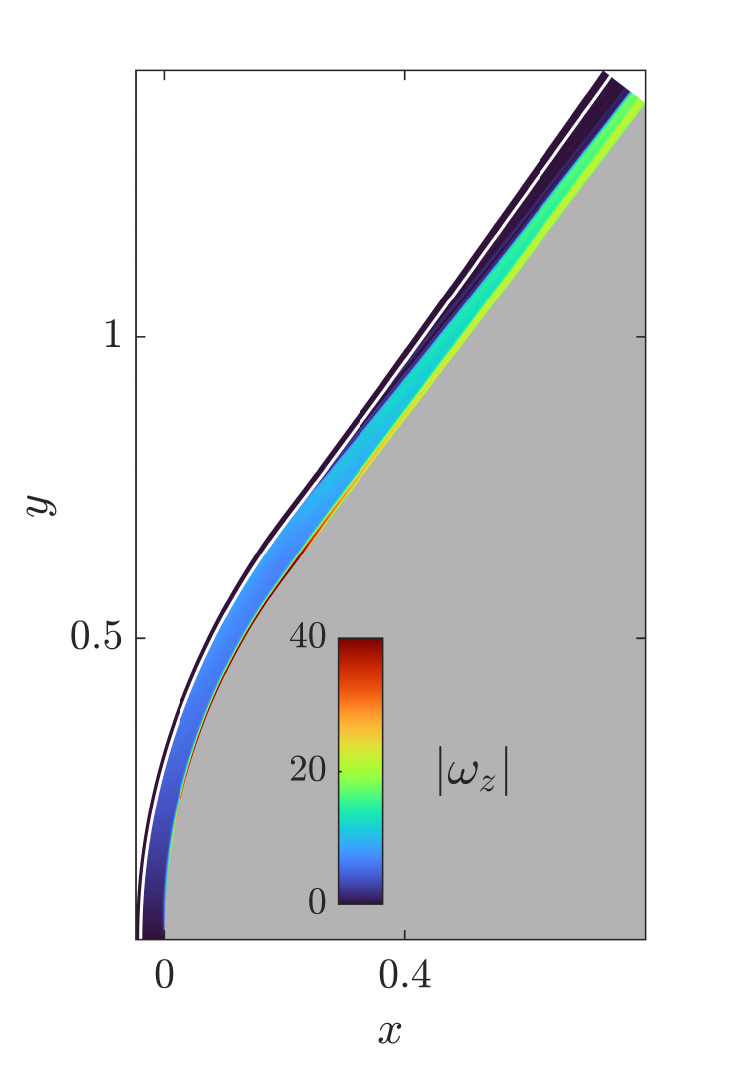}
    \caption{}
    \label{fig:base_flow_vort}
  \end{subfigure}
  \begin{subfigure}[t]{0.42\textwidth}
    \includegraphics[width=\textwidth]{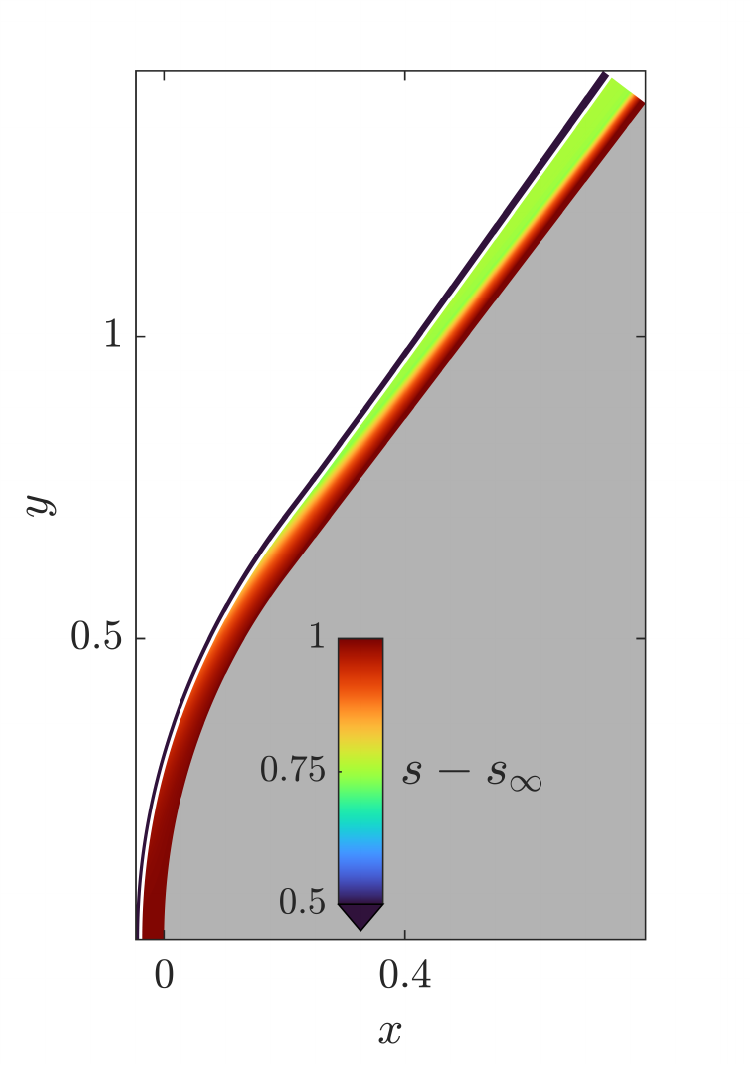}
    \caption{}
    \label{fig:base_flow_entropy}
  \end{subfigure}
  \begin{subfigure}[t]{0.42\textwidth}
    \includegraphics[width=\textwidth]{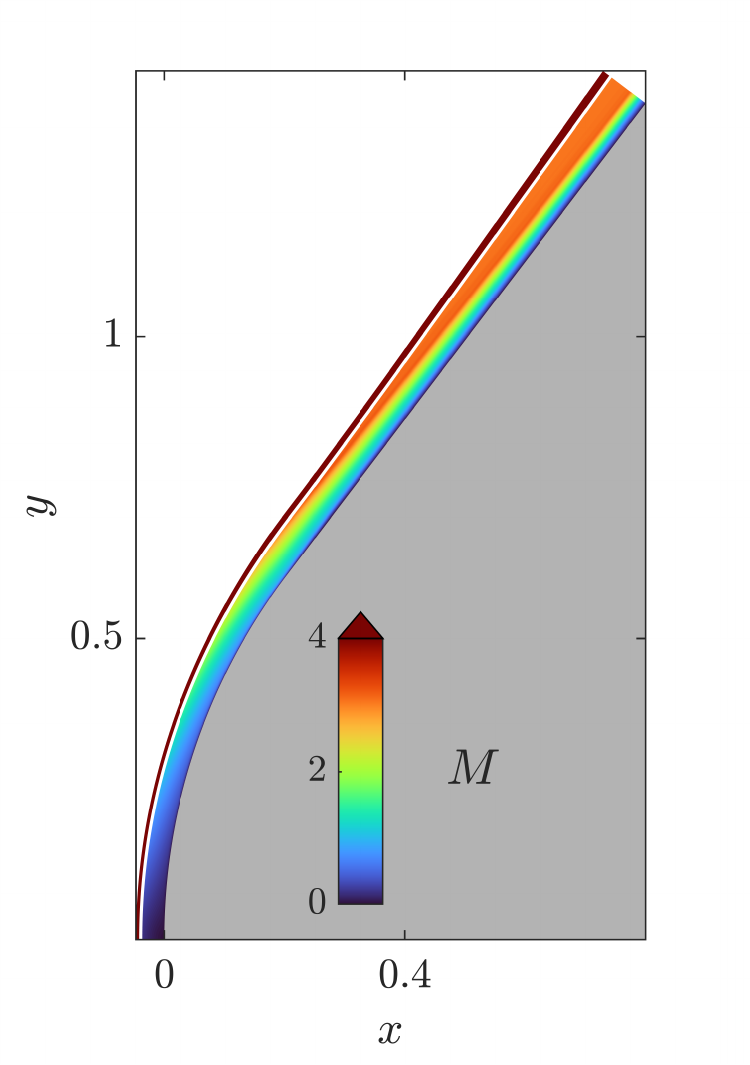}
    \caption{}
    \label{fig:base_flow_Mach}
  \end{subfigure}
  \caption{Steady base flow computed with the freestream conditions
    from table~\ref{tab:freestream} at $Re_\infty = 285\,000$ and
    $M_\infty = 28.7$. The capsule is colored in gray. The bow shock
    is indicated by a solid white line.  (\textit{a})
    Vorticity.  (\textit{b}) Entropy. (\textit{c}) Mach number. All variables are
    non-dimensionalized as described in \S\ref{sec:ns}, while entropy
    is non-dimensionalized with the freestream entropy $s_\infty$.}
  \label{fig:base_flow}
\end{figure}

We first consider the steady base flow obtained with the freestream
conditions listed in table~\ref{tab:freestream} at $Re_\infty =
285\,000$. The corresponding fields are shown in
figure~\ref{fig:base_flow}. A salient feature of this case is the
small shock standoff distance: the computed post-shock density exceeds
the freestream density by more than a factor of 20, so the post-shock
layer is thin. Figure~\ref{fig:base_flow} also shows a pronounced
post-shock entropy layer accompanied by strong tangential shear. As
the flow turns around the forebody, this shear--entropy layer moves
toward the wall and later interacts with the near-wall region.

The response to freestream forcing is best interpreted as a
two-stage process. First, the bow shock transmits and amplifies the
incoming disturbance. Second, the transmitted disturbance convects
within the post-shock domain and undergoes additional growth by
extracting energy from the base flow. To quantify these contributions,
let $\dot{E}_{PS}(t)$ denote the surface-integrated disturbance-energy
flux evaluated immediately downstream of the shock. We define the net
increase of disturbance-energy flux across the shock as
\begin{equation*}
  \Delta\dot{E}_{S}(t)
  =
  \dot{E}_{PS}(t)-\dot{E}_{\infty}(t).
\end{equation*}
The cycle-averaged decomposition used below is
\begin{equation*}
  \overline{G}_{S}^{\mathrm{opt}}
  =
  \frac{\overline{\Delta\dot{E}}_{S}}
       {\overline{\dot{E}}_{\infty}},
  \qquad
  \overline{G}_{D}^{\mathrm{opt}}
  =
  \frac{\overline{E}}
       {\overline{\Delta\dot{E}}_{S}\,T^{\mathrm{ref}}},
  \qquad
  \overline{G}_{T}^{\mathrm{opt}}
  =
  \frac{\overline{E}}
       {\overline{\dot{E}}_{\infty}\,T^{\mathrm{ref}}}
  =
  \overline{G}_{S}^{\mathrm{opt}}\,\overline{G}_{D}^{\mathrm{opt}}.
\end{equation*}
Here $\overline{E}$ is the cycle-averaged disturbance energy stored in
the post-shock domain, and $T^{\mathrm{ref}}$ is the reference time
defined in \S\ref{sec::freestream_receptivity_implementation}.
Thus, $\overline{G}_{S}^{\mathrm{opt}}$ measures the net shock-induced increase of
disturbance-energy flux relative to the incident flux, whereas
$\overline{G}_{D}^{\mathrm{opt}}$ measures the storage and convective amplification
inside the post-shock domain relative to that net transmitted input.

Table~\ref{tab:modes_shock_post_shock_amplification} lists the
cycle-averaged gains for the five leading optimal freestream forcing
modes. For the leading mode, the shock contributes $\overline{G}_{S}^{\mathrm{opt}} =
422$ and the downstream evolution contributes $\overline{G}_{D}^{\mathrm{opt}} =
1\,870$, yielding a total cycle-averaged gain $\overline{G}_{T}^{\mathrm{opt}} = 7.9
\times 10^5$. The large response is therefore not produced by shock
transmission alone: it results from the combination of strong shock
amplification and subsequent convective growth within the post-shock
layer.
\begin{table}
  \begin{center}
\def~{\hphantom{0}}
  \begin{tabular}{cccc}
      Mode & $\overline{G}_{S}^{\mathrm{opt}}$ & $\overline{G}_{D}^{\mathrm{opt}}$ & $\overline{G}_{T}^{\mathrm{opt}}$ \\[6pt] \midrule
       1 & 422 & 1870 & 790\,000\\
       2 & 445 & ~951 & 423\,000\\
       3 & 465 & ~692 & 322\,000\\
       4 & 489 & ~630 & 308\,000\\
       5 & 533 & ~386 & 206\,000\\
  \end{tabular}
  \caption{Cycle-averaged decomposition of the gain into
    shock-transmission and downstream-amplification contributions for
    the five leading optimal freestream forcing modes at $Re_\infty =
    285\,000$ and $M_\infty = 28.7$.}
  \label{tab:modes_shock_post_shock_amplification}
  \end{center}
\end{table}

\begin{figure}[]
\centering
\includegraphics[width=0.55\textwidth]{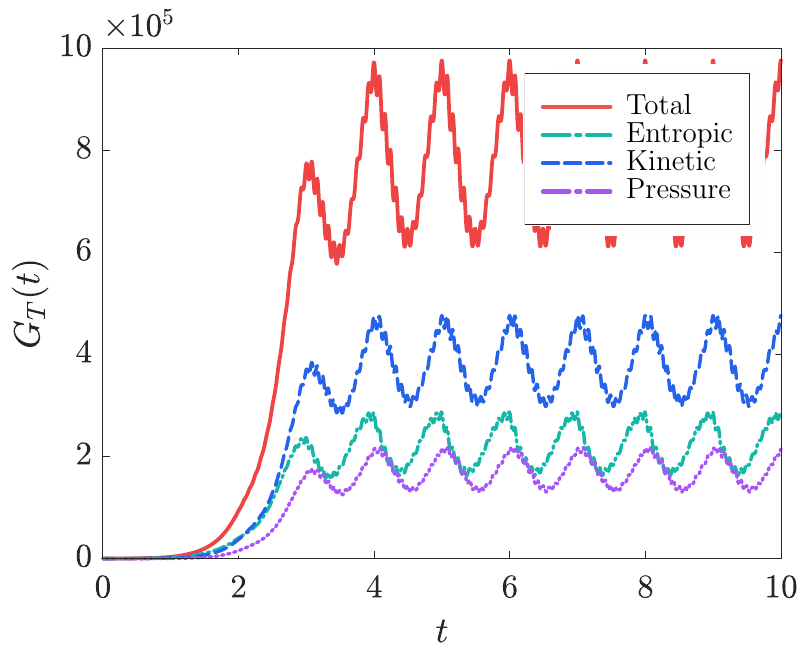}
\caption{Instantaneous total gain $G_T(t)$ for the leading optimal
  freestream forcing mode. After the initial transient, the response
  approaches a time-periodic state whose maximum-over-phase gain is
  $G_{T,\max}^{\mathrm{opt}} = 9.7\times10^5$. Base flow
  computed with freestream conditions from table~\ref{tab:freestream}
  at $Re_\infty = 285\,000$ and $M_\infty = 28.7$. Time is
  non-dimensionalized as described in \S\ref{sec:ns}.}
\label{fig:freestream_energy_growth_first}
\end{figure}

Figure~\ref{fig:freestream_energy_growth_first} shows the
instantaneous gain $G_T(t)$ for the leading optimal forcing
mode. After an initial transient, the response becomes
time-periodic. The maximum gain is
$G_{T,\max}^{\mathrm{opt}} = 9.7\times 10^5$, whereas the
corresponding cycle-averaged gain is the value reported in
table~\ref{tab:modes_shock_post_shock_amplification}.  In the periodic
regime, the post-shock disturbance energy, when partitioned according
to the Chu-energy norm (equation~\ref{eq:chu}), is dominated by the kinetic term, which
accounts for approximately $50\%$ of the total, while the pressure and
entropy terms contribute roughly $25\%$ each.

The Chu-energy partition of the optimal upstream forcing is reported
in table~\ref{tab:re100k_modes}. The five leading optimal modes
consist of approximately $73\%$ pressure energy, $26\%$ entropy
energy, and less than $2\%$ kinetic energy. This partition is computed
from the Chu energy norm of equation~\ref{eq:chu}, evaluated
immediately upstream of the bow shock. While the Chu norm is closely
related to the Kov\'asznay modal decomposition, the two are not
equivalent. For the Kov\'asznay modes, LIA predicts transmitted
disturbance-energy fluxes that scale as $O(M_1^2)$ in the strong-shock
limit, where $M_1$ denotes the pre-shock Mach number
\citep{mckenzie1968interaction}. By contrast, incident vortical
Kov\'asznay modes exhibit only order-unity kinetic-energy amplification
across the shock \citep{ribner1954convection}. Because
$M_1=M_\infty=28.7$ in the present case, this shock-transmission
scaling favors incident disturbances with acoustic or entropic
character over vortical disturbances.  Hence, the
optimization selects an upstream trace with little vortical-energy
content and strong acoustic/entropic content, which is the part of the
admissible disturbance space expected from LIA to couple most
efficiently through the bow shock. 

\begin{table}
  \begin{center}
\def~{\hphantom{0}}
  \begin{tabular}{cccc}
      Mode & Pressure (\%) & Entropic (\%) & Kinetic (\%) \\[3pt] \midrule
       1 & 73.2 & 25.6 & 1.2\\
       2 & 73.3 & 25.5 & 1.2\\
       3 & 73.0 & 25.7 & 1.3\\
       4 & 72.7 & 25.8 & 1.5\\
       5 & 72.7 & 25.8 & 1.5\\
  \end{tabular}
  \caption{Energy composition immediately upstream of the bow shock
    for the five leading optimal freestream-forcing modes, expressed
    in terms of the Chu-energy norm at $Re_\infty = 285\,000$ and
    $M_\infty = 28.7$.}
  \label{tab:re100k_modes}
  \end{center}
\end{table}
\begin{figure}[]
\centering
\begin{subfigure}[t]{0.55\textwidth}
\includegraphics[width=\textwidth]{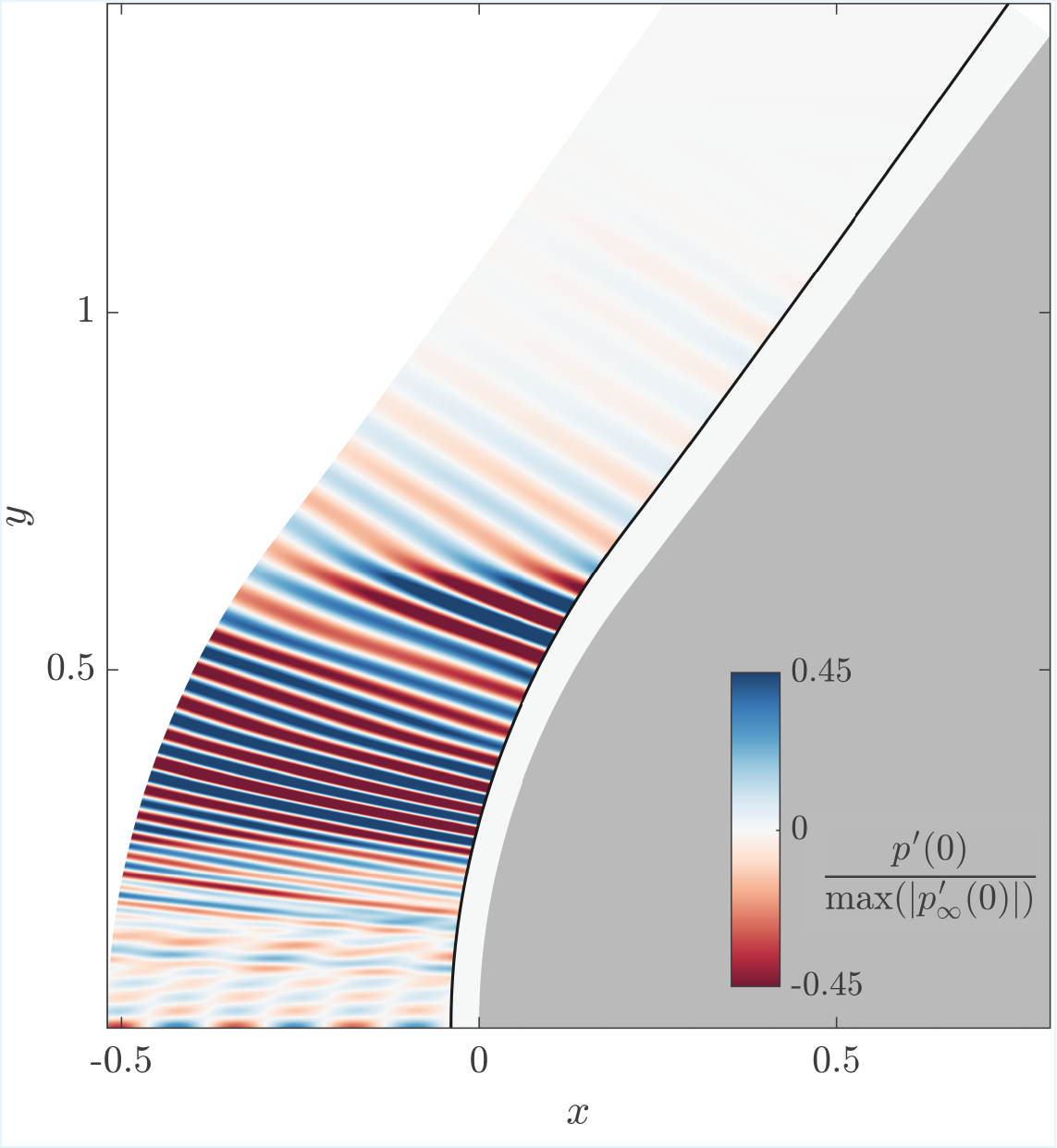}
\caption{}
\label{fig:freestream_dist_first_0}
\end{subfigure}
\begin{subfigure}[t]{0.45\textwidth}
\includegraphics[width=\textwidth]{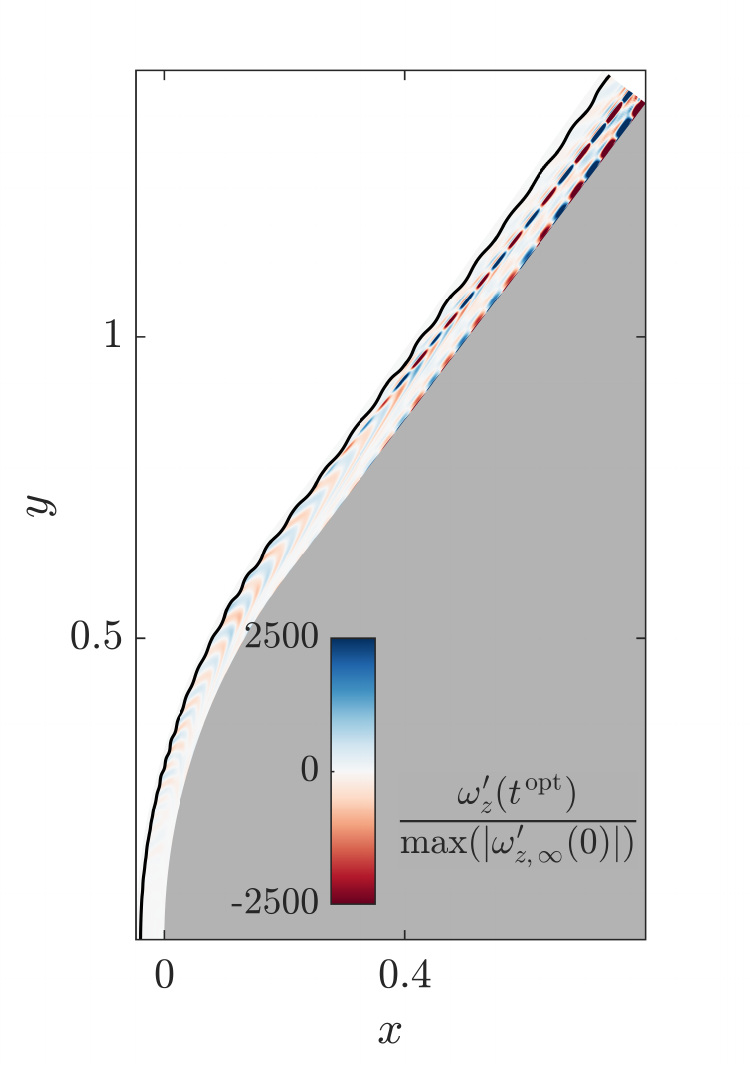}
\caption{}
\label{fig:freestream_dist_first_1}
\end{subfigure}%
\begin{subfigure}[t]{0.45\textwidth}
\includegraphics[width=\textwidth]{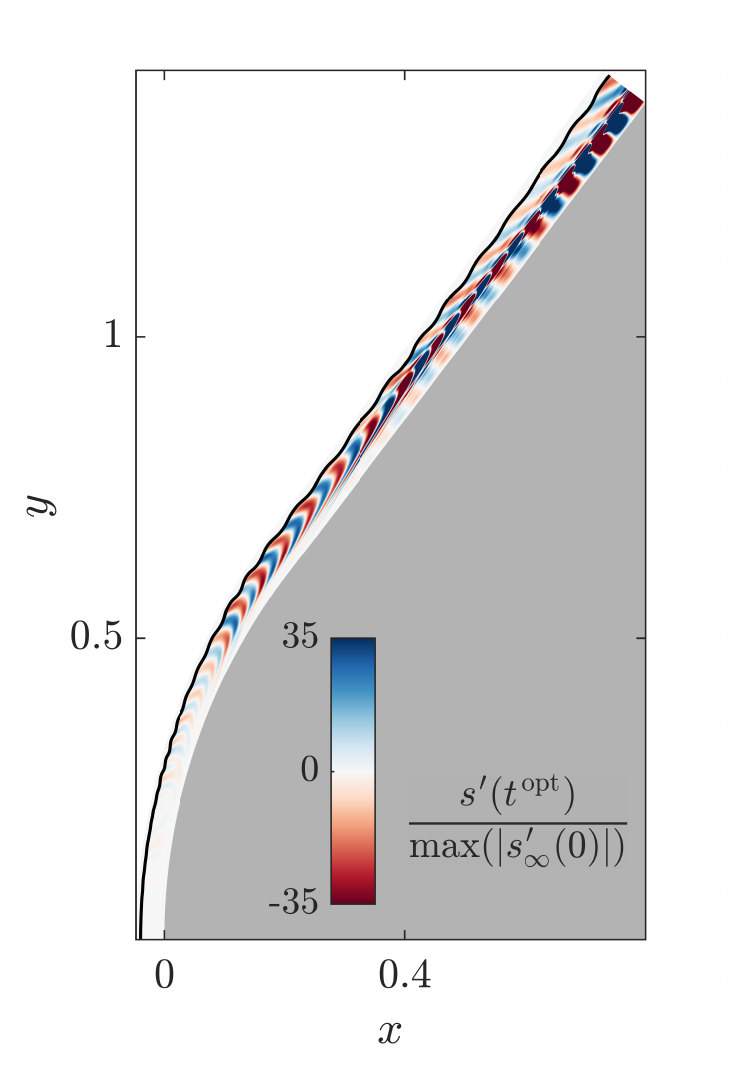}
\caption{}
\label{fig:freestream_dist_first_2}
\end{subfigure}
\caption{Leading optimal freestream forcing mode. Base flow computed
  with freestream conditions from table~\ref{tab:freestream}. The
  fitted-shock location is indicated by the black line. $Re_\infty
  = 285\,000$ and $M_\infty = 28.7$. (\textit{a}) Pressure at
  $t=0$. (\textit{b}) Vorticity at the instant of peak gain on the
  limit cycle. (\textit{c}) Entropy at the same instant. All variables
  are non-dimensionalized as described in \S\ref{sec:ns}.}
\label{fig:freestream_dist_first}
\end{figure}

The spatial structure of the leading mode is shown in
figure~\ref{fig:freestream_dist_first}. Since the optimal freestream
disturbance is available only immediately upstream of the fitted bow
shock, the wave packet shown in
figure~\ref{fig:freestream_dist_first_0} is reconstructed by
projecting the optimal disturbance onto Kov\'asznay modes. The
reconstruction is nearly exact, with an accuracy close to $100\%$.
Further details are provided in Appendix~\ref{app:Kovasznay}.  At
$t=0$ the reconstructed forcing is arranged so that the transmitted
packet enters the shear--entropy layer near the region where the shock
curvature varies most rapidly
(figure~\ref{fig:freestream_dist_first_0}). As the packet convects
downstream, it develops both vorticity and entropy signatures
(figures~\ref{fig:freestream_dist_first_1} and
\ref{fig:freestream_dist_first_2}) that amplify along the
shear--entropy layer. The later interaction with the near-wall region
is secondary: it reflects the advection of the layer toward the wall
and does not require a separate classical boundary-layer instability.

\subsection{Energetics of the optimal disturbance}
\label{sec::budget}

To identify the source of the downstream growth, we examine the
kinetic and entropic energy budgets associated with the Chu norm
integrated over the post-shock region $V_D$. We denote $\left\langle f
\right\rangle_D \equiv \int_{V_D} f\,dV$, and define the kinetic
energy contribution to Chu's norm as $E^k = \left\langle 1/2\rho_0
u'_i u'_i \right\rangle_D$. The corresponding kinetic-energy budget is
\begin{equation}
  \frac{dE^k}{dt} 
  =
  \left\langle \mathcal{A}^k \right\rangle_D
  +
  \left\langle \mathcal{P}^k \right\rangle_D
  +
  \left\langle \Pi_d^k \right\rangle_D
  +
  \left\langle \mathcal{T}^k \right\rangle_D
  +
  \left\langle \mathcal{D}^k \right\rangle_D,
  \label{eq:kinetic-budget-main}
\end{equation}
where $\mathcal{A}^k$ is advection of perturbation kinetic energy by
the base flow, $\mathcal{P}^k$ is production by base-flow velocity
gradients, $\Pi_d^k$ is pressure--dilatation, $\mathcal{T}^k$ is the
combined pressure and viscous transport, and $\mathcal{D}^k$ is
viscous dissipation.  The entropic budget for $E^s= \left\langle
(\gamma_0^*-1)p_0/(2\gamma_0^*) (s'/R_{g,0})^2 \right\rangle_D$ is
\begin{equation}
  \frac{dE^s}{dt} 
  =  
  \left\langle \mathcal{A}^s \right\rangle_D
  +
  \left\langle \mathcal{P}^s \right\rangle_D
  +
  \left\langle \mathcal{T}^s \right\rangle_D
  +
  \left\langle \mathcal{D}^s \right\rangle_D
  +
  \left\langle \mathcal{S}^s \right\rangle_D ,
  \label{eq:entropic-budget-main}
\end{equation}
where $\mathcal{A}^s$ is base-flow advection of entropic energy,
$\mathcal{P}^s$ is production by perturbation advection of the
base-flow entropy gradient, $\mathcal{T}^s$ is heat-flux transport,
$\mathcal{D}^s$ is diffusive dissipation, and $\mathcal{S}^s$ is the
source associated with viscous heating.  For compactness, the brackets
$\langle\cdot\rangle_D$ are omitted below when referring to
domain-integrated budget terms. The detailed derivation of these
budgets is given in Appendix~\ref{sec:energy-budgets}.

Figure~\ref{fig:freestream_budgets_first} shows that, once the
disturbance is inside the post-shock domain, the dominant positive
term is the production $\mathcal{P}^k$. In the domain-integrated
budget, this production is balanced primarily by advective removal
$\mathcal{A}^k$ and transport $\mathcal{T}^k$. Viscous dissipation is
negligible at this Reynolds number, and pressure dilatation makes only
a minor contribution to the kinetic-energy balance.
\begin{figure}[]
\centering
\begin{subfigure}[t]{0.49\textwidth}
\includegraphics[width=\textwidth]{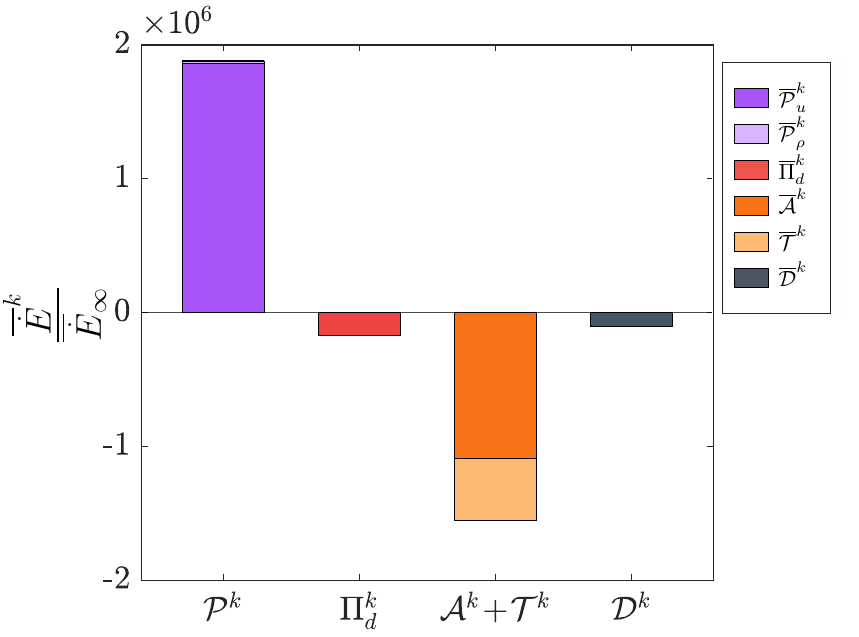}
\caption{}
\label{fig:freestream_budgets_kinetic_bar}
\end{subfigure}
\begin{subfigure}[t]{0.49\textwidth}
\includegraphics[width=\textwidth]{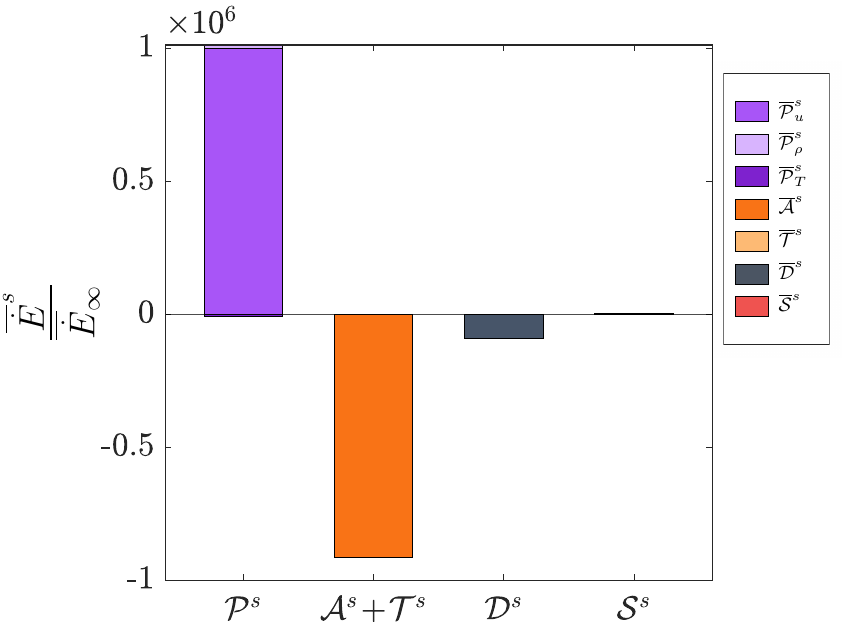}
\caption{}
\label{fig:freestream_budgets_entropic_bar}
\end{subfigure}
\begin{subfigure}[t]{0.49\textwidth}
\includegraphics[width=\textwidth]{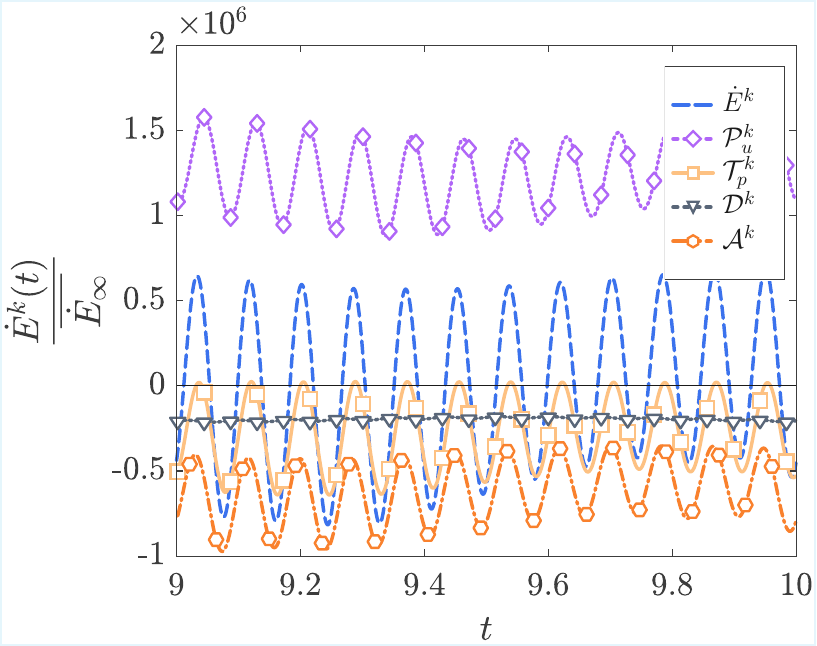}
\caption{}
\label{fig:freestream_budgets_kinetic}
\end{subfigure}
\begin{subfigure}[t]{0.49\textwidth}
\includegraphics[width=\textwidth]{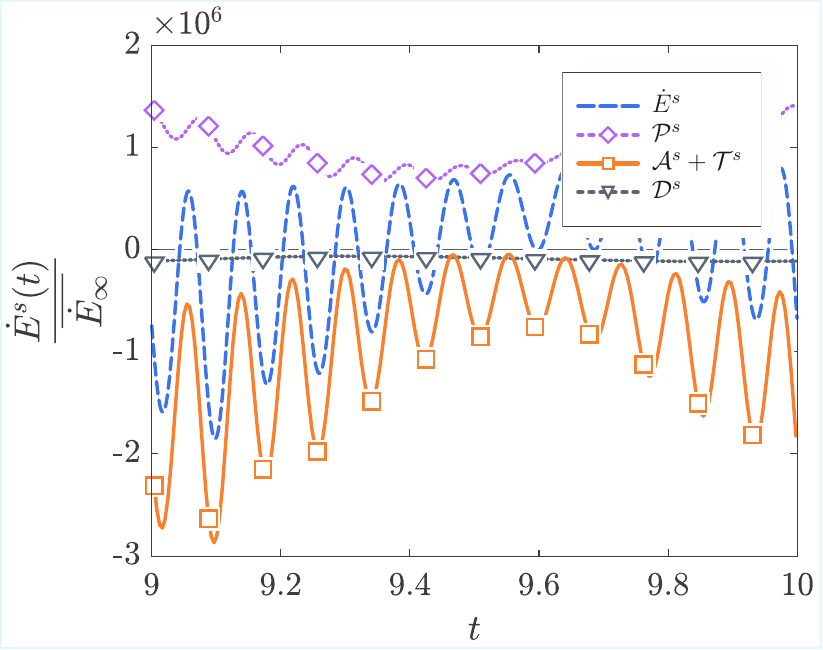}
\caption{}
\label{fig:freestream_budgets_entropic}
\end{subfigure}
\caption{Post-shock domain-integrated kinetic and entropic Chu-energy
  budgets for the leading optimal freestream forcing mode. The base
  flow is computed using freestream conditions from
  table~\ref{tab:freestream} at $Re_\infty = 285\,000$ and $M_\infty =
  28.7$. Energy rates are normalized by the cycle-averaged freestream
  disturbance-energy flux crossing the shock,
  $\overline{\dot{E}_{\infty}}$. Budget term definitions appear in
  Appendix~\ref{sec:energy-budgets}.  (\textit{a}) Spatially and
  time-averaged mean kinetic energy budget over one period.
  (\textit{b}) Entropy budget corresponding to panel (\textit{a}).
  (\textit{c}) Spatially averaged kinetic budget terms as functions of
  time over one period.  (\textit{d}) Corresponding entropy budget
  terms as in panel (\textit{c}).}
\label{fig:freestream_budgets_first}
\end{figure}

\begin{figure}[]
\centering
\begin{subfigure}[t]{0.49\textwidth}
\includegraphics[width=\textwidth]{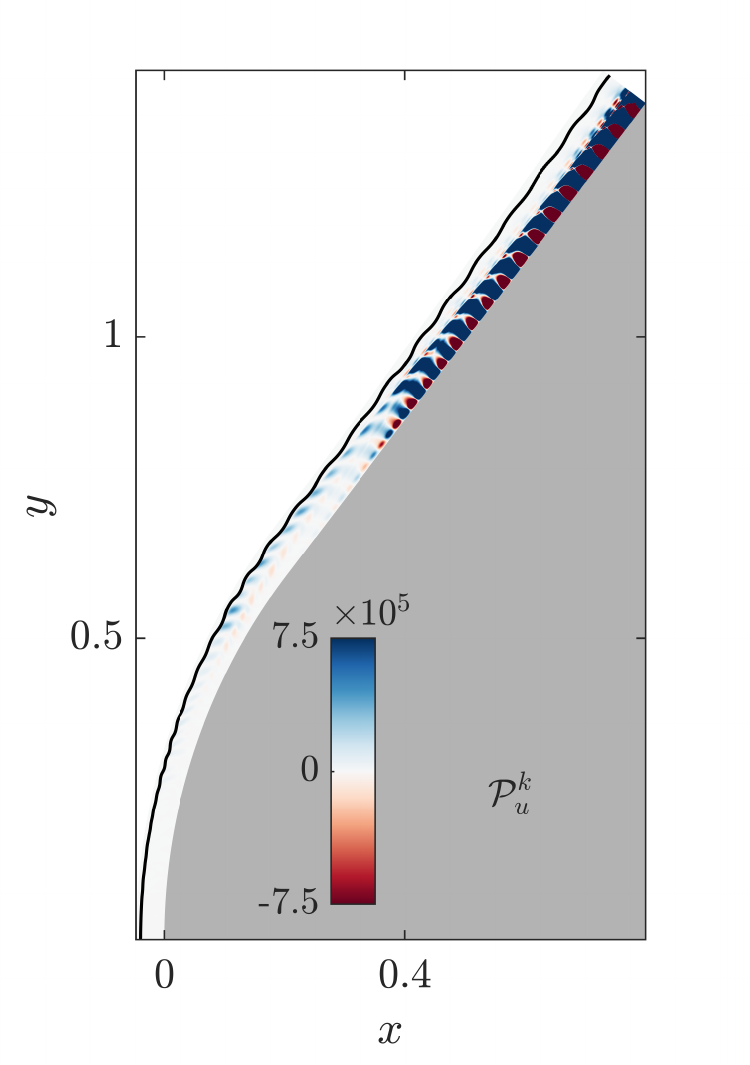}
\caption{}
\label{fig:freestream_budgets_plot_first_0}
\end{subfigure}
\begin{subfigure}[t]{0.49\textwidth}
\includegraphics[width=\textwidth]{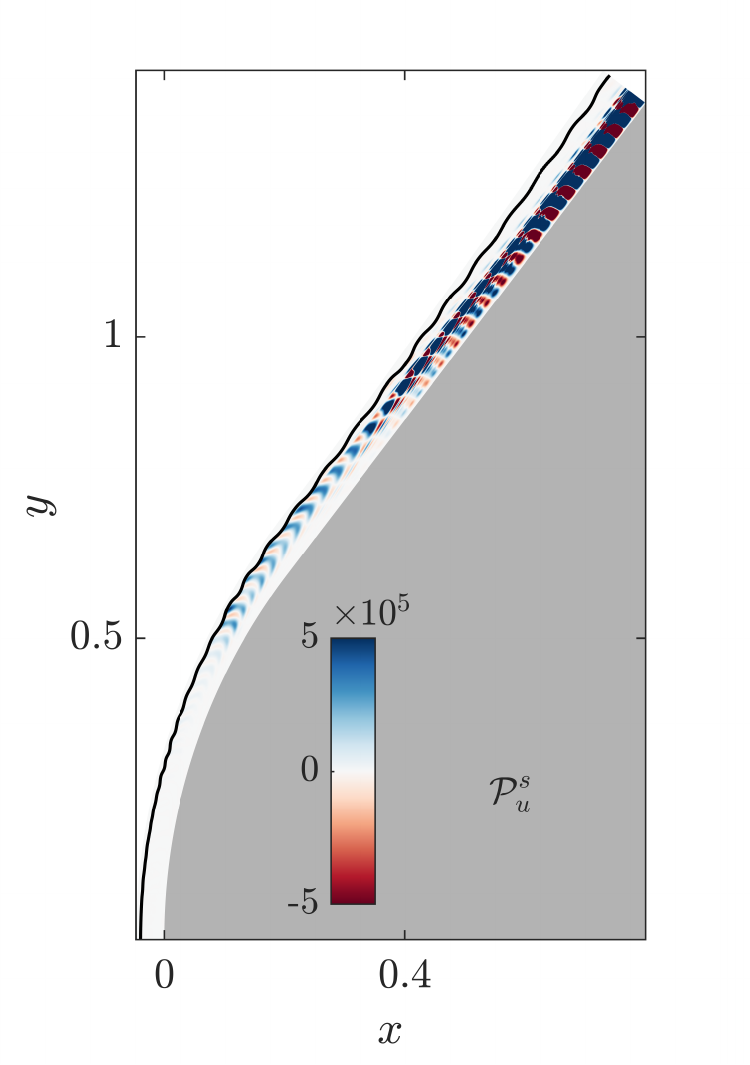}
\caption{}
\label{fig:freestream_budgets_plot_first_1}
\end{subfigure}
\caption{Contours of the dominant production terms in the kinetic and
  entropic budgets for the leading optimal freestream forcing mode,
  shown at the instant of maximum gain during one period. The base
  flow is computed using freestream conditions from
  table~\ref{tab:freestream} with $Re_\infty = 285\,000$ and $M_\infty
  = 28.7$. Both $\mathcal{P}^k_u$ and $\mathcal{P}^s_u$ are normalized
  by $\overline{\dot{E}_{\infty}}/V_D$. (\textit{a}) Reynolds-stress
  production term in the kinetic budget: $\mathcal{P}^k_u$.
  (\textit{b}) Entropic production term due to velocity fluctuations:
  $\mathcal{P}^s_u$.}
\label{fig:freestream_budgets_plot_first}
\end{figure}

The dominant local source of kinetic energy is the Reynolds-stress
production term,
\begin{equation*}
  \mathcal{P}^k_u
  = - \rho_0 \, u'_i u'_j \,\frac{\partial u_{i,0}}{\partial x_j}.
\end{equation*}
Positive $\mathcal{P}^k_u$ therefore corresponds to extraction of mean
kinetic energy by the fluctuations. The spatial distribution of this
term, shown in figure~\ref{fig:freestream_budgets_plot_first_0}, is
concentrated in the post-shock shear--entropy layer and intensifies as
the packets convect downstream. Near the wall, the term exhibits both
signs because the no-slip condition restricts the fluctuation field,
but the strongest positive contribution remains outside the boundary
layer, in the outer part of the post-shock layer.

\begin{figure}[]
\centering
\includegraphics[width=0.6\textwidth]{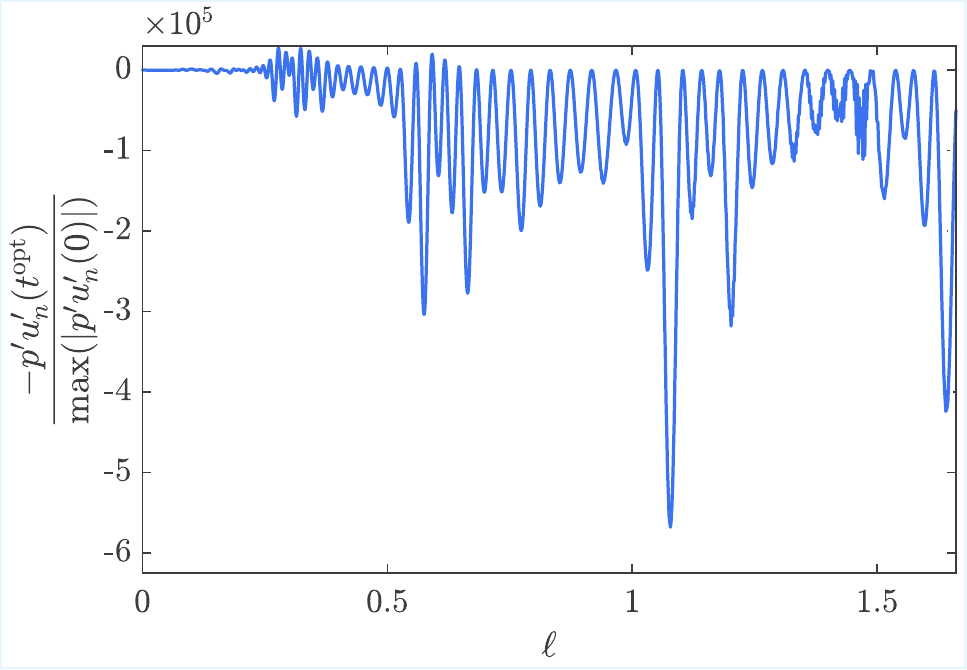}
\caption{Shock integrand of the pressure-work term, $-p' u'_n$,
  evaluated immediately downstream of the bow shock, where $u'_n$ is
  the disturbance velocity component normal to the shock. The
  horizontal axis is the nondimensional arc length $l/R$ measured
  along the shock from the axis of symmetry. The result is for the
  leading optimal freestream forcing mode at the instant of peak gain
  on the limit cycle $t^{\text{opt}}$. Base flow computed with
  freestream conditions from table~\ref{tab:freestream} at $Re_\infty
  = 285\,000$ and $M_\infty = 28.7$.}
\label{fig:freestream_budgets_plot_first_2}
\end{figure}

The first dominant negative contribution in the domain-integrated
kinetic-energy budget is the advective flux,
\begin{equation*}
  \mathcal{A}^k
  = - \rho_0 \, u_{j,0} \, \frac{\partial}{\partial x_j}
  \left( \tfrac{1}{2} u'_i u'_i \right),
\end{equation*}
whose negative peaks occur whenever a disturbance packet leaves the
computational domain. Under continuous freestream forcing, the
long-time response is periodic, so successive packets produce a
sequence of such peaks. The second dominant negative contribution is
the pressure-transport term,
\begin{equation*}
  \mathcal{T}^k_p
  = \frac{\partial}{\partial x_j}\left(-u'_j p'\right).
\end{equation*}
Because $\mathcal{T}^k_p$ is a divergence term, its domain integral
reduces to a boundary flux. The wall and the symmetry axis do not
contribute because $u'_n = 0$ there; thus, the dominant contribution
arises at the bow shock, with a smaller contribution at the outflow.
Figure~\ref{fig:freestream_budgets_plot_first_2} shows the shock
integrand, $-u'_n p'$, immediately downstream of the shock. With the
outward normal convention used in the domain integral, its
predominantly negative sign indicates that pressure transport removes
kinetic energy from the post-shock perturbation field through the
fitted-shock boundary. Conversely, the disturbance field performs a
substantial amount of work on the bow shock, causing it to become
corrugated. As discussed later, this introduces a strong feedback
mechanism. This effect is present because shock displacement is
retained in the linearization of the bow shock. In the fixed-shock
approximation adopted in the preliminary study of
\citet{anton2025bow}, this mechanism is suppressed.

The entropic budget has the same overall structure: production
dominates and is balanced mainly by advection. Its leading source term
is
\begin{equation*}
  \mathcal{P}^s_u = -\frac{(\gamma_0^*-1)p_0}{\gamma_0^*R_{g,0}^2}
  u_j' s'\frac{\partial s_0}{\partial x_j}.
\end{equation*}
This term shows that entropic amplification is driven by
advection of the base-flow entropy gradient by the velocity
fluctuations amplified in the shear--entropy
layer. Figure~\ref{fig:freestream_budgets_plot_first_1} confirms that
$\mathcal{P}^s_u$ is collocated with $\mathcal{P}^k_u$: wherever the
shear layer amplifies velocity fluctuations, the same fluctuations
also amplify the entropic component.  A key difference from the
kinetic budget is that the entropic budget has no comparably strong
shock-pressure-work sink. The transport term $\mathcal{T}^s$ and
diffusive dissipation are both small, so entropic energy continues to
increase until the packet is advected out of the domain.

The resulting picture is as follows: shock transmission produces an
optimal freestream disturbance dominated by acoustic and entropy
components of the Chu energy partition, and the post-shock
shear--entropy layer amplifies this response convectively. The
disturbances then impinge back on the bow shock, generating shock
corrugation. For the present optimal response, a separate classical
boundary-layer mode is not needed to account for the dominant computed
gain.

\subsection{Scaling analysis of the gain}
\label{sec::scaling}

We investigate the scaling laws governing the gain as a function of
$Re_\infty$, $M_\infty$, and $\gamma_2^*$. The analysis is divided
into two steps: the scaling of the shock-transmission gain and that of
the post-shock convective gain. We also investigate the nonlinear
feedback between post-shock disturbances and bow shock
corrugation. 

\subsubsection{Shock-transmission gain}
\label{sec::scaling_GS}

We first examine the dependence of the cycle-averaged,
shock-transmission gain $\overline{G}_{S}^{\mathrm{opt}}$ on $Re_\infty$
at fixed $M_\infty = 28.7$. The results are reported in
table~\ref{tab:Re_shock_post_shock_amplification}. The shock gain is
nearly independent of Reynolds number, with
$\overline{G}_{S}^{\mathrm{opt}} \approx 4 \times 10^2$ throughout the
range considered. This weak dependence is expected because the leading
amplification across the bow shock is governed by the linearized
Rankine--Hugoniot jump conditions, which are inviscid. At fixed
upstream thermochemical state and capsule geometry, Reynolds-number
effects can therefore enter $\overline{G}_{S}^{\mathrm{opt}}$ only through
small viscous changes to the base flow and shock shape.
\begin{table}
  \begin{center}
\def~{\hphantom{0}}
  \begin{tabular}{cccc}
      $Re_\infty$ & $\overline{G}_{S}^{\mathrm{opt}}$ & $\overline{G}_{D}^{\mathrm{opt}}$ & $\overline{G}_{T}^{\mathrm{opt}}$ \\[6pt] \midrule
       ~~10\,000 & 452 & ~9.64 & ~~~4 360\\
       ~~20\,000 & 450 & ~33.1 & ~~14 900\\
       ~~50\,000 & 439 & ~362 & ~159 000\\
       ~100\,000 & 428 & 1189 & ~509 000\\
       ~200\,000 & 422 & 1635 & ~690 000\\
       ~285\,000 & 422 & 1870 & ~790 000\\
       ~500\,000 & 421 & 2143 & ~902 000\\
  \end{tabular}
  \caption{Cycle-averaged optimal gain across the shock and within
    the post-shock flow at different Reynolds numbers. The values are
    computed after the response has reached its asymptotic periodic
    state. The reported gains correspond to the optimal freestream
    disturbance that maximizes $G_T^{\text{opt}}$ at
    $M_\infty = 28.7$.}
  \label{tab:Re_shock_post_shock_amplification}
  \end{center}
\end{table}

\begin{figure}[]
    \centering
    \begin{subfigure}[t]{0.49\textwidth}
    \includegraphics[width=\linewidth]{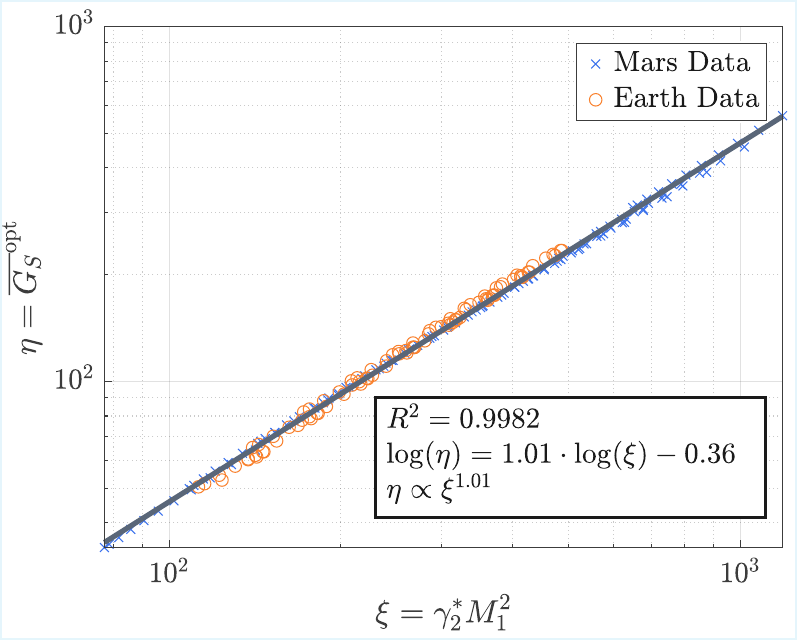}
    \caption{}
    \label{fig:shock_scalings}
    \end{subfigure}
    \begin{subfigure}[t]{0.49\textwidth}
    \includegraphics[width=\textwidth]{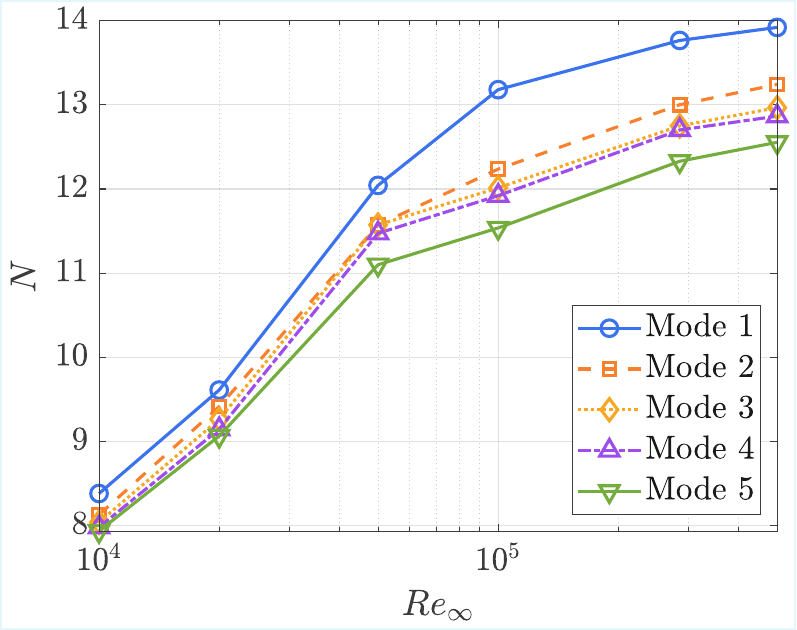}
    \caption{}
    \label{fig:energy_gain_Re}
\end{subfigure}
    \caption{ Energy gains of the optimal disturbance. (\textit{a})
      Optimal gain across the shock,
      $\overline{G}_{S}^{\mathrm{opt}}$, expressed in terms of the
      scaling parameters, with a linear fit overlaid. Freestream
      conditions are varied to probe the scaling of the shock
      gain. (\textit{b}) Energy-gain logarithm, $N=\ln
      G_{T,\max}^{\mathrm{opt}}$, as a function of $Re_\infty$
      with the remaining freestream conditions fixed to those of
      table~\ref{tab:freestream}.}
\end{figure}

The Mach-number scaling of $\overline{G}_{S}^{\mathrm{opt}}$ is
explained by classical LIA theory~\citep{mckenzie1968interaction}, in
which the ratio of post-shock disturbance energy flux, $\Delta
\dot{E}_{S}$, to the freestream disturbance energy flux,
$\dot{E}_\infty$, scales as $\Delta \dot{E}_{S}/\dot{E}_\infty =
\mathcal{O}(M_1^2)$ for high Mach numbers. One correction added
relative to the classical theory is the inclusion of the effective
$\gamma_2^*$, yielding the scaling $\overline{G}_{S}^{\mathrm{opt}}
\sim \gamma_2^* M_1^2$.  Figure~\ref{fig:shock_scalings} presents the
energy amplification across the shock for the optimal disturbances at
different flight conditions, compared with the theoretical scaling
$\gamma_2^* M_1^2$. The collapse of both the Mars and Earth data onto
this predicted scaling supports the dominant role of Mach number in
the shock-induced amplification of acoustic and entropic disturbances.
%
\subsubsection{Post-shock convective gain}
\label{sec::scaling_G_D}

The post-shock gain scaling is computed for a fixed capsule geometry
(see \S\ref{sec:problem-setup}).  As reported in
table~\ref{tab:Re_shock_post_shock_amplification}, the downstream gain
increases monotonically with $Re_\infty$, at a rate that decreases
with increasing Reynolds number. This behavior is made explicit in
figure~\ref{fig:energy_gain_Re}, which shows the energy-gain logarithm
of the maximum steady-state gain, defined as $N=\ln
G_{T,\max}^{\mathrm{opt}}$. The progressive weakening of the
$Re_\infty$ dependence suggests that the downstream amplification
approaches a high-Reynolds-number, convectively-dominated limit, in
which finite-$Re_\infty$ effects enter only as a viscous and thermal
diffusive correction to an otherwise inviscid growth process.

To study this observation, we construct a scaling argument for the
downstream gain that isolates the inviscid and viscous
contributions. The point of departure is the energy-budget analysis of
\S\ref{sec::budget}, which establishes two facts that are central to
what follows. First, the kinetic part of the Chu energy is the largest
single contribution to the optimal response. Second, within the
kinetic-energy budget the leading production mechanism is
Reynolds-stress work against the mean shear of the shock-induced
shear--entropy layer. Guided by these observations, we write a local
Chu-energy balance for a convecting wave packet in which inviscid
shear production is balanced by viscous dissipation,
\begin{equation}
\frac{1}{E}\frac{D E}{D t}
\;\sim\;
\chi\,\frac{\partial U_t}{\partial y_n}
\;-\;
\frac{k_{\rm eff}^{2}}{Re_\infty},
\label{eq:local_energy_scaling}
\end{equation}
where $E$ denotes a local Chu-energy density along the wave-packet
trajectory, $y_n$ is a cross-layer coordinate, and $U_t$ is the
tangential base-flow velocity. The coefficient $\chi=O(1)$ represents
the correlation between the velocity fluctuations and the mean shear,
and $k_{\rm eff}$ is the characteristic wavenumber of the downstream
packet. The first term in (\ref{eq:local_energy_scaling}) represents
kinetic production due to Reynolds stresses ($\mathcal{P}_u^k$),
whereas the second term represents viscous damping ($\mathcal{D}^k$).

The production term is controlled by the shear across the
shear--entropy layer $\partial U_t/\partial y_n \sim \Delta
U_t/\delta_s$, where $\delta_s$ is its characteristic thickness. For
the fixed capsule geometry and the family of freestream states
considered here, variations of $\Delta U_t$ are weaker than the
changes associated with the layer thickness and are absorbed into the
fitted constant below. In the hypersonic blunt-body limit, the
thickness of the shock-generated shear--entropy layer is taken to
scale with the shock standoff distance $d$. We therefore use
$\delta_s/R \sim d/R \sim (\rho_2/\rho_1)^{-1}$, where $\rho_1$ and
$\rho_2$ are the pre- and post-shock densities of the corresponding
normal-shock state. Thus larger compression produces a thinner shock
layer and a steeper shear--entropy layer. Integrating the production
part of (\ref{eq:local_energy_scaling}) over the residence time of the
packet therefore gives
\begin{equation}
\int
\chi
\frac{\partial U_t}{\partial y_n}\,dt
\sim
\frac{\rho_2/\rho_1}{C},
\label{eq:production_scaling}
\end{equation}
where $C$ is a positive constant containing the geometric factors
including shock obliquity and curvature, the residence length, the
convective speed, the proportionality between $\delta_s$ and $d$, and
the production-efficiency factor $\chi$.

The Reynolds-number correction follows from the dissipative term in
\eqref{eq:local_energy_scaling}. We model the characteristic
wavenumber selected by the optimal packet as $k_{\rm eff}\sim
Re_\infty^{1/4}$, where the exponent $1/4$ is chosen so that the
viscous correction follows the Reynolds-number dependence observed in
the computed optimal gains.  This scaling suggests that the global
input--output problem selects an intermediate disturbance scale, and
is reminiscent of mixed viscous--inviscid scale selections that arise
in other stability and receptivity analyses
\citep{mack1984boundary,schmid2012stability}.

Combining (\ref{eq:production_scaling}) and the viscous correction from 
(\ref{eq:local_energy_scaling}), the downstream amplification
factorizes into an inviscid and a viscous contribution,
\begin{equation}
\overline{G}_{D}^{\mathrm{opt}}
\;\sim\;
\overline{G}_{D,I}^{\mathrm{opt}}\,
\overline{G}_{D,V}^{\mathrm{opt}}
\;\sim\;
\exp\!\left[\frac{\rho_2/\rho_1}{C}\right]
\exp\!\left[-\frac{B}{\sqrt{Re_\infty}}\right],
\label{eq:GD_scaling_Re_density}
\end{equation}
where $C$ and $B$ are treated as constants for a fixed body geometry
and a fixed class of optimal disturbance paths. The first exponential
represents the inviscid convective production within the compressed
shear--entropy layer; the second encodes the leading viscous
correction.
\begin{figure}[]
    \centering
    \begin{subfigure}[t]{0.49\textwidth}
        \includegraphics[width=\linewidth]{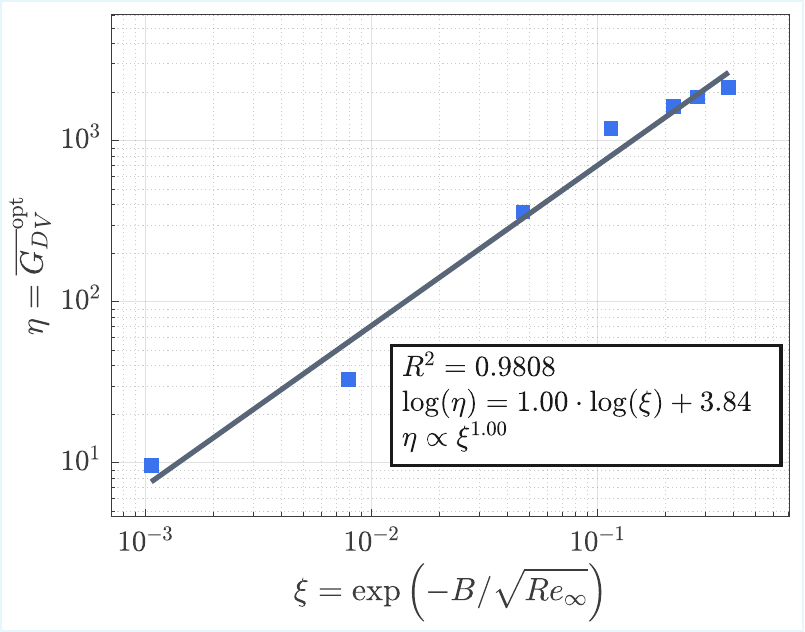}
        \caption{}
        \label{fig:downstream_viscous_scalings}
    \end{subfigure}
    \begin{subfigure}[t]{0.49\textwidth}
        \includegraphics[width=\linewidth]{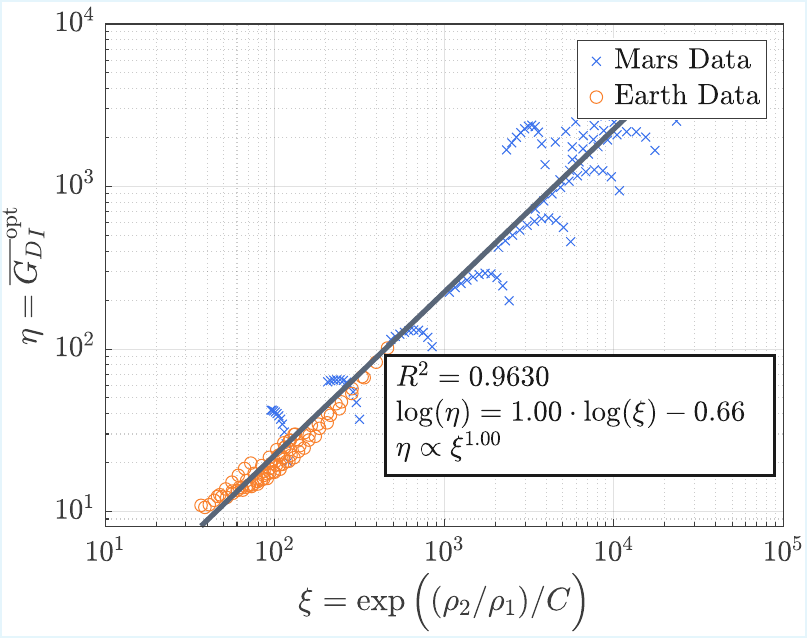}
        \caption{}
        \label{fig:downstream_inviscid_scalings}
    \end{subfigure}
    \caption{Determination of the viscous, $\overline{G}_{D,V}^{\mathrm{opt}}$,
      and inviscid, $\overline{G}_{D,I}^{\mathrm{opt}}$, contributions
      to the downstream optimal gain $\overline{G}_{D}^{\mathrm{opt}}$.
      Energy gains are plotted against the corresponding scaling
      parameters. Mars and Earth data are both obtained from the
      present stability-analysis computations; linear fits are
      superimposed. (\textit{a}) Viscous scaling
      $\overline{G}_{D,V}^{\mathrm{opt}}$: freestream conditions are
      fixed to those of table~\ref{tab:freestream}, and $Re_\infty$
      is varied; the best-fit constant is $B=685$. (\textit{b})
      Inviscid scaling $\overline{G}_{D,I}^{\mathrm{opt}}$: the
      Reynolds number is fixed at $Re_\infty = 285\,000$, while the
      freestream Mach number and atmospheric composition (Earth and
      Mars) are varied; the best-fit constant is $C=2.23$.}
    \label{fig:scalings_gains}
\end{figure}
%
Equation~(\ref{eq:GD_scaling_Re_density}) is tested in two
complementary sweeps designed to isolate each factor. We first fix the
freestream conditions of table~\ref{tab:freestream} and vary
$Re_\infty$ to determine $B$ within
$\overline{G}_{D,V}^{\mathrm{opt}}$. The results are shown in
figure~\ref{fig:downstream_viscous_scalings}.
%
We next fix $Re_\infty = 285\,000$ and vary the freestream Mach number
and the atmospheric composition (Earth and Mars), so that only the
inviscid factor $\overline{G}_{D,I}^{\mathrm{opt}}$ is isolated. The
resulting collapse, shown in
figure~\ref{fig:downstream_inviscid_scalings}, is reasonable given the
simplifying assumptions underlying (\ref{eq:production_scaling}), and
supports the interpretation of $\rho_2/\rho_1$ as the controlling
inviscid parameter.

The compression ratio itself carries a non-trivial Mach-number
dependence that is worth making explicit. For a calorically perfect
gas, the Rankine--Hugoniot relations give
\begin{equation}
\frac{\rho_2}{\rho_1}
= \frac{(\gamma+1) M_{n,1}^{2}}{(\gamma-1) M_{n,1}^{2} + 2},
\label{eq:comp_ratio}
\end{equation}
with $M_{n,1}$ the shock-normal Mach number, so that $\rho_2/\rho_1$
and hence $\overline{G}_{D,I}^{\mathrm{opt}}$ saturate at a finite
high-Mach-number limit. In the high-enthalpy regime relevant to
planetary entry, however, vibrational excitation and dissociation
reduce the effective post-shock specific-heat ratio and allow for
substantially larger compression ratios. The downstream gain can
therefore continue to grow with $M_\infty$ through the thermochemical
dependence of $\rho_2/\rho_1$ on the post-shock state. This mechanism
provides a direct route by which imperfect-gas effects enhance the
downstream amplification beyond the perfect-gas bound.

\begin{figure}[]
    \centering
    \includegraphics[width=0.5\linewidth]{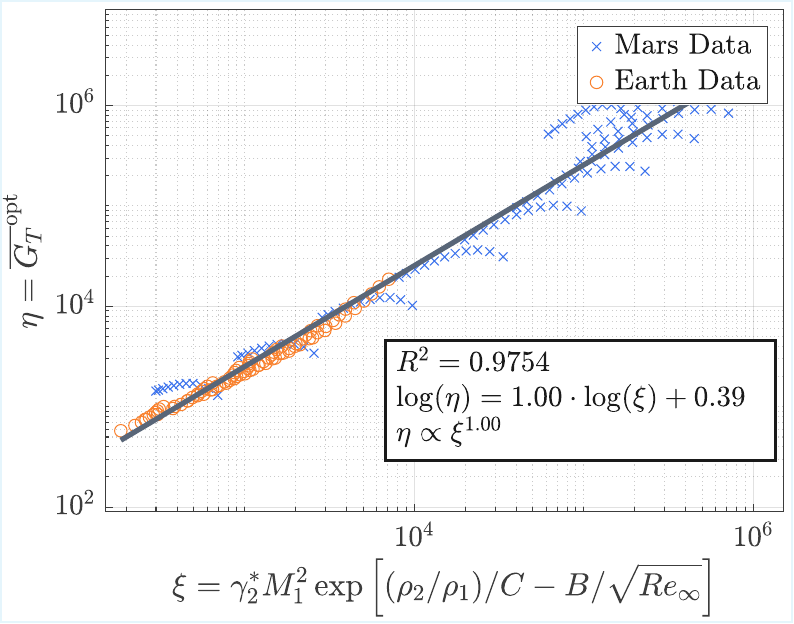}
    \caption{Total cycle-averaged optimal gain
      $\overline{G}_{T}^{\mathrm{opt}}$ expressed in terms of the
      scaling parameter in \eqref{eq:GT}. Both Mars and Earth data are
      obtained from the present receptivity-analysis computations. A
      linear fit is superimposed. The fitted constants are $C=2.23$
      and $B=685$.}
    \label{fig:total_scalings}
\end{figure}

Finally, combining the two stages of the amplification process yields
a total scaling factor of
\begin{equation}
\overline{G}_{T}^{\mathrm{opt}}
\sim
\gamma_2^* M_1^2\exp\left[
\frac{\rho_2/\rho_1}{C}
-
\frac{B}{\sqrt{Re_\infty}}
\right].
\label{eq:GT}
\end{equation}
Figure~\ref{fig:total_scalings} presents the cycle-averaged total gain
$\overline{G}_{T}^{\mathrm{opt}}$ obtained with the optimal freestream
disturbances, plotted against the right-hand side of \eqref{eq:GT}.
The constants fitted from the present numerical experiments are
$C=2.23$ and $B=685$. Despite the simplifying assumptions used in the
derivation, the data collapse is reasonable. This agreement supports
the interpretation that shock transmission and shear-driven production
are the dominant amplification mechanisms in this regime.

\subsection{Feedback via shock corrugation}
\label{sec::corrugation_feedback_effect}

The gains discussed above quantify the open-loop, linear amplification
of a prescribed freestream disturbance. By construction, this approach
cannot account for the finite-amplitude feedback through which the
amplified downstream disturbance modifies the shock profile itself. In
this section, we show that the feedback loop:
$$ \text{shear--entropy layer} \to \text{bow-shock corrugation} \to
\text{shear--entropy layer},$$
reinforces the instability over short times.

The feedback mechanism can be understood in terms of the vorticity
deposited by the curved bow shock. For a uniform upstream flow, the
magnitude of the out-of-plane vorticity generated immediately behind a
curved shock has the strong-shock scaling~\citep{hornung1998gradients}
\begin{equation}
    \frac{|\omega_{2,z}|R}{U_\infty}
    \simeq
    |\cos\beta|\,|\kappa|R
    \left(1-\frac{\rho_1}{\rho_2}\right)^{\!2}
    \frac{\rho_2}{\rho_1},
    \label{eq:shock_vorticity}
\end{equation}
where $z$ is the direction normal to the meridional plane, $\kappa$ is
the local shock curvature, and $\beta$ is the angle between the local
shock tangent and the freestream direction, so that $\beta=\pi/2$ for
a locally normal shock. This deposited vorticity is then advected
downstream and forms the post-shock shear--entropy layer in which the
dominant Reynolds-stress production term, $\mathcal{P}^k_u \sim
\partial U_t / \partial y_n \sim |\omega_{2,z}|$, amplifies the
disturbance energy that contributes to $G_D$. This estimate is
consistent with the baseline scaling from the previous section, where
the same shear was approximated as $\Delta U_t/\delta_s$. Here,
$|\omega_{2,z}|$ is used to retain the additional dependence on shock
corrugation.

At that point, the feedback process can be decomposed into three
stages:
\begin{itemize}
\item Perturbations transmitted through the bow shock enter the
  shear--entropy layer and are amplified as they convect downstream.
  Their pressure component, $p'$, impinges on the bow shock and
  produces a finite corrugation. The level of corrugation is
  especially high for bow shocks with a small standoff distance, as is
  typical of the large post-shock compression ratios encountered under
  the high-enthalpy Martian-entry conditions considered here.

\item The corrugated shock modifies the local shock-jump conditions.
  Because the vorticity deposited by a curved shock depends on both
  the local curvature and the local shock angle, even a modest
  corrugation changes the vorticity injected into the post-shock
  region. The additional vorticity advects downstream and reinforces
  the shear--entropy layer, strengthening the downstream amplification
  of disturbances.

\item The amplified shear--entropy-layer fluctuations generate a
  larger pressure imprint on the shock, which increases the shock
  corrugation even more and the loop closes. From the scaling analysis
  of $G_D$ above, for which $\partial U_t / \partial y_n \sim
  |\omega_{2,z}|$, and assuming that the vorticity is simply advected
  and conserved along streamlines, we expect $$\Delta N = \ln\left(
  \frac{G_D^{\text{feedback}}} {G_D } \right) = \mathcal{O}\left(
  \frac{|\omega_{2,z}|^{\text{corrugated}}}
       {|\omega_{2,z}|}-1\right),$$ where $G_D^{\text{feedback}}$ is
       the feedback-modified gain after one advection time of the
       disturbances.
\end{itemize}

\begin{figure}
    \centering
    \includegraphics[width=0.5\linewidth]{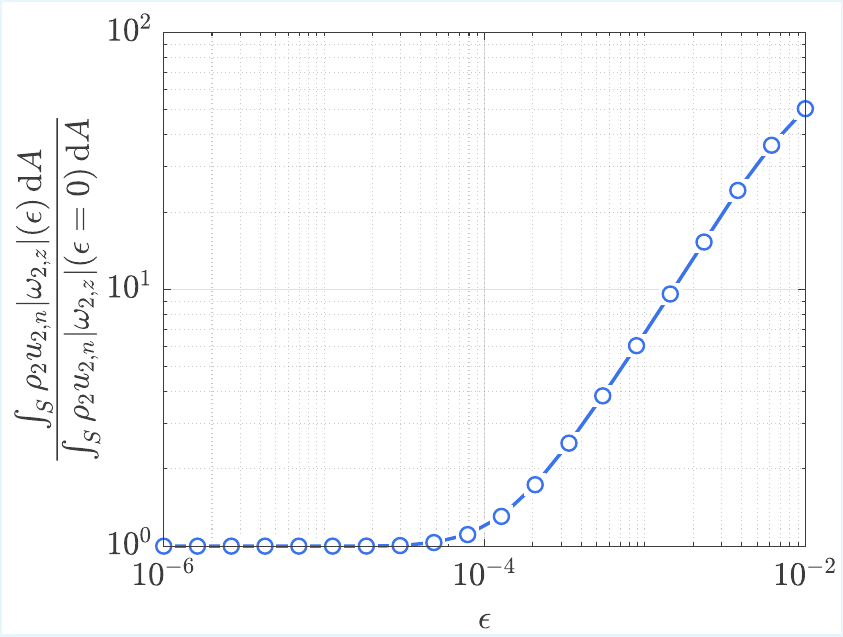}
    \caption{Increase in the absolute vorticity flux across the bow shock
      induced by finite-amplitude shock corrugation. The plotted ratio
      compares the vorticity injection of the baseline shock, computed
      with the freestream conditions from table~\ref{tab:freestream},
      with that obtained after imposing a finite shock disturbance of
      amplitude $\epsilon$. The imposed shock disturbance has the shape
      of the corrugation induced by the leading optimal freestream mode
      in \S\ref{sec::Optimal_disturbance}.}
    \label{fig:corrugation_shock}
\end{figure}

To assess the strength of the feedback, we perturb the bow shock using
the optimal shock-displacement shape obtained in
Section~\ref{sec::Optimal_disturbance}, scaled to a finite amplitude
$\epsilon$, and recompute the post-shock azimuthal vorticity injection for the
resulting corrugated-shock geometry, denoted by
$J_\omega(\epsilon) = \int_S \rho_2 u_{2,n} |\omega_{2,z}| dA$. Figure~\ref{fig:corrugation_shock} shows the
variation in the flux of absolute vorticity across the shock as a
function of $\epsilon$. If $J_\omega$ is taken as an integral proxy
for the vorticity supplied to the shear layer, such that
$|\omega_{2,z}(\epsilon)| / |\omega_{2,z}(0)|  \sim J_\omega(\epsilon) / J_\omega(0) $, the results
indicate that, even for corrugation amplitudes as small as $1\%$ of
the capsule radius, the vorticity content of the mean shear increases
by a factor of $50$. This corresponds to $\Delta N = \ln \left(
G_D^{\text{feedback}} / G_D \right) \approx \mathcal{O}(10)$, whose
large magnitude suggests a strong coupling induced by bow-shock
corrugation.

The implications are twofold. First, the linear gains reported in
Section~\ref{sec::Optimal_disturbance} should be regarded as open-loop
gains: once the disturbance reaches an amplitude at which the induced
shock corrugation is no longer negligible, the production term
$\mathcal{P}^k_u$ in the kinetic budget acquires an additional
self-sustaining contribution that is absent from the linearized
operator. Second, the feedback loop provides a plausible route to a
secondary instability and to an accelerated transition scenario
through the breakdown of the bow shock itself, before transition
occurs in either the boundary layer or the shear--entropy layer.

\subsection{Velocity--altitude energy-gain maps for EDL vehicles}
\label{sec::path}

We next map the energy-gain logarithm $N$ over velocity--altitude
space for Mars and Earth entry conditions. Motivated by the scaling in
\eqref{eq:GT}, we assume that the Reynolds numbers in this range are
large enough that variations in $Re_\infty$ produce only secondary
corrections to $N$. Accordingly, all computations in this map are
performed at $Re_\infty=285\,000$. The velocity--altitude sweeps
assume chemical and thermal equilibrium. Nonequilibrium effects are
expected where the chemical or vibrational relaxation times become
comparable to the shock-layer residence time; however, as discussed in
Appendix~\ref{sec::validity_equilibrium}, the largest amplification
factors occur in the region where the equilibrium approximation is
justified.

\begin{figure}[]
\centering
\begin{subfigure}[t]{0.49\textwidth}
\includegraphics[width=\textwidth]{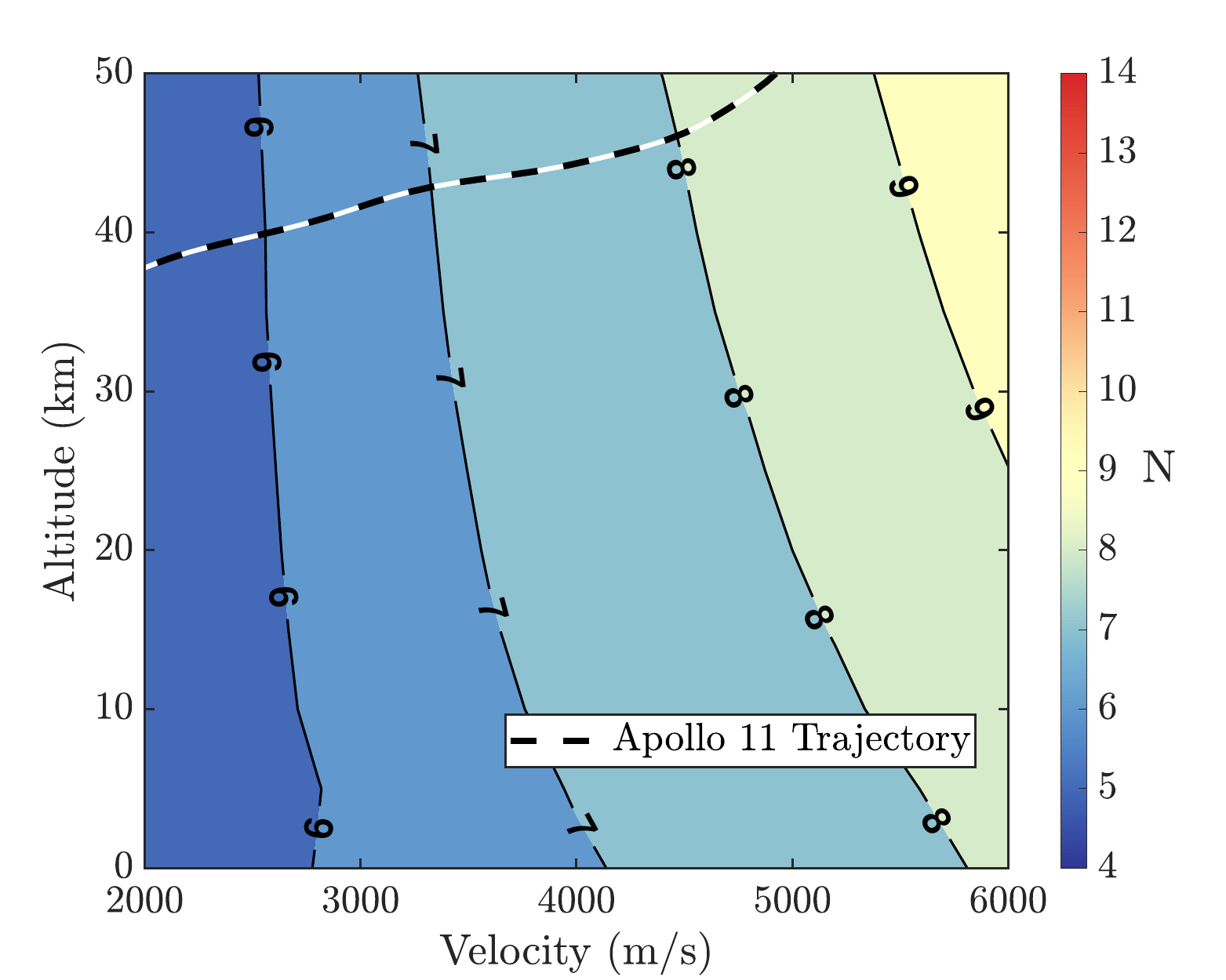}
\caption{}
\label{fig:h_vs_velocity_Earth}
\end{subfigure}
\begin{subfigure}[t]{0.49\textwidth}
\includegraphics[width=\textwidth]{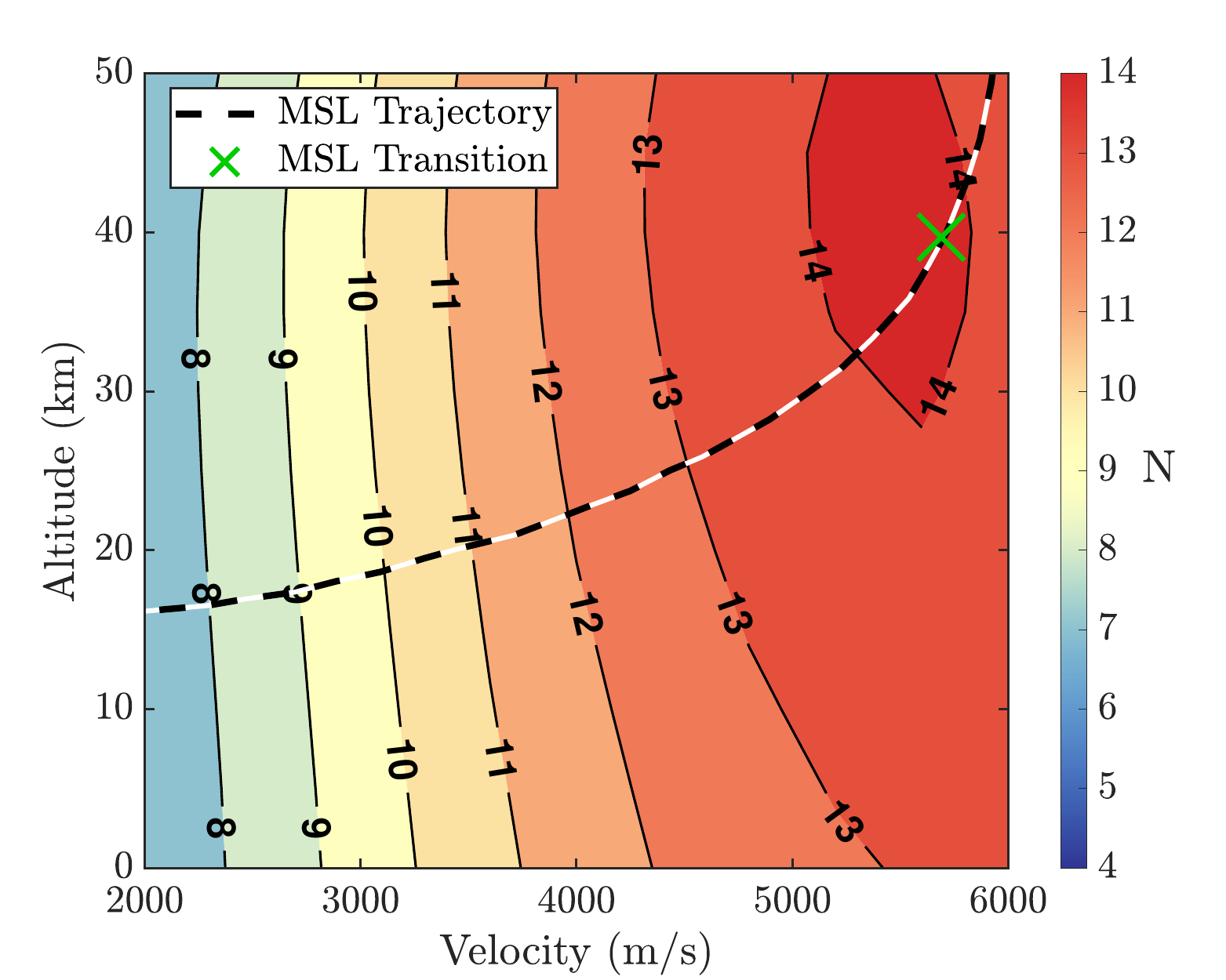}
\caption{}
\label{fig:h_vs_velocity_Mars}
\end{subfigure}
\caption{Energy-gain logarithm
  $N=\ln G_{T,\max}^{\mathrm{opt}}$ over a range of velocities
  and altitudes. The Reynolds number is fixed at $Re_\infty =
  285\,000$, and all computations assume chemical and thermal
  equilibrium. For reference, the figure also shows the entry
  trajectories of the MSL (Mars entry,
  \citet{edquist2007aerothermodynamic}) and Apollo~11 (Earth entry,
  \citet{manders1970apollo11}) capsules. The transition point inferred
  from the MSL flight reconstruction \citep{bose2014reconstruction} is
  also indicated by a green cross. (\textit{a}) Earth atmosphere.
  (\textit{b}) Mars atmosphere.}
\label{fig:h_vs_velocity}
\end{figure}

Figure~\ref{fig:h_vs_velocity} presents $N$ over velocity--altitude
space for both Earth and Mars atmospheres. For reference, the figure
also depicts the entry trajectories of the MSL (Mars entry) and
Apollo~11 (Earth entry) capsules. The energy-gain logarithms for
re-entry in the Earth atmosphere (figure~\ref{fig:h_vs_velocity_Earth})
are clearly smaller than those for the Mars atmosphere
(figure~\ref{fig:h_vs_velocity_Mars}). The reduced disturbance
amplification in Earth's atmosphere, relative to Mars, can be
attributed to two factors. First, the density ratio across the shock is
smaller in Earth's atmosphere because diatomic nitrogen and oxygen
store less vibrational energy than the triatomic carbon dioxide that
dominates the Martian atmosphere, resulting in lower compression
ratios. Second, for the atmospheric states considered here, the lower
molecular weight of Earth's atmosphere and its thermodynamic state give
larger sound speeds than in the corresponding Martian cases, leading
to lower Mach numbers at comparable entry velocities. These combined
effects result in significantly smaller values of the scaling
parameter from \eqref{eq:GT} for Earth entry conditions.

The results for the Martian atmosphere also reveal a pronounced,
localized region of peak amplification at high velocities and
altitudes. Remarkably, the MSL entry trajectory passes directly
through this critical region. The transition location inferred from the
MSL flight aeroheating reconstruction \citep{bose2014reconstruction},
which occurred on the leeside at approximately $64\,\mathrm{s}$ after
atmospheric interface, coincides with the region of maximum
amplification. A similar observation applies to the
Mars~2020/Perseverance mission, which used a geometrically similar
aeroshell and followed a comparable entry profile. The reconstructed
MEDLI2 aeroheating data indicate transition-associated heat-flux
augmentation in the same region of velocity--altitude space
characterized here by elevated amplification factors
\citep{edquist2022mars,alpert2022inverse}.

\subsection{Nonlinear WMLES of the MSL configuration}
\label{sec::non_linear_effects}

To test whether the amplification mechanism identified by the linear
receptivity analysis can proceed to nonlinear breakdown, wall-modeled
large-eddy simulations (WMLES) were performed for the MSL entry
vehicle under a representative Mars-entry condition. The purpose of
the WMLES is not to replace the receptivity analysis, but to determine
whether a three-dimensional nonlinear flow develops the same
shock/shear--entropy-layer instability and the same qualitative
heating footprint.  The capsule geometry was taken from
\citet{edquist2007aerothermodynamic}, and the vehicle was simulated at
an angle of attack $\alpha=17.3^\circ$, characteristic of the flight
attitude near the transition event. This angle is also consistent with
the effective angle of attack used in the receptivity analysis. The
freestream state is that of table~\ref{tab:freestream}, which
corresponds to the region of the MSL trajectory in which the
reconstructed aerothermal data indicate a strong increase in leeside
heating \citep{bose2014reconstruction}.  Similar leeside heating
augmentation was also inferred for the geometrically related
Mars~2020/Perseverance entry capsule \citep{edquist2022mars,
  alpert2022inverse}.

The simulations were carried out with the charLES solver using a
second-order finite-volume discretization with entropy-preserving
properties. The unresolved turbulent stresses were modeled with the
Vreman subgrid-scale model~\citep{vreman2004eddy}, coupled to an
equilibrium, algebraic wall model~\citep{kawai2012wall}.  The wall was
modeled as isothermal, with $T_w=700\,\mathrm{K}$, representative of
reconstructed heat-shield temperatures near the transition region
during MSL entry~\citep{bose2014reconstruction}. The nondimensional
transport coefficients were prescribed by the power laws
$\mu^*=(T/T_\infty)^{3/4}$ and $k^*=(T/T_\infty)^{3/4}$.

The gas was modeled as a calorically perfect ideal gas with a
prescribed constant effective specific-heat ratio $\gamma^*$. This
closure is a simplification with respect to the receptivity analysis,
as a constant $\gamma^*$ cannot reproduce the spatially varying
composition and temperature-dependent thermodynamics of an equilibrium
Mars-entry shock layer. Instead, $\gamma^*$ is used below as a
controlled parameter that changes the post-shock compressibility and
density ratio. The rationale for varying $\gamma^*$ follows from the
gain scaling derived in \S\ref{sec::scaling}. At fixed geometry and
Reynolds number, the leading inviscid part of the
freestream-to-shock-layer amplification scales as $\gamma_2^* M_1^2
\exp[(\rho_2/\rho_1)/C]$, up to the viscous correction in
\eqref{eq:GT}. For a calorically perfect gas, the normal-shock density
ratio from \eqref{eq:comp_ratio} shows that
$\rho_2/\rho_1\to(\gamma^*+1)/(\gamma^*-1)$ in the strong-shock limit
$M_{n,1}\gg1$ \citep{batchelor2000introduction}. Thus, lowering
$\gamma^*$ increases the post-shock compression ratio and, through the
exponential factor in the gain scaling, increases the downstream
convective amplification. In the strong-shock approximation, the
corresponding inviscid control parameter is
\[
  \Pi(\gamma^*)
  \equiv
  \gamma^* M_1^2
  \exp\left[
    \frac{1}{C}
    \frac{\gamma^*+1}{\gamma^*-1}
  \right].
\]
Using the value $C=2.23$ obtained in \S\ref{sec::scaling}, decreasing
$\gamma^*$ from $1.25$ to $1.20$ increases $\Pi$ by a factor of
approximately $2.3$ at fixed $M_1$. The two constant-$\gamma^*$ cases
therefore provide a controlled nonlinear test of the same compression-
and curvature-driven amplification mechanism identified by the linear
analysis. For reference, the equilibrium-gas computations discussed in
the receptivity analysis yield spatially varying post-shock values of
$\gamma^*$, with values as low as approximately $1.08$ for Mars-entry
conditions and $1.17$ for Earth-entry conditions over the range
considered here.
\begin{figure}[]
    \centering
    \includegraphics[width=0.85\linewidth]{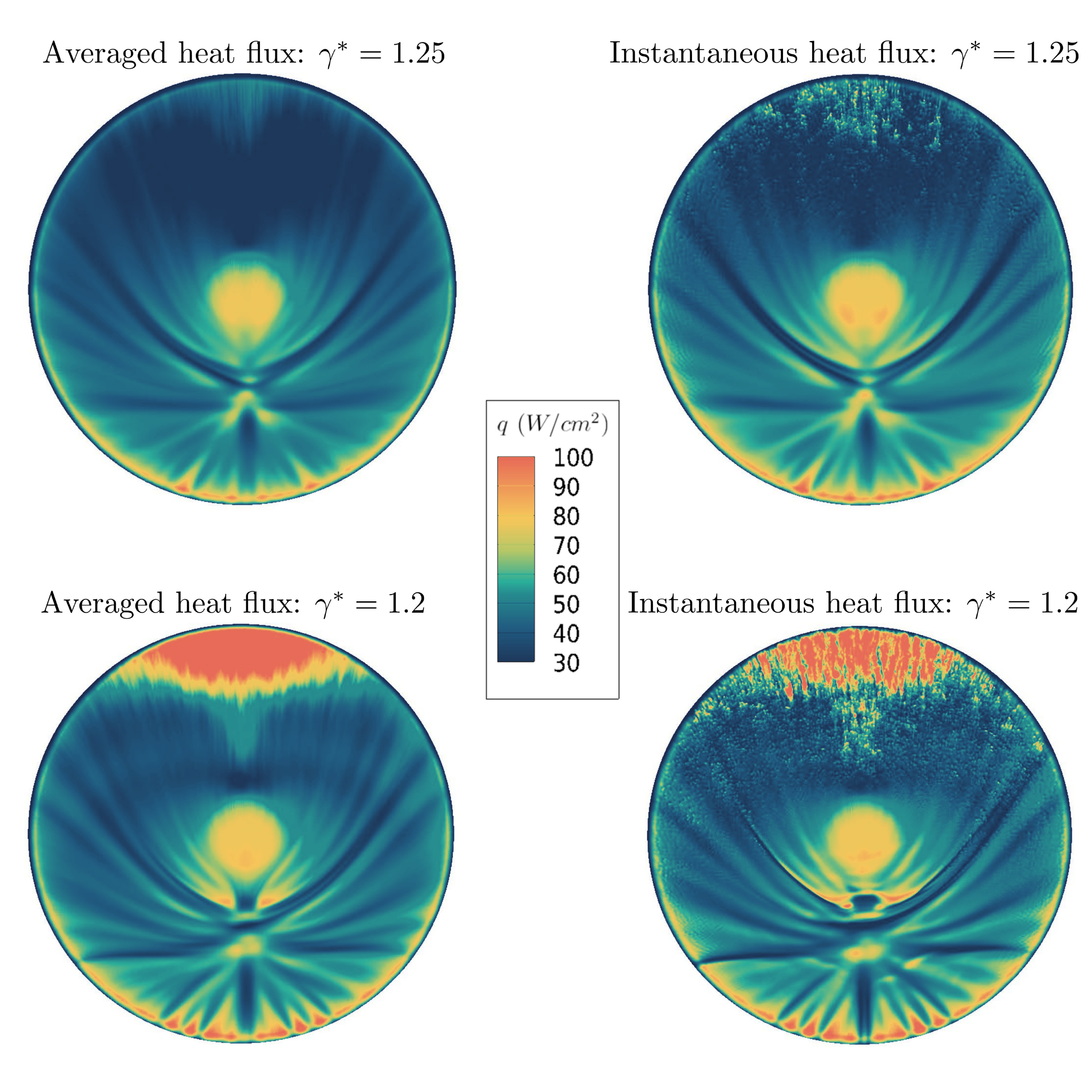}
    \caption{Time-averaged and instantaneous wall heat flux on the MSL
      heat shield obtained from WMLES for two values of the constant
      effective specific-heat ratio $\gamma^*$.  Freestream conditions
      from table~\ref{tab:freestream} for a calorically perfect ideal
      gas at $\alpha=17.3^\circ$. The leeside corresponds to the upper
      half of the capsule surface.}
    \label{fig:heat_flux_MSL}
\end{figure}

The sensitivity of the nonlinear flow to $\gamma^*$ is shown in
figure~\ref{fig:heat_flux_MSL}. For $\gamma^*=1.25$, the heat-flux
field contains weak streaky modulations near the downstream portion of
the leeside, but no large localized heat-flux spike develops. By
contrast, the $\gamma^*=1.20$ case exhibits a strong, localized
increase in leeside heat flux, visible in both the instantaneous and
time-averaged fields. No deterministic freestream disturbance was
prescribed in these simulations. The initial disturbance amplitude is
supplied by the small numerical, grid-scale, and round-off
perturbations present in the computation. Consequently, the WMLES
should be interpreted as evidence that the flow is nonlinearly
susceptible to this instability route, rather than as a calibrated
prediction of a transition threshold for a specified freestream
disturbance environment.

The region of strongest computed heating lies on the leeside shoulder,
consistent with the surface region where post-flight MSL
reconstructions inferred the largest transition-associated heating
augmentation~\citep{bose2014reconstruction}. The comparison with
flight data should nevertheless be interpreted
qualitatively rather than as a direct reconstruction of the measured
heat-flux histories. The constant-$\gamma^*$ perfect-gas model, the
simplified transport closure, the wall model, and the omission of
ablation, pyrolysis, radiation, and finite-rate thermochemistry
preclude a one-to-one comparison of the absolute heat-flux magnitude.
The relevant observation concerns the underlying mechanism:
when the compression-controlled amplification is increased, the WMLES
produces a localized nonlinear breakdown on the leeside and a
corresponding localized heating footprint, consistent with the leeside
transition inferred from MSL and Mars~2020/Perseverance flight data
\citep{bose2014reconstruction, edquist2022mars, alpert2022inverse}.

\begin{figure}[]
\centering
\begin{subfigure}[t]{0.33\textwidth}
\includegraphics[width=\textwidth]{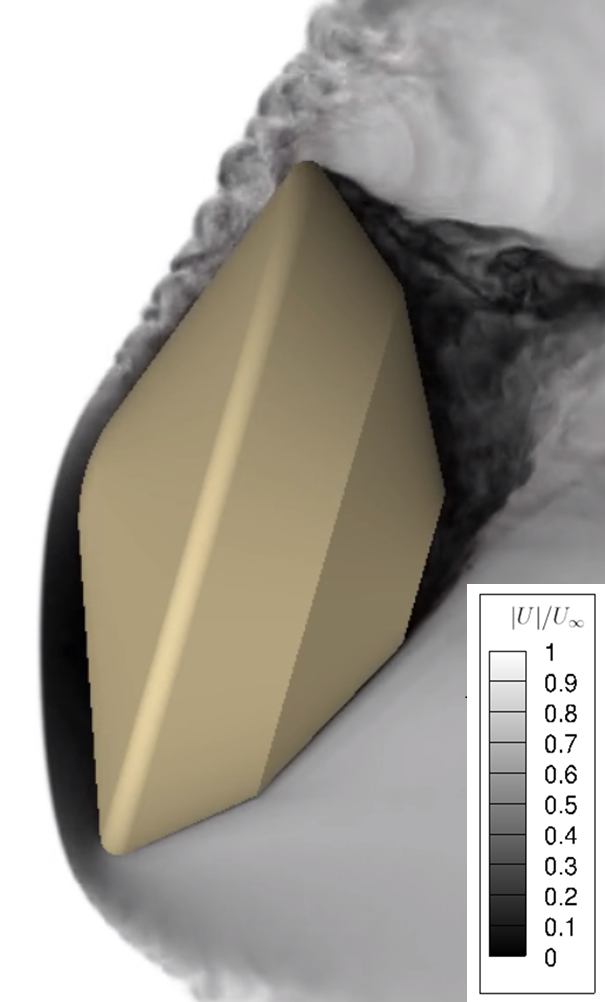}
\caption{}
\label{fig:velocity_WMLES}
\end{subfigure}
\begin{subfigure}[t]{0.58\textwidth}
\includegraphics[width=\textwidth]{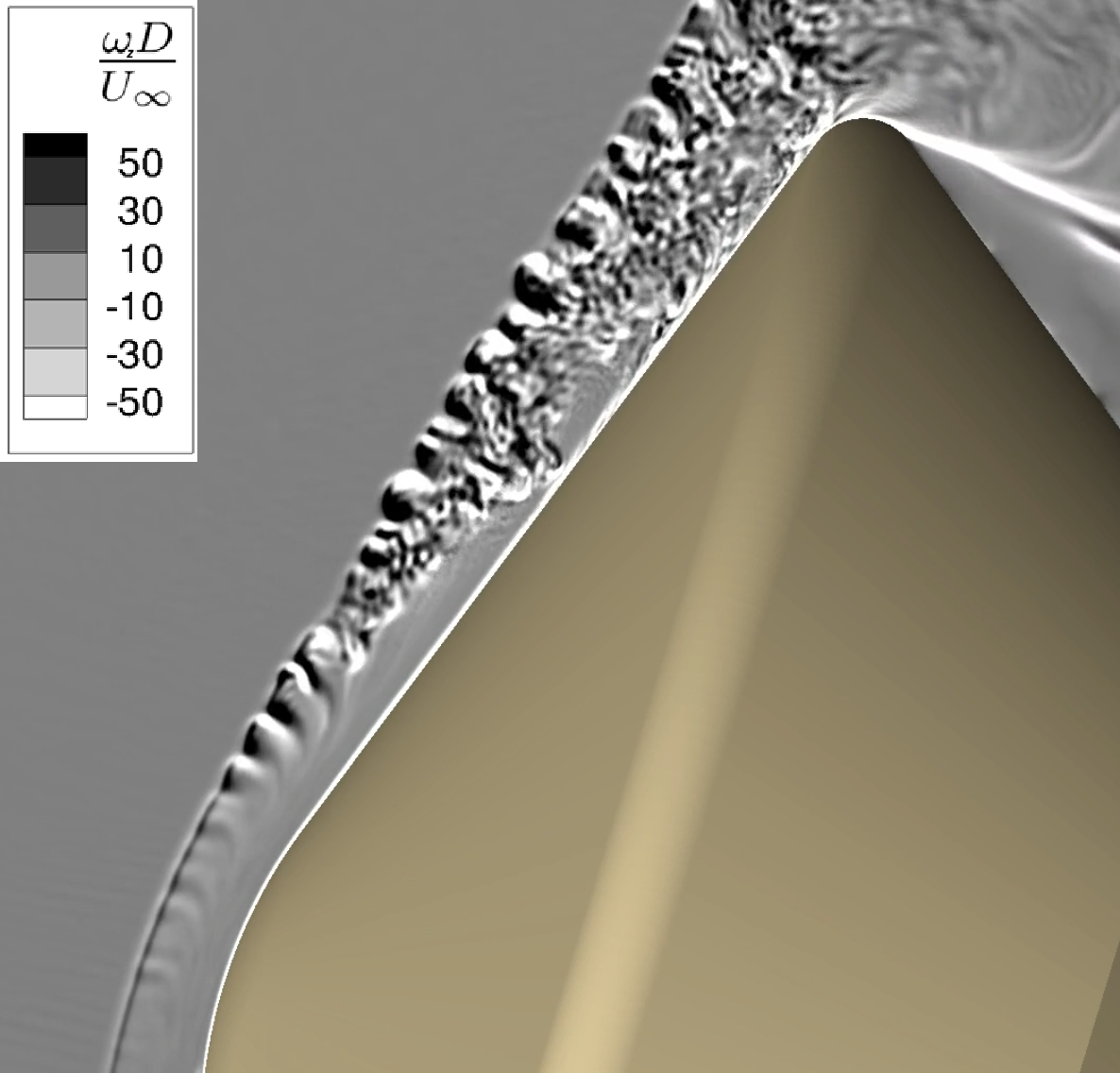}
\caption{}
\label{fig:vort_WMLES}
\end{subfigure}
\caption{WMLES of the MSL configuration at a representative point on
  the entry trajectory
  \citep{edquist2007aerothermodynamic}. Freestream conditions from
  table~\ref{tab:freestream} for a calorically perfect ideal gas with
  $\gamma^*=1.20$ at $\alpha=17.3^\circ$. (\textit{a}) Velocity
  magnitude. (\textit{b}) Out-of-plane vorticity component. The view
  is focused on the leeside of the MSL heat shield.}
\label{fig:fields_WMLES}
\end{figure}

The instantaneous WMLES fields in figure~\ref{fig:fields_WMLES}
clarify the mechanism producing the elevated leeside heat flux. The
velocity magnitude (figure~\ref{fig:velocity_WMLES}) and out-of-plane
vorticity component (figure~\ref{fig:vort_WMLES}) show coherent
disturbances forming in the shear--entropy layer immediately behind
the curved bow shock. These disturbances amplify as they convect
downstream, roll up, and eventually break down. The resulting turbulent
mixing increases wall-normal transport of high-enthalpy fluid toward
the heat shield, producing the localized heat-flux rise. Thus the
dominant nonlinear route is the breakdown of the shock-generated
shear--entropy layer, rather than the independent onset of a classical
boundary-layer instability at the wall.

The same mechanism also explains why the strongest transition
signature is localized on the leeside. In both the WMLES and the
flight reconstructions, no comparable heat-flux spike appears on the
windward side. The asymmetry can be interpreted using the curved-shock
vorticity scaling in \eqref{eq:shock_vorticity}, which shows that
vorticity deposition increases with shock curvature, shock obliquity,
and compression ratio.  At this angle of attack, the leeside branch of
the bow shock is more oblique and more strongly curved than the
windward branch. The windward shock, by contrast, is closer to normal
incidence and has smaller curvature. The leeside shock therefore
generates a stronger shear--entropy layer, providing the energetic
base state on which the amplification mechanism identified in
\S\ref{sec::scaling} acts. This interpretation is also consistent with
the high-density-ratio shock-layer instability picture of
\citet{hornung2001shock}.

To connect the nonlinear calculation with the linear theory, a
freestream-receptivity calculation was performed at the same
freestream state and with the same constant-$\gamma^*$ gas model. The
linear solver used in the preceding sections assumes an axisymmetric
base flow, and hence cannot represent the full three-dimensional MSL
flow at $\alpha=17.3^\circ$. As discussed in
\S\ref{sec:problem-setup}, an effective angle construction is used to
match the local forebody inclination seen by the leeside flow,
yielding a comparable shock obliquity and curvature along that branch,
the two geometric factors entering \eqref{eq:shock_vorticity}. It does
not reproduce the azimuthal curvature or the full three-dimensional
shock shape, and the comparison should be understood as a test of the
local instability mechanism rather than as a one-to-one reproduction
of the complete MSL flow.

\begin{figure}[]
\centering
\begin{subfigure}[t]{0.50\textwidth}
\includegraphics[width=\textwidth]{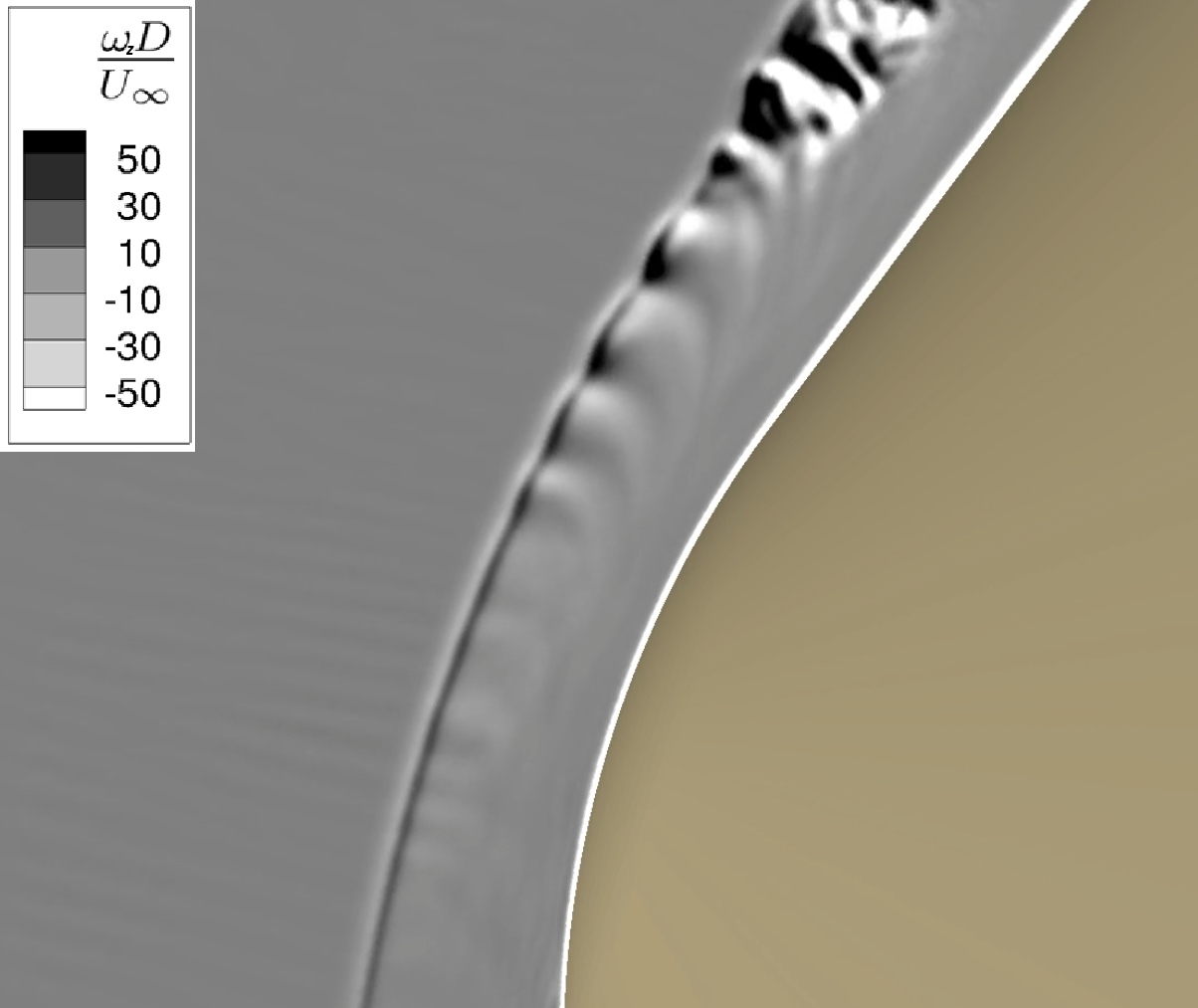}
\caption{}
\label{fig:WMLES_zoom_vort}
\end{subfigure}
\begin{subfigure}[t]{0.245\textwidth}
\includegraphics[width=\textwidth]{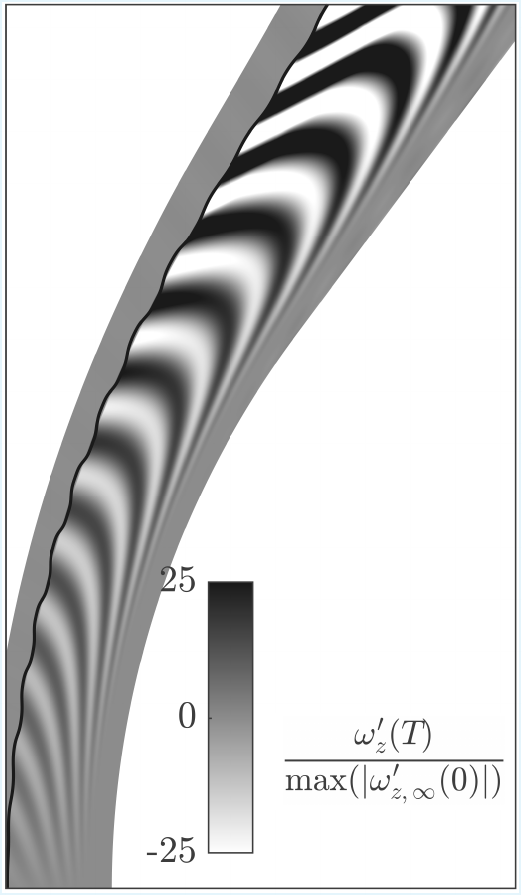}
\caption{}
\label{fig:linear_vort}
\end{subfigure}
\caption{Out-of-plane vorticity component in the incipient region of
  the instability.  Comparison between the WMLES field and the leading
  freestream-receptivity response.  Both calculations use the
  freestream conditions from table~\ref{tab:freestream} and a
  calorically perfect ideal gas with $\gamma^*=1.20$. The WMLES uses
  the actual MSL geometry at $\alpha=17.3^\circ$; the linear
  receptivity calculation uses an axisymmetric blunt cone with an
  effective half-angle of $\theta=52.7^\circ$. (\textit{a}) WMLES.
  (\textit{b}) Linear freestream-receptivity response.}
\label{fig:linear_WMLES_comparison}
\end{figure}

Figure~\ref{fig:linear_WMLES_comparison} compares the out-of-plane
vorticity from the early WMLES field with the leading linear
freestream-receptivity response. The corresponding cycle-averaged gain
is $\overline{G}_{T}^{\mathrm{opt}}=4.92\times10^4$ for
freestream receptivity. The leading linear response displays vortical
structures in the same shock/shear--entropy-layer region as the
incipient WMLES disturbance, before nonlinear saturation dominates.

Both approaches exhibit the same sequence of events: disturbances are
generated in the region where the bow shock is curved, they are
amplified while convecting through the shear--entropy layer, and they
corrugate the shock as the pressure field couples back to the shock
motion. Remarkably, the nonlinear WMLES displays the feedback
mechanism anticipated from the analysis in
\S\ref{sec::corrugation_feedback_effect}. Once the bow shock develops
appreciable corrugation, the additional curvature visible in
figure~\ref{fig:WMLES_zoom_vort} injects further vorticity into the
post-shock region. These streaks reinforce the local mean shear and
feed the production term $\mathcal{P}^k_u$ identified in
\S\ref{sec::budget}. This closes a finite-amplitude loop between shock
corrugation, modified vorticity injection, mean-shear amplification,
and disturbance growth.  The WMLES result suggests that, beyond the
linear regime, this self-reinforcing cycle can contribute to the rapid
breakdown observed in the simulation and is a plausible candidate for
the secondary instability pathway that precedes transition in this
configuration.

\section{Discussion}

The heat-flux rise interpreted as leeside transition during the MSL
entry occurred near $Re_\infty = 2.85\times 10^{5}$
\citep{bose2014reconstruction,edquist2007aerothermodynamic}, well
before the trajectory reached its maximum nose-radius Reynolds number,
$Re_\infty \simeq 8.8\times 10^{5}$, at approximately 85 seconds after
atmospheric interface. By that later time, the abrupt leeside
heat-flux spike had largely subsided and the measured values were
closer to the windside levels. This chronology does not rule out
boundary-layer instability mechanisms, but it is difficult to
reconcile with a transition criterion controlled primarily by
$Re_\infty$ or $Re_\theta$ alone. The present results point instead to
a gain scale that depends on Reynolds number through the
finite-$Re_\infty$ viscous correction derived in
\S\ref{sec::scaling_G_D}, namely
\[
\mathcal{S}(Re_\infty,M_1,\gamma_2^*,\rho_2/\rho_1)
=
\gamma_2^* M_1^2
\exp\!\left[
\frac{\rho_2/\rho_1}{C}
-
\frac{B}{\sqrt{Re_\infty}}
\right],
\]
where $C$ and $B$ are fixed, for the present geometry and disturbance
family, by the scaling analysis of \S\ref{sec::scaling}. At the large
Reynolds numbers relevant to the MSL transition point, the dominant
trajectory dependence of $\mathcal{S}$ is governed by $M_1$,
$\gamma_2^*$ and the compression ratio $\rho_2/\rho_1$, rather than by
$Re_\infty$ alone. The fact that the maximum of this
Mach--thermochemical gain scale occurs near the reconstructed MSL
transition location and the corresponding Mars~2020/Perseverance
velocity--altitude region supports the interpretation that the
observed leeside event was influenced by the
bow-shock/shear--entropy-layer receptivity mechanism identified here.

A separate question is whether the optimal freestream disturbances
computed in the linear analysis are likely to be realized in free
flight. The leading optimal mode should be interpreted as the most
amplified member of an admissible disturbance space, and therefore as
an upper bound on linear energy amplification. Real atmospheric
disturbances will instead contain a broadband distribution of
acoustic, entropic and vortical components with different projections
onto the optimal input directions. Nevertheless, the present spectrum
is not sharply isolated.  At $Re_\infty = 2.85\times 10^{5}$ and the
freestream conditions of table~\ref{tab:freestream}, the first ten
eigenvalues of the cycle-averaged gain operator have logarithmic
cycle-averaged gains $\overline{N}_j=\ln
\overline{G}_{T,j}^{\,\infty}$ exceeding 11, and the first hundred
have $\overline{N}_j>9.6$, where the subscript $j$ denotes the ordered
optimal input. Thus the large amplification is not confined to a
single precisely tuned disturbance.  The linear calculation does not,
by itself, set the absolute transition threshold; that threshold
depends on the incident disturbance amplitudes, their spectral
content, and the nonlinear breakdown process. It does show, however,
that relatively small upstream acoustic and entropic components can be
converted into a large post-shock response under the critical
Mars-entry conditions.

The ground-test database used in the design and interpretation of
MSL-class blunt capsules provides an important but nontrivial point of
comparison. Experiments in AEDC Tunnel 9, CUBRC LENS facilities,
Caltech T5, and the NASA Langley Mach 6 and Mach 10 tunnels reported
transition and enhanced leeside heating on MSL/CEV-type geometries
\citep{hollis2005transition,hollis2007turbulent,hollis2010blunt,
  liechty2006mars,wright2006modeling}. This database motivated the
engineering smooth-body criterion $Re_\theta \approx 200$ used in MSL
design and subsequent blunt-body transition correlations
\citep{edquist2007aerothermodynamic, hollis2010blunt}. In terms of the
present mechanism, however, the relevant inviscid part of the gain
scale is
\[
\mathcal{S}_0
=
\gamma_2^* M_1^2
\exp\!\left(\frac{\rho_2/\rho_1}{C}\right),
\]
with the finite-Reynolds-number correction supplied by the additional
factor $\exp[-B/\sqrt{Re_\infty}]$. The values of $\mathcal{S}_0$ in
many of the ground-test conditions are one to two orders of magnitude
smaller than those at the MSL and 2020/Perseverance flight points
where enhanced leeside heating was reconstructed. This difference
explains why a direct Reynolds-number extrapolation from tunnel data
to flight can be misleading for this instability.

The lower gain scale in those facilities does not imply that the
ground-test transition measurements are irrelevant. Rather, it means
that the transition threshold in a facility depends on both the
intrinsic amplification and the incident disturbance environment.
These facilities are not quiet in the sense relevant to receptivity:
their freestream disturbance spectra can contain acoustic radiation
from turbulent wall boundary layers, driver- or nozzle-generated
fluctuations, and other facility-dependent acoustic, vortical and
entropic components \citep{schneider2001effects}.  Because the optimal
inputs in the present receptivity analysis are primarily acoustic and
entropic, a larger facility-noise level can compensate for a smaller
value of $\mathcal{S}_0$ and still trigger transition. The existing
tunnel data should therefore be interpreted as evidence that MSL-class
geometries are susceptible to leeside transition, but not as a direct
measurement of the free-flight threshold. Future experiments aimed
specifically at this mechanism would benefit from quiet or otherwise
well-characterized freestream conditions, together with diagnostics of
the incident acoustic and entropic spectra.

The projectile experiments of \citet{hornung2001shock} are especially
relevant in this context because they reduced the usual tunnel-wall
noise pathway by launching a body into a quiescent test gas. Their
propane cases were chosen to approach the Newtonian, high-compression
limit, for which vibrational and chemical relaxation can produce large
values of $\rho_2/\rho_1$. They reported two representative propane
conditions: one at 2.26 km s$^{-1}$ with $\rho_2/\rho_1 \simeq 13$,
and another at 2.70 km s$^{-1}$ with $\rho_2/\rho_1 \simeq 20$.  Using
the present inviscid gain scale and assuming $C=2.23$, the ratio
between the two cases is
\[
\frac{\mathcal{S}_{0,2}}{\mathcal{S}_{0,1}}
=
\frac{\gamma_{2,2}^*}{\gamma_{2,1}^*}
\left(\frac{M_{1,2}}{M_{1,1}}\right)^2
\exp\!\left[
\frac{(\rho_2/\rho_1)_2-(\rho_2/\rho_1)_1}{C}
\right].
\]
If the upstream sound speed and post-shock effective specific-heat
ratio are comparable between the two propane cases, this gives
\[
\frac{\mathcal{S}_{0,2}}{\mathcal{S}_{0,1}} \approx
\left(\frac{2.70}{2.26}\right)^2 \exp\!\left(\frac{20-13}{2.23}\right)
\simeq 33.
\]
This estimate is consistent with the qualitative experimental
observation that the higher-compression case developed a much more
disturbed shock layer, including visible shock oscillations in the
schlieren images. Similar shock corrugation appears in the present
linear receptivity calculation and in the early stages of the WMLES
solution in figure~\ref{fig:linear_WMLES_comparison}.

\section{Conclusions}

The Mars Science Laboratory and Perseverance entries exhibited
localized leeside heating increases that are consistent with
transition and remain difficult to predict with existing blunt-body
transition criteria \citep{bose2014reconstruction,edquist2022mars,
  alpert2022inverse}. This study investigated a candidate mechanism
for those events by combining shock-fitted freestream receptivity
analysis, energy-budget diagnostics, scaling arguments,
velocity--altitude gain maps, and wall-modeled large-eddy
simulations. The central result is that, under high-enthalpy
Mars-entry conditions, the detached bow shock and the post-shock
shear--entropy layer constitute a high-gain receptivity pathway
through which linear mechanisms amplify freestream disturbances by
orders of magnitude. These amplified disturbances corrugate the bow
shock, establishing a feedback mechanism that further drives the
instability. Corresponding nonlinear simulations show that the
resulting shock-layer response can evolve toward breakdown and
localized heating.

The optimal freestream inputs are dominated by acoustic and entropic
components, with only a small kinetic contribution.  This preference
is consistent with classical linear interaction theory: in the
strong-shock limit, acoustic and entropic incident disturbances
produce transmitted thermodynamic and energy-flux perturbations that
increase as $\mathcal{O}(M_1^2)$, whereas the
kinetic-energy-containing vortical component is not amplified by the
same Mach-number factor \citep{mckenzie1968interaction,
  ribner1954convection}.  The bow shock therefore acts as a selective
amplifier of the disturbance components that are most effective at
forcing the post-shock layer. The curved shock also deposits vorticity
and entropy gradients \citep{hornung1998gradients}, creating the
shear--entropy layer in which the downstream amplification occurs.

The physical mechanism can be summarized as a three-step
process. First, the incoming acoustic and entropic components are
amplified as they cross the fitted bow shock. Second, the transmitted
disturbance packet enters the post-shock shear--entropy layer and
extracts energy from the base flow. The dominant kinetic-energy
production is the Reynolds-stress work against the mean shear
$\mathcal{P}_u^k$, while the accompanying entropy gradient produces
the leading entropic source term, $\mathcal{P}_u^s$.  Third, the
amplified disturbance field perturbs and corrugates the bow shock,
closing the shock-layer feedback. For the open-loop linear gain, the
scaling reduces to two multiplicative factors: shock transmission and
post-shock convective amplification.  The corresponding cycle-averaged
asymptotic gain scale is
\[
\overline{G}_T^{\mathrm{opt}}
\sim
\gamma_2^* M_1^2
\exp\!\left[
\frac{\rho_2/\rho_1}{C}
-
\frac{B}{\sqrt{Re_\infty}}
\right],
\]
with $C=2.23$ and $B=685$ for the present geometry and class of
optimal disturbance paths. This expression retains the inviscid
Mach-number and thermochemical dependence through
$\gamma_2^*M_1^2\exp[(\rho_2/\rho_1)/C]$, while also including the
leading viscous correction inferred from the Reynolds-number sweep.

At the representative Mars-entry condition, the computed optimal total
gains are of order $10^{6}$
($\overline{G}_T^{\mathrm{opt}}=7.9\times10^{5}$ and
$G_{T,\max}^{\mathrm{opt}}=9.7\times10^{5}$ in the baseline
case). In the energy-gain convention used here, $N\equiv \ln G$, these
values correspond to $N\simeq 14$. The velocity--altitude maps show
that the largest values of the gain scale occur near the reconstructed
MSL transition point and in the same region of velocity--altitude
space where Mars~2020/Perseverance reconstructions also indicate
transition-associated heat-flux augmentation. By contrast, Earth-entry
conditions produce substantially smaller values of the same scale
because the lower compression ratios and lower Mach numbers reduce
both the shock transmission and post-shock convective
amplification. This comparison supports the interpretation that the
observed Mars-entry behavior is not controlled by Reynolds number
alone, but by the coupled dependence on Mach number, thermochemistry
and shock-layer compression.

The nonlinear WMLES calculations provide an independent consistency
check on the linear mechanism. In the MSL geometry at representative
Mars-entry conditions, reducing the effective specific-heat ratio
$\gamma^*$ increases the post-shock compression ratio and hence the
gain scale. The simulations then develop strong shear--entropy-layer
disturbances on the leeside, followed by nonlinear breakdown and a
localized heat-flux spike. The location and flow topology of this
breakdown are qualitatively consistent with the flight heat-flux
reconstructions and with the leading structures predicted by the
linear receptivity analysis. 

Overall, the study identifies a bow-shock-mediated transition route
for high-enthalpy blunt-body entry flows. The mechanism is distinct
from classical wall-localized boundary-layer routes: the dominant
amplification begins at the shock and in the outer post-shock
shear--entropy layer, and only later leads to near-wall turbulent
heating. The resulting scaling provides a compact, physically
interpretable gain indicator for identifying regimes in which this
pathway is likely to be important. Its use in
thermal-protection-system design should be paired with estimates of
the freestream disturbance environment and with additional validation
across capsule geometries, wall thermal conditions and thermochemical
models.

\section*{Acknowledgements}
The authors gratefully acknowledge Professor Hans G. Hornung for
valuable discussions on entropy layer instabilities. We also thank
Professor Joseph E. Shepherd for generously sharing thermochemical
data files.

\section*{Funding}
This work is supported by an Early Career Faculty grant from NASA’s Space Technology Research Grants Program
(grant \#80NSSC23K1498).

\section*{Declaration of interests}
The authors report no conflict of interest.

\section*{Author ORCIDs}
A. Ant\'on-\'Alvarez, https://orcid.org/0000-0002-6434-4211; \\ A. Lozano-Dur\'an, https://orcid.org/0000-0001-9306-0261

\clearpage

\bibliography{main}

\clearpage

\appendix

\section{Validity of thermochemical models}
\label{sec::validity_equilibrium}

This appendix assesses the range of Mars- and Earth-entry conditions
for which the equilibrium assumptions introduced in
\S\ref{sec:thermochem} are valid. We first examine the chemical- and
thermal-equilibration time scales. Equilibrium models presume that the
characteristic times of chemical relaxation and internal-energy
relaxation are small compared with the flow time scales of interest.
We therefore examine finite-rate chemistry and vibrational relaxation.
Electronic excitation is not assigned a separate relaxation time in the
present estimates, because its contribution is negligible over the
temperature range considered here; when it is present in the equilibrium
thermodynamic database, it is included only in the combined
vibrational--electronic energy bookkeeping below.

The chemical relaxation time $\tau_{\mathrm{chem}}$ is defined here as
an effective one-time-scale measure of the approach of a reacting
mixture to its local equilibrium composition. We use the departure of
the translational--rotational energy from its equilibrium value as a
scalar progress variable. Assuming first-order relaxation,
\begin{equation}
\frac{e_{T\text{-}R}(t)-e_{T\text{-}R}^{\mathrm{eq}}}
     {e_{T\text{-}R}(0)-e_{T\text{-}R}^{\mathrm{eq}}}
=
\exp\!\left(-\frac{t}{\tau_{\mathrm{chem}}}\right).
\end{equation}
The relaxation time $\tau_{\mathrm{chem}}$ is recovered from the time
$t_5$ at which the normalized departure from equilibrium first falls
to $5\%$:
\begin{equation}
\left.
\frac{e_{T\text{-}R}(t)-e_{T\text{-}R}^{\mathrm{eq}}}
     {e_{T\text{-}R}(0)-e_{T\text{-}R}^{\mathrm{eq}}}
\right|_{t=t_5}
=0.05
\quad\Longrightarrow\quad
\tau_{\mathrm{chem}}(P,T,\bm{X})
=
-\frac{t_5}{\ln(0.05)} .
\label{eq:tau-chem}
\end{equation}
We obtain $t_5$ by integrating an adiabatic, constant-pressure reactor
in Cantera~\citep{cantera}, using the thermochemical data set of
\citet{shepherd_cti_mech}, from the initial state $(P,T,\bm{X})$ and
recording the first crossing of the $5\%$ threshold.

The vibrational relaxation time $\tau_{\mathrm{vib}}$ is estimated
from the Millikan--White correlation~\citep{millikan1963systematics}.
For the Earth-atmosphere cases, we use the dominant self-collision
relaxation of $\mathrm{N}_2$, with
$\theta=3\,350~\mathrm{K}$ and
$\mu^{\mathrm{red}}_{\mathrm{N}_2\text{-}\mathrm{N}_2}
=14~\mathrm{g\,mol^{-1}}$. For the Mars-atmosphere cases,
$\mathrm{CO}_2$ is dominant but has three characteristic vibrational
frequencies and four vibrational degrees of freedom. As a conservative
estimate of the slowest relaxation process, we use the asymmetric
stretching mode, with $\theta=3\,380~\mathrm{K}$ and
$\mu^{\mathrm{red}}_{\mathrm{CO}_2\text{-}\mathrm{CO}_2}
=22~\mathrm{g\,mol^{-1}}$. This dominant-species approximation neglects
mixture corrections to vibrational relaxation and is less accurate for
polyatomic molecules than for diatomic nitrogen; the latter limitation
is well documented for $\mathrm{CO}_2$ kinetics
\citep{armenise2018effect}.

Computing the chemical and vibrational relaxation times requires an
estimate of the post-shock state $(P,T,\bm{X})$. Although the actual
post-shock state varies along the curved bow shock, we use the state
immediately downstream of the corresponding normal shock as a
stagnation-line estimate. This construction should therefore be
interpreted as an order-of-magnitude criterion for the forebody shock
layer. Across the shock, vibrational excitation, electronic excitation
and chemical reactions are assumed frozen, while translational and
rotational modes are taken to be equilibrated. The composition is
therefore equal to the freestream composition immediately behind the
shock. Under these assumptions, the jump conditions reduce to the
Rankine--Hugoniot relations for a calorically perfect gas with
$\gamma_{T\text{-}R}=1+R_g/c_{v,T\text{-}R}$.

Next, we assess the validity of the calorically perfect gas
approximation. Starting from the frozen post-shock Rankine--Hugoniot
($\mathrm{RH}$) state, we compute the corresponding equilibrium state
with Cantera~\citep{cantera} and the thermochemical data set of
\citet{shepherd_cti_mech}. The equilibrium state used to define $E_R$
is obtained at fixed density and fixed specific internal energy. The
fraction of the frozen post-shock translational--rotational thermal
energy transferred to vibrational--electronic and chemical energy
storage during this relaxation quantifies the departure from
calorically perfect behavior.

Assuming translational--rotational equipartition, the specific
translational--rotational energy is
\begin{equation}
e_{T\text{-}R}=c_{v,T\text{-}R}T,
\qquad
\frac{c_{v,T\text{-}R}}{R_u}
=
\frac{1}{W}\sum_i X_i\frac{Z_i}{2},
\end{equation}
where $R_u$ is the universal gas constant, $X_i$ are mole fractions,
$W=\sum_i X_iW_i$ is the mixture molar mass, with all molar masses
expressed in consistent units, and $Z_i$ is the number of active
translational--rotational degrees of freedom of species $i$.

Because vibrational excitation, electronic excitation and chemical
reactions are frozen across the shock, the shock-generated thermal
energy initially resides in the translational--rotational pool, $e_{T\text{-}R}^{\mathrm{RH}}$. We
therefore normalize the redistributed energy by
$e_{T\text{-}R}^{\mathrm{RH}}$. Conservation of total internal energy
during the fixed-density relaxation implies
$\Delta e_{T\text{-}R}
=-\Delta(e_{V\text{-}E}+e_{\mathrm{chem}})$, so the relaxed energy
fraction is
\begin{align}
E_R
&\equiv
\frac{
e_{T\text{-}R}^{\mathrm{RH}}-e_{T\text{-}R}^{\mathrm{eq}}
}{
e_{T\text{-}R}^{\mathrm{RH}}
}
=
\frac{\Delta(e_{V\text{-}E}+e_{\mathrm{chem}})}
     {e_{T\text{-}R}^{\mathrm{RH}}}
\notag\\
&=
\frac{\displaystyle
\frac{T^{\mathrm{RH}}}{W^{\mathrm{RH}}}
\sum_i X_i^{\mathrm{RH}}\frac{Z_i}{2}
-
\frac{T^{\mathrm{eq}}}{W^{\mathrm{eq}}}
\sum_i X_i^{\mathrm{eq}}\frac{Z_i}{2}}
{\displaystyle
\frac{T^{\mathrm{RH}}}{W^{\mathrm{RH}}}
\sum_i X_i^{\mathrm{RH}}\frac{Z_i}{2}} .
\end{align}
Thus, $E_R$ represents the fraction of the frozen post-shock
translational--rotational thermal energy that is redistributed to
vibrational--electronic and chemical modes at equilibrium.
Mass-specific energies are used throughout because the number of moles
is not conserved during chemical reactions, whereas mass is.

\begin{figure}[]
\centering
\begin{subfigure}[c]{0.49\textwidth}
\includegraphics[width=\textwidth]{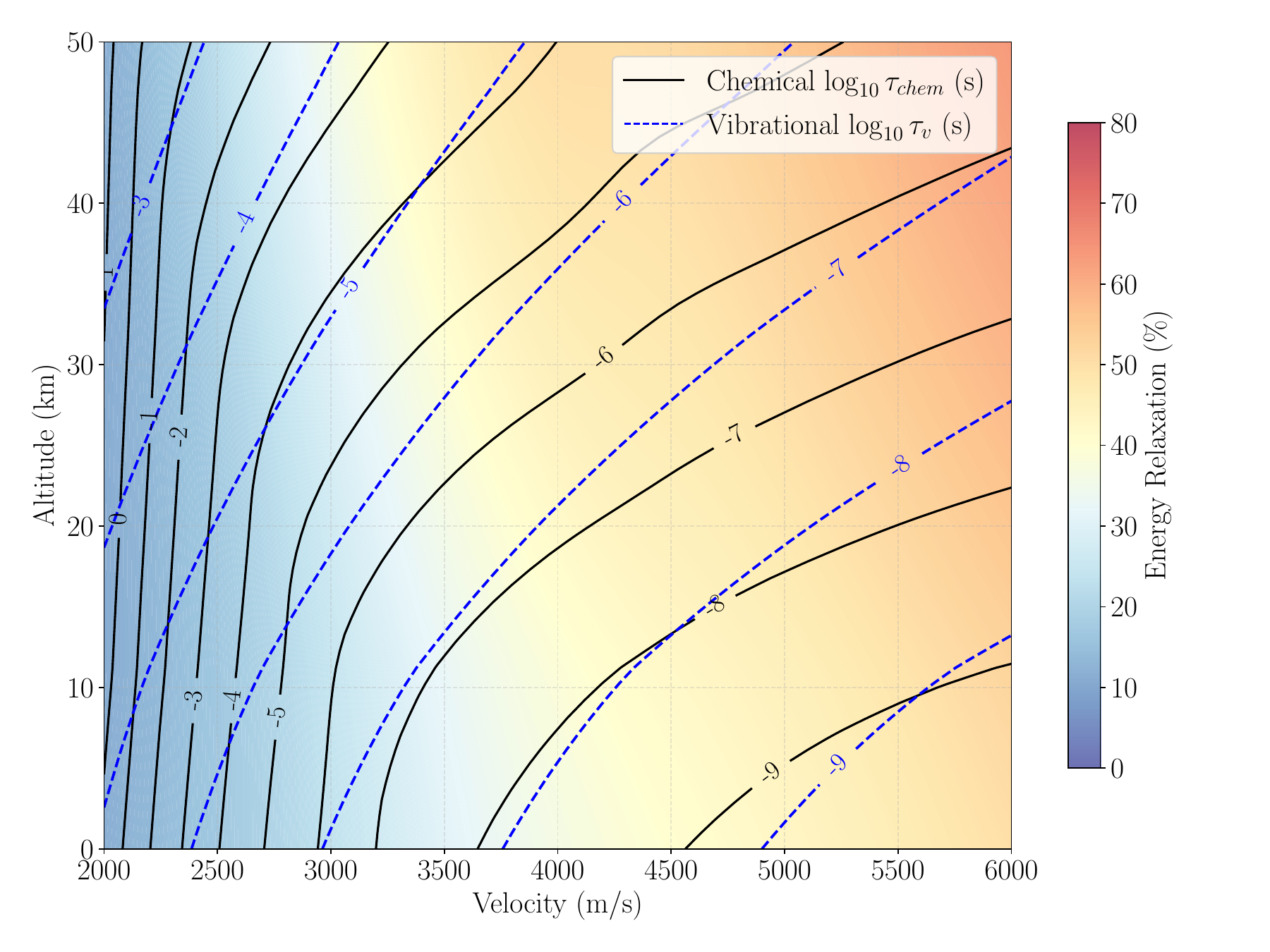}
\caption{}
\label{fig:tau_Earth}
\end{subfigure}
\begin{subfigure}[c]{0.49\textwidth}
\includegraphics[width=\textwidth]{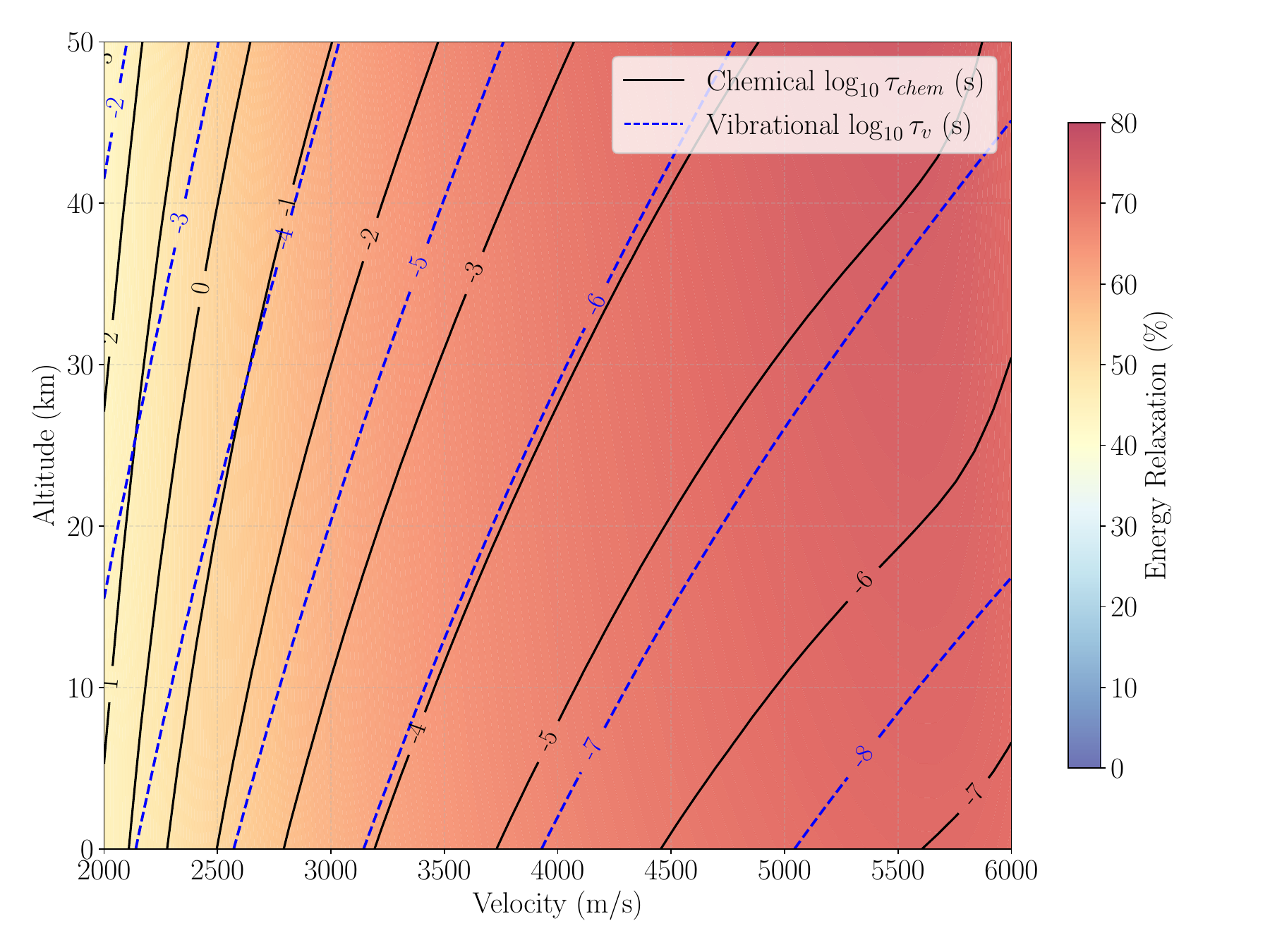}
\caption{}
\label{fig:tau_Mars}
\end{subfigure}
\caption{Chemical and vibrational relaxation times for Earth and Mars
  atmospheres. Contours are shown as $\log_{10}(\tau/\mathrm{s})$.
  The background color shows the energy-relaxation fraction $E_R$,
  expressed as a percentage of the frozen post-shock
  translational--rotational thermal energy. (\textit{a}) Earth
  atmosphere: $X_{\mathrm{N}_2}=0.7812$,
  $X_{\mathrm{O}_2}=0.2095$, $X_{\mathrm{Ar}}=0.0093$.
  (\textit{b}) Mars atmosphere: $X_{\mathrm{CO}_2}=0.9556$,
  $X_{\mathrm{N}_2}=0.0270$, $X_{\mathrm{Ar}}=0.0160$,
  $X_{\mathrm{O}_2}=0.0014$.}
\label{fig:tau}
\end{figure}

The vibrational relaxation time $\tau_{\mathrm{vib}}$, chemical
relaxation time $\tau_{\mathrm{chem}}$, and energy-relaxation fraction
$E_R$ are shown in figures~\ref{fig:tau_Earth} and
\ref{fig:tau_Mars} for the Earth and Mars atmospheres, respectively.
The Martian atmosphere is modeled following \citet{nasagrc_mars},
while the Earth atmosphere is modeled using the U.S. Standard
Atmosphere~\citep{usstdatm1976}. The first observation is that the
calorically perfect gas approximation remains valid to higher
velocities in the Earth atmosphere than in the Martian atmosphere. In
the Martian atmosphere, contributions to $E_R$ become significant at
lower velocities because the dominant species, $\mathrm{CO}_2$,
activates vibrational-energy storage at lower temperatures than
$\mathrm{N}_2$. For example, over the altitude range shown,
Earth-entry states with velocities below approximately
$2\,000~\mathrm{m\,s^{-1}}$ have $E_R\lesssim 10\%$, so a calorically
perfect model is often reasonable. In the Martian atmosphere, states
near $2\,000~\mathrm{m\,s^{-1}}$ can have $E_R$ of order $40\%$ over
the same range, indicating that vibrational and chemical energy
storage may already be important. In particular, excitation of the
$\mathrm{CO}_2$ bending mode provides the dominant contribution to
$E_R$ at lower velocities because this mode has a low characteristic
vibrational temperature.

\begin{figure}[]
\centering
\begin{subfigure}[t]{0.49\textwidth}
\includegraphics[width=\textwidth]{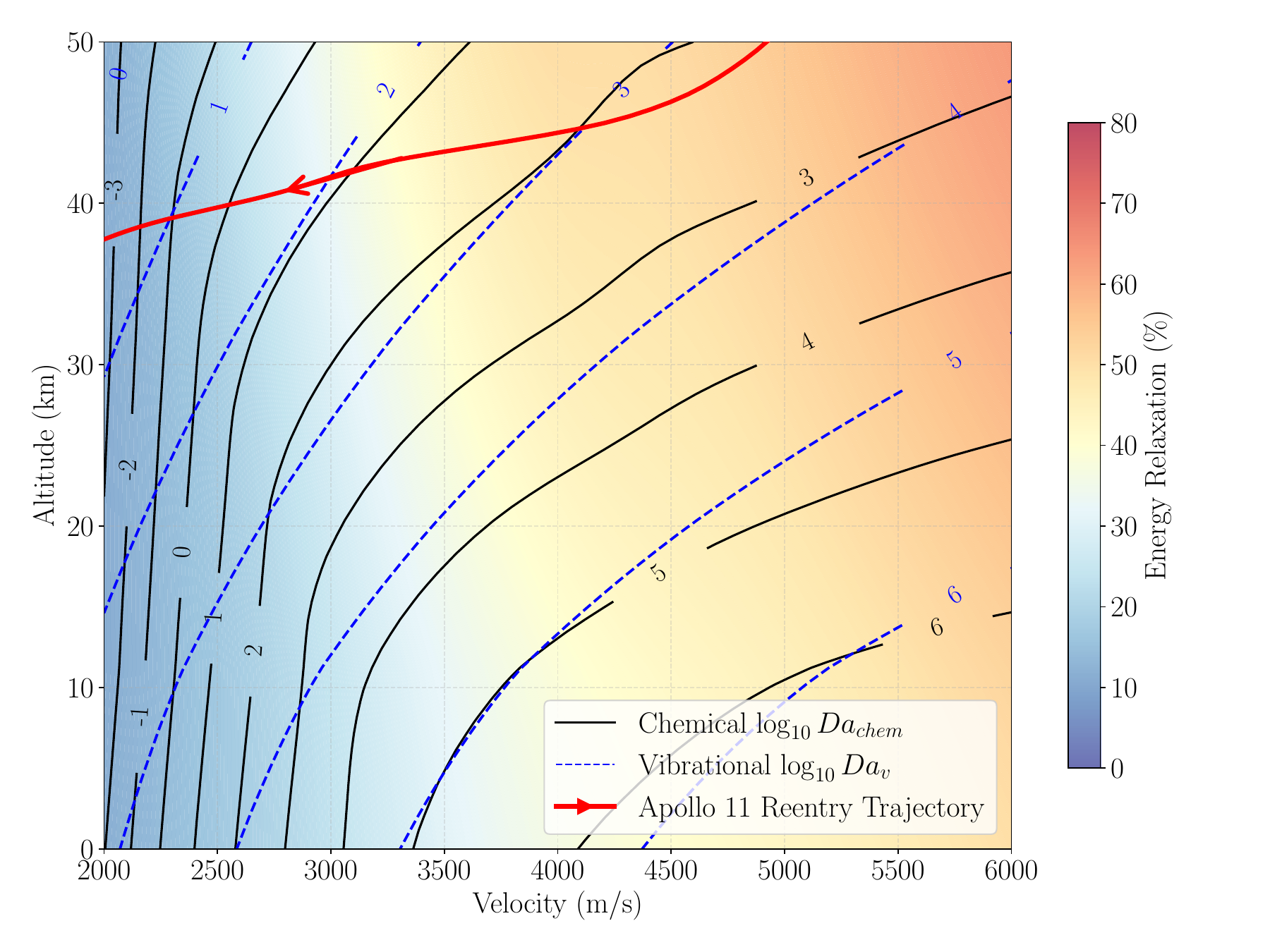}
\caption{}
\label{fig:da_Earth}
\end{subfigure}
\begin{subfigure}[t]{0.49\textwidth}
\includegraphics[width=\textwidth]{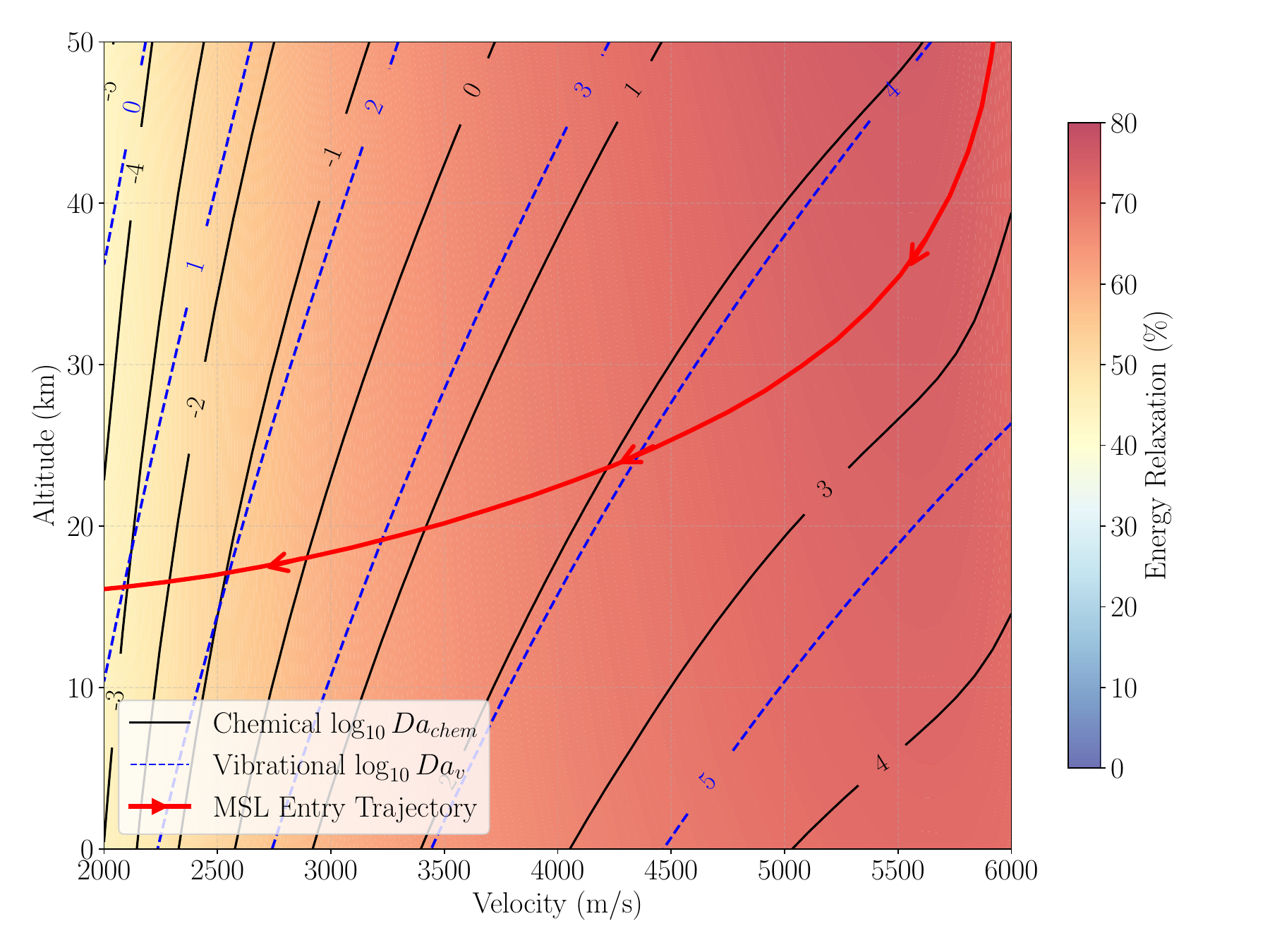}
\caption{}
\label{fig:da_MSL}
\end{subfigure}
\caption{Contour lines show Damk\"ohler numbers for Earth and Mars
  atmospheres as $\log_{10}(Da)$. Red curves indicate the Apollo~11
  Earth re-entry trajectory and the MSL Mars-entry trajectory. The
  background color shows the energy-relaxation fraction $E_R$,
  expressed as a percentage of the frozen post-shock
  translational--rotational thermal energy. (\textit{a}) Earth
  atmosphere: $X_{\mathrm{N}_2}=0.7812$,
  $X_{\mathrm{O}_2}=0.2095$, $X_{\mathrm{Ar}}=0.0093$.
  (\textit{b}) Mars atmosphere: $X_{\mathrm{CO}_2}=0.9556$,
  $X_{\mathrm{N}_2}=0.0270$, $X_{\mathrm{Ar}}=0.0160$,
  $X_{\mathrm{O}_2}=0.0014$.}
\label{fig:Da}
\end{figure}

Regarding the chemical and thermal equilibrium assumptions, the
relaxation times must be compared with a representative residence time
of the gas in the shock layer. The chemical and vibrational relaxation
times in the Earth-atmosphere cases are much shorter than in the
Martian cases over most of the plotted range, primarily because the
post-shock Earth-atmosphere states are denser and molecular collisions
are therefore more frequent. To compare relaxation and flow time
scales along representative trajectories, we consider the Apollo~11
Earth re-entry trajectory~\citep{manders1970apollo11} and the MSL
Mars-entry trajectory~\citep{edquist2007aerothermodynamic}. We define
a body-scale post-shock flow time
\begin{equation}
\tau_{\mathrm{flow}}
=
\frac{R}{U_2^{\mathrm{RH}}},
\end{equation}
where $R$ is the capsule nose radius and $U_2^{\mathrm{RH}}$ is the
post-shock velocity from the corresponding normal-shock
Rankine--Hugoniot estimate. This gives
\begin{equation}
Da_{\mathrm{chem}}
=
\frac{\tau_{\mathrm{flow}}}{\tau_{\mathrm{chem}}},
\qquad
Da_{\mathrm{vib}}
=
\frac{\tau_{\mathrm{flow}}}{\tau_{\mathrm{vib}}}.
\end{equation}
These Damk\"ohler numbers, together with the Apollo~11 and MSL
trajectories, are shown in figure~\ref{fig:Da}.

For the Earth-entry case, figure~\ref{fig:da_Earth} shows that, for
velocities above approximately $2\,500~\mathrm{m\,s^{-1}}$,
vibrational and chemical relaxation are much faster than the
body-scale flow time, so thermal and chemical equilibrium are
reasonable approximations. At higher altitudes and velocities near
$2\,000~\mathrm{m\,s^{-1}}$, the relaxation times become comparable
to the flow time. In that regime, however, $E_R$ is less than $10\%$,
so the departure from a calorically perfect gas remains small. We
therefore infer that thermal and chemical equilibrium are reasonable
approximations along the Apollo~11 re-entry trajectory over the range
considered here.

We next consider the MSL entry into the Martian atmosphere
(figure~\ref{fig:da_MSL}). For velocities around
$3\,000~\mathrm{m\,s^{-1}}$, the chemical and vibrational relaxation
times start to become comparable to the body-scale flow time. In this
regime, $E_R$ remains substantial, with values of order $50\%$, so
thermochemical nonequilibrium should be retained for accurate
aerothermodynamic modeling. Of this energy fraction, most of the
energy is transferred to vibrational degrees of freedom for
post-shock temperatures below approximately $3\,000~\mathrm{K}$.
Although vibrational relaxation is faster than chemical relaxation,
$Da_{\mathrm{vib}}$ is already of order unity near
$2\,000~\mathrm{m\,s^{-1}}$, so thermal nonequilibrium cannot be
neglected there. At higher velocities,
$U_\infty>4\,000~\mathrm{m\,s^{-1}}$, the equilibrium approximation
becomes increasingly appropriate, especially for vibrational
relaxation. This high-velocity region is the most relevant one for the
present stability analysis, because it overlaps the portion of the
Mars-entry trajectories where large aeroheating rates and
transition-associated heat-flux augmentation are reported
\citep{bose2014reconstruction,edquist2022mars}.

The critical Mars-entry region in which transition-associated heating
was inferred for MSL and Mars 2020/Perseverance, and from which the
representative state in table~\ref{tab:freestream} is drawn,
corresponds to velocities above $5\,000~\mathrm{m\,s^{-1}}$ and
altitudes near $30~\mathrm{km}$. In this regime, the Damk\"ohler
numbers in the Martian atmosphere (figure~\ref{fig:da_MSL}) lie in
the ranges $Da_{\mathrm{vib}}\approx 10^{4}\text{--}10^{5}$ and
$Da_{\mathrm{chem}}\approx 10^{2}\text{--}10^{3}$. The relaxation
times are therefore short compared with the body-scale residence time,
supporting the use of thermochemical equilibrium in the stability
calculations, although chemical relaxation remains slower than
vibrational relaxation. As a check, several computations were
performed within the critical gain region using the finite-rate
chemistry model developed in \citet{AntonAlvarez2026HYMOR}, and no
changes in the optimal gains were found at a level that would affect
the conclusions. These computations were, however, considerably more
expensive: the thin thermochemical relaxation layer immediately behind
the shock, in which translational--rotational energy is redistributed
into vibrational, electronic and chemical energy modes, required a
much finer local grid and a correspondingly smaller time step. Given
the substantial increase in computational cost and the negligible
impact on the amplification factors, thermochemical equilibrium was
adopted throughout the subsequent analysis.

\section{Energy budgets}
\label{sec:energy-budgets}

This appendix derives the transport equations for the kinetic and
entropic contributions to Chu's norm. Acoustic energy is omitted
because it plays no leading role in the cases analyzed in the main
text.

We decompose the flow variables into a steady base state and
infinitesimal perturbations,
\begin{align*}
\rho &= \rho_0 + \rho', \qquad |\rho'| \ll \rho_0, \\
u_i &= u_{i,0} + u_i', \qquad |u_i'| \ll |u_{i,0}|, \\
p &= p_0 + p', \qquad |p'| \ll p_0, \\
T &= T_0 + T', \qquad |T'| \ll T_0, \\
s &= s_0 + s', \qquad |s'| \ll s_0,
\end{align*}
where $\rho$, $u_i$, $p$, $T$, and $s$ denote density, velocity,
pressure, temperature, and entropy, respectively. The base flow is
steady, so $\partial(\cdot)_0/\partial t = 0$.

\subsection{Kinetic energy budget}

The kinetic contribution to Chu's norm is
\begin{equation*}
E^k = \int \frac{1}{2}\rho_0 u_i'u_i'\,dV.
\end{equation*}
Retaining terms up to second order in the perturbation amplitudes
gives the familiar compressible kinetic-energy
budget~\citep{sagaut2008homogeneous},
\begin{equation}
\rho_0\frac{\partial}{\partial t}\left(\frac{1}{2}u_i'u_i'\right)
= \mathcal{A}^k + \mathcal{P}^k + \Pi_d^k + \mathcal{T}^k + \mathcal{D}^k,
\label{eq:kinetic-budget}
\end{equation}
where
\begin{itemize}
\item Advection:
\begin{equation*}
\mathcal{A}^k = -\rho_0 u_{j,0}\frac{\partial}{\partial x_j}\left(\frac{1}{2}u_i'u_i'\right),
\end{equation*}
\item Production:
\begin{equation*}
\mathcal{P}^k = -\rho_0u_i'u_j'\frac{\partial u_{i,0}}{\partial x_j}
- \rho' u_i' u_{j,0}\frac{\partial u_{i,0}}{\partial x_j},
\end{equation*}
which represents transfer from the mean flow to the perturbations.
\item Pressure--dilatation:
\begin{equation*}
\Pi_d^k = p'\frac{\partial u_i'}{\partial x_i},
\end{equation*}
\item Transport:
\begin{equation*}
\mathcal{T}^k = \frac{\partial}{\partial x_j}\left(-u_j'p' + u_i'\tau_{ij}'\right),
\end{equation*}
\item Dissipation:
\begin{equation*}
\mathcal{D}^k = -\tau_{ij}'\frac{\partial u_i'}{\partial x_j}.
\end{equation*}
\end{itemize}
The linearized viscous stress is
\begin{equation*}
\tau_{ij}' =
\frac{1}{Re_\infty}
\left[
\mu^{*\prime}
\left(
\frac{\partial u_{i,0}}{\partial x_j}
+\frac{\partial u_{j,0}}{\partial x_i}
-\frac{2}{3}\delta_{ij}
\frac{\partial u_{k,0}}{\partial x_k}
\right)
+\mu_0^*
\left(
\frac{\partial u_i'}{\partial x_j}
+\frac{\partial u_j'}{\partial x_i}
-\frac{2}{3}\delta_{ij}
\frac{\partial u_k'}{\partial x_k}
\right)
\right].
\end{equation*}

\subsection{Entropic energy budget}

The entropic contribution to Chu's norm is
\begin{equation*}
E^s = \int \frac{(\gamma_0^*-1)p_0}{2\gamma_0^*}\left(\frac{s'}{R_{g,0}}\right)^2 dV.
\end{equation*}
To obtain its transport equation, we start from the entropy equation for a viscous compressible flow,
\begin{equation}
\rho T\frac{Ds}{Dt} = -\frac{\partial q_j}{\partial x_j} + \Phi,
\label{eq:entropy}
\end{equation}
where the heat flux is
\begin{equation*}
q_j = -k\frac{\partial T}{\partial x_j},
\end{equation*}
and the Rayleigh dissipation function is
\begin{equation*}
\Phi = \tau_{ij}\frac{\partial u_i}{\partial x_j}
= \mu\left(\frac{\partial u_i}{\partial x_j} + \frac{\partial u_j}{\partial x_i} - \frac{2}{3}\delta_{ij}\frac{\partial u_k}{\partial x_k}\right)\frac{\partial u_i}{\partial x_j}.
\end{equation*}

Linearizing~\eqref{eq:entropy} about the steady base flow gives
\begin{equation*}
\rho_0T_0\left(\frac{\partial s'}{\partial t} + u_{j,0}\frac{\partial s'}{\partial x_j}\right)
+ \left(\rho_0T_0u_j' + \rho'T_0u_{j,0} + \rho_0T'u_{j,0}\right)\frac{\partial s_0}{\partial x_j}
= -\frac{\partial q_j'}{\partial x_j} + \Phi'.
\end{equation*}
Multiplying by $s'$ and rearranging yields
\begin{equation*}
\rho_0T_0\left[\frac{\partial}{\partial t}\left(\frac{s'^2}{2}\right) + u_{j,0}\frac{\partial}{\partial x_j}\left(\frac{s'^2}{2}\right)\right]
= -\left(\rho_0T_0u_j' + \rho'T_0u_{j,0} + \rho_0T'u_{j,0}\right)s'\frac{\partial s_0}{\partial x_j}
- \frac{\partial}{\partial x_j}(s'q_j') + \frac{\partial s'}{\partial x_j}q_j' + s'\Phi'.
\end{equation*}
Scaling this result to match Chu's norm gives
\begin{equation}
\frac{(\gamma_0^*-1)p_0}{2\gamma_0^*}\frac{\partial}{\partial t}\left(\frac{s'^2}{R_{g,0}^2}\right)
= \mathcal{A}^s + \mathcal{P}^s + \mathcal{T}^s + \mathcal{D}^s + \mathcal{S}^s,
\label{eq:entropy-budget}
\end{equation}
where
\begin{itemize}
\item Advection:
\begin{equation*}
\mathcal{A}^s = -\frac{(\gamma_0^*-1)p_0}{\gamma_0^*R_{g,0}^2}
 u_{j,0}\frac{\partial}{\partial x_j}\left(\frac{s'^2}{2}\right),
\end{equation*}
\item Production:
\begin{equation*}
\mathcal{P}^s = -\frac{(\gamma_0^*-1)p_0}{\gamma_0^*R_{g,0}^2}
\left(u_j' + \frac{\rho'}{\rho_0}u_{j,0} + \frac{T'}{T_0}u_{j,0}\right)
 s'\frac{\partial s_0}{\partial x_j},
\end{equation*}
which contains velocity-, density-, and temperature-induced production.
\item Transport:
\begin{equation*}
\mathcal{T}^s = -\frac{(\gamma_0^*-1)}{\gamma_0^*R_{g,0}}
\frac{\partial}{\partial x_j}(s'q_j'),
\end{equation*}
\item Dissipation:
\begin{equation*}
\mathcal{D}^s = \frac{(\gamma_0^*-1)}{\gamma_0^*R_{g,0}}
\left(\frac{\partial s'}{\partial x_j}q_j'\right),
\end{equation*}
\item Source:
\begin{equation*}
\mathcal{S}^s = \frac{(\gamma_0^*-1)}{\gamma_0^*R_{g,0}}(s'\Phi').
\end{equation*}
\end{itemize}

\section{Additional details on Reynolds-number dependence}
\label{sec:Re-dependence}

\begin{figure}[]
\centering
\begin{subfigure}[t]{0.49\textwidth}
\includegraphics[width=\textwidth]{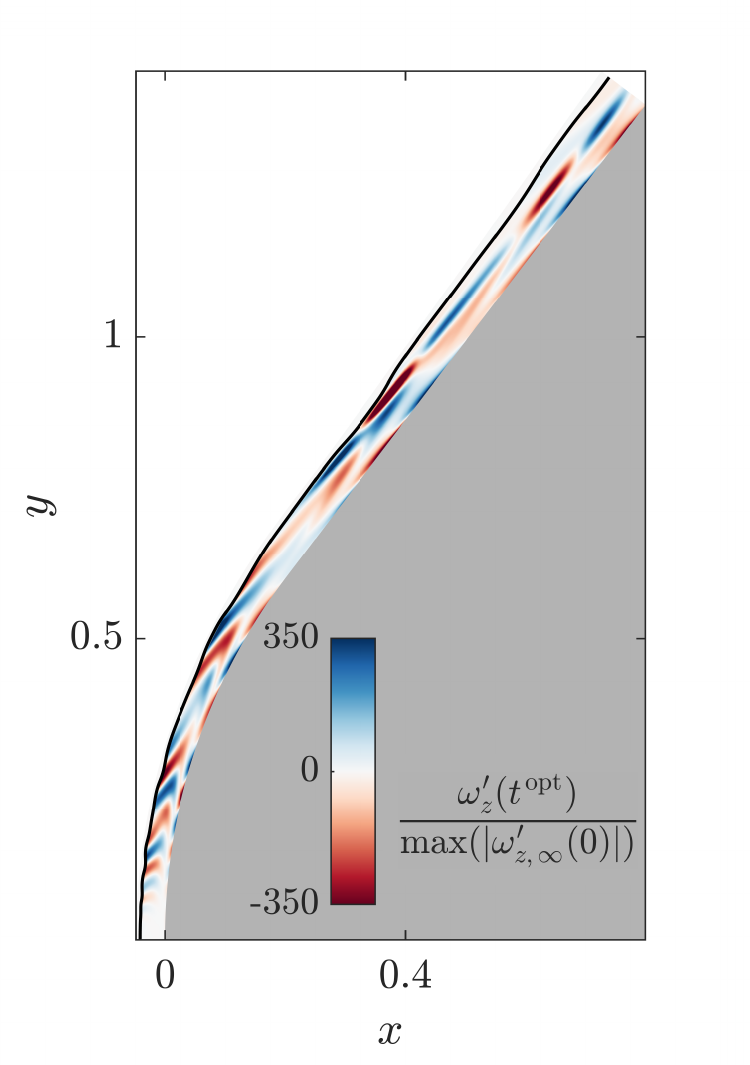}
\caption{}
\label{fig:Re_sweep_disturbances_20000_1}
\end{subfigure}
\begin{subfigure}[t]{0.49\textwidth}
\includegraphics[width=\textwidth]{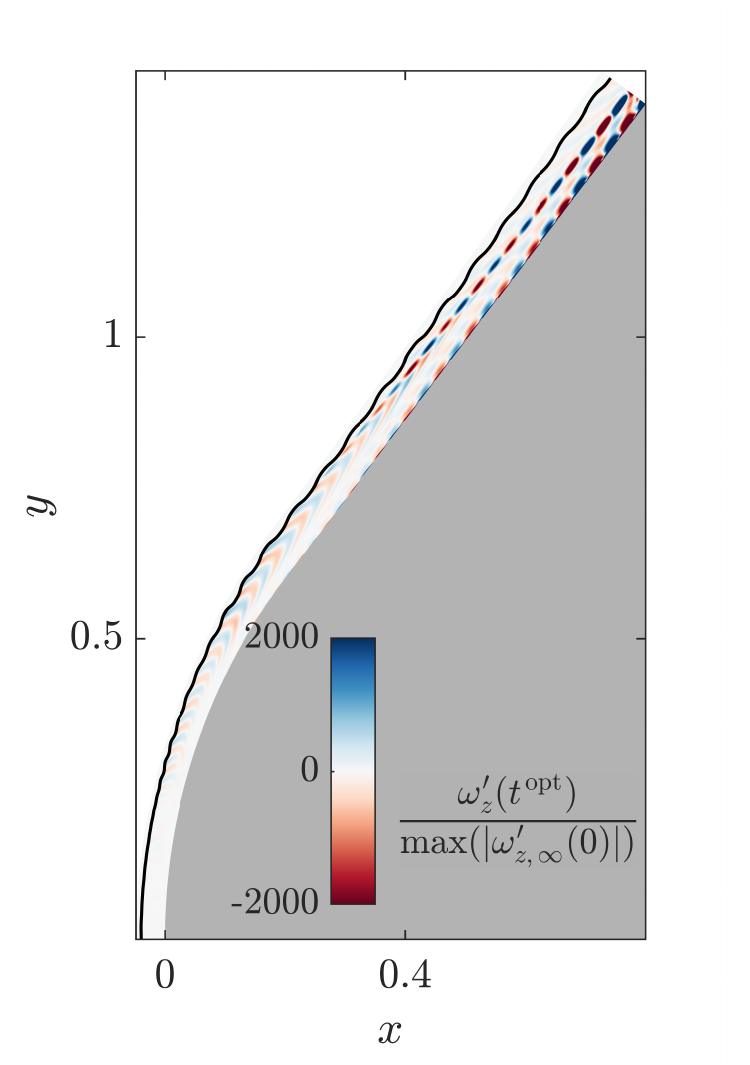}
\caption{}
\label{fig:Re_sweep_disturbances_100000_1}
\end{subfigure}
\caption{Optimal disturbance that gives maximum steady-state energy
  growth $G_T^{\text{opt}}$ for various Reynolds numbers
  $Re_\infty$. Base flow computed with freestream conditions from
  table~\ref{tab:freestream}. $M_\infty = 28.7$. (\textit{a})
  Vorticity at the instant of peak gain on the limit cycle, $Re_\infty
  = 20\,000$. (\textit{b}) Vorticity at the same instant, $Re_\infty =
  100\,000$.}
\label{fig:Re_sweep_disturbances}
\end{figure}

The underlying mechanics of the scaling in the downstream region are
tied to the viscous dissipation of the optimal disturbances. Since
dissipation scales as $k^2$, where $k$ is the disturbance wavenumber,
higher Reynolds numbers support the sustenance of higher-frequency
freestream disturbances for which viscous terms remain
negligible. This effect is evident in figure
\ref{fig:Re_sweep_disturbances}, which reveals that the characteristic
frequency of the optimal disturbance increases noticeably between
$Re_\infty = 20\,000$ and $Re_\infty = 100\,000$, enabling
high-frequency excitation of both the shock and the entropy layer.
\begin{figure}[]
\centering
\begin{subfigure}[t]{0.49\textwidth}
\includegraphics[width=\textwidth]{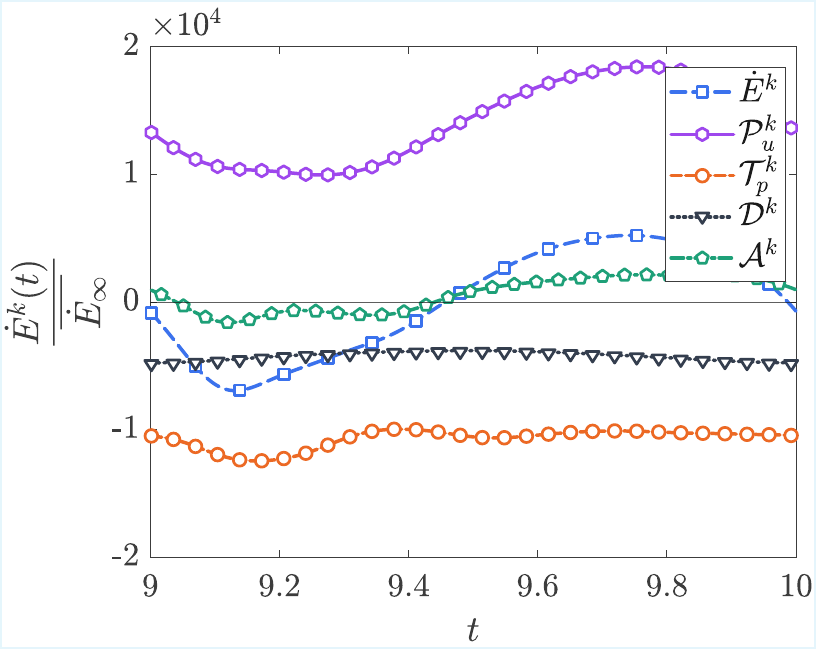}
\caption{}
\label{fig:Re_sweep_disturbances_20000_1_b}
\end{subfigure}
\begin{subfigure}[t]{0.49\textwidth}
\includegraphics[width=\textwidth]{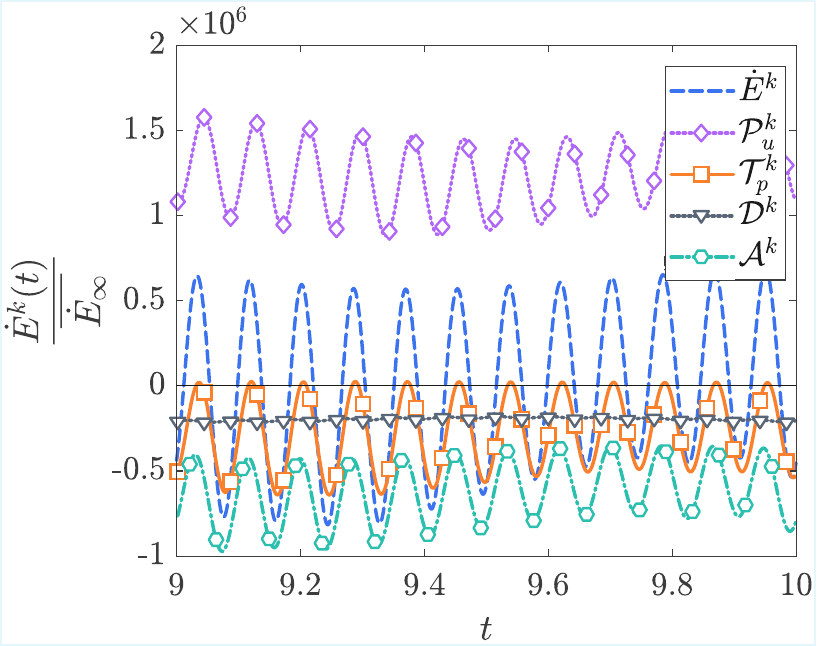}
\caption{}
\label{fig:Re_sweep_disturbances_100000_1_b}
\end{subfigure}
\caption{Most important terms from the kinetic energy budget for the
  optimal disturbance that gives maximum energy growth
  $G_T^{\text{opt}}$. Budgets compared across two different
  Reynolds numbers $Re_\infty = 20\,000$ and $100\,000$. Base flow
  computed with freestream conditions from
  table~\ref{tab:freestream}. $M_\infty = 28.7$. Energy rates are
  non-dimensionalized with the mean freestream disturbance energy flux
  that crosses the shock wave, $\overline{\dot{E}_\infty}$.
  Definitions and derivations of the budget terms are provided in
  Appendix~\ref{sec:energy-budgets}. (\textit{a}) Kinetic energy
  budget, $Re_\infty = 20\,000$. (\textit{b}) Kinetic energy budget,
  $Re_\infty = 100\,000$.}
\label{fig:budgets_Re}
\end{figure}

To further clarify the physical mechanisms underlying these Reynolds
number scalings, the most relevant terms of the kinetic energy budgets
are compared in figure~\ref{fig:budgets_Re} for $Re_\infty = 20\,000$
and $Re_\infty = 100\,000$. At $Re_\infty = 20\,000$, viscous
dissipation $\mathcal{D}^k$ constitutes a significant contribution to
the budget, directly competing with production and limiting the net
energy growth. As the Reynolds number increases, dissipation
diminishes in magnitude; however, the decrease is more gradual than
one might anticipate on the basis of Reynolds number scaling alone.
The reason is that the characteristic frequency of the optimal
disturbance also increases with $Re_\infty$ (as shown in
figure~\ref{fig:Re_sweep_disturbances}), which partially offsets the
reduction in viscous effects by introducing finer spatial scales that
are more susceptible to dissipation.

Despite this frequency shift, Reynolds stress production
$\mathcal{P}^k_u$ remains the dominant source term in the kinetic
energy budget at all Reynolds numbers investigated. This production
continues to be counteracted primarily by the pressure work associated
with shock oscillations, $\mathcal{T}^k_p$. At higher Reynolds
numbers, however, wave packets accumulate substantially more energy
before exiting the computational domain. This is reflected in the
advection term $\mathcal{A}^k$, which exhibits increasingly pronounced
negative spikes at higher $Re_\infty$ as energetic wave packets are
advected out of the domain.

\section{Kov\'asznay reconstruction of the optimal freestream disturbance}
\label{app:Kovasznay}

The freestream-receptivity formulation in
\S\ref{sec::freestream_receptivity_implementation} optimizes the
incident perturbation only through its trace immediately upstream of
the fitted bow shock.  No upstream domain is solved, and no relation
between the temporal frequency and the streamwise wavenumber is
imposed during the optimization.  The purpose of the present appendix
is to describe the \textit{a posteriori} reconstruction used to
interpret the optimal shock trace as a superposition of Kov\'asznay
modes of the uniform freestream.

\paragraph{Derivation of modes.}
The uniform upstream flow is in the calorically perfect gas regime,
with constant state
$(\rho_\infty,U_\infty,p_\infty,a_\infty,\gamma_\infty^*)$, so that the
freestream sound speed $a_\infty$ and the Kov\'asznay decomposition are
well defined.  We introduce coordinates aligned with the freestream,
$x_\parallel=\bm{e}_\parallel\cdot\bm{x}$ and
$x_\perp=\bm{e}_\perp\cdot\bm{x}$, where
$\bm{e}_\parallel=\bm{u}_\infty/U_\infty$.  In the present reference
frame, the $x$-axis is aligned with the freestream direction and the
$y$-axis is the transverse direction, so that $x_\parallel=x$,
$x_\perp=y$, $u'_\parallel=u'$ and $u'_\perp=v'$.  The Kov\'asznay modes
are written in the HYMOR/Chu variables
\[
  \bm{\psi}'=[p',u',v',S']^T,
  \qquad
  S'=\frac{s'}{R_{g,\infty}},
\]
using the plane-wave ansatz
\begin{equation}
  \bm{\psi}'(\bm{x},t)
  =
  \widehat{\bm{\psi}}\,
  \exp\left\{i\left(\alpha x_\parallel+\beta x_\perp-\omega t\right)\right\},
  \label{eq:kovasznay_ansatz}
\end{equation}
where $\alpha$ and $\beta$ are the freestream-parallel and transverse
wavenumbers, $\bm{k}=\alpha\bm{e}_\parallel+\beta\bm{e}_\perp$ is the
wavevector, and the Doppler-shifted frequency is
$\Omega=\omega-U_\infty\alpha$.

In the uniform inviscid freestream, the entropy and vortical modes are
convected with the mean flow, so $\Omega=0$ and hence
$\alpha_c=\omega/U_\infty$.  Writing
$K_c=(\alpha_c^2+\beta^2)^{1/2}$ and
$\bm{k}_c=\alpha_c\bm{e}_\parallel+\beta\bm{e}_\perp$, the entropic and
vortical modes take the vector form
\begin{align}
  \bm{b}^{\,e}(\bm{x},t;\omega,\beta)
  &=
  \begin{bmatrix}
  0\\
  0\\
  0\\
  1
  \end{bmatrix}
  \exp\left\{i\left(\alpha_c x_\parallel
  +\beta x_\perp-\omega t\right)\right\},
  \label{eq:kovasznay_entropy_mode}
  \\
  \bm{b}^{\,v}(\bm{x},t;\omega,\beta)
  &=
  \begin{bmatrix}
  0\\
  -\beta/K_c\\
  \alpha_c/K_c\\
  0
  \end{bmatrix}
  \exp\left\{i\left(\alpha_c x_\parallel
  +\beta x_\perp-\omega t\right)\right\}.
  \label{eq:kovasznay_vortical_mode}
\end{align}
The first vector represents a pure convected entropy disturbance.  The
second has zero pressure and entropy perturbations and is solenoidal,
$\bm{k}_c\cdot\widehat{\bm{u}}=0$.

The acoustic mode is isentropic, so the perturbation pressure and
density obey the frozen sound-speed relation $p'=a_\infty^2\rho'$.  The
linearized uniform-flow momentum and continuity equations then give
\[
  \rho_\infty \Omega\,\widehat{\bm{u}}=\bm{k}\,\widehat p,
  \qquad
  \Omega^2=a_\infty^2(\alpha^2+\beta^2).
\]
Substituting $\Omega=\omega-U_\infty\alpha$ into the dispersion
relation, the acoustic values of $\alpha$ for a prescribed pair
$(\omega,\beta)$ are the roots of
\begin{equation}
  (U_\infty^2-a_\infty^2)\alpha^2
  -2U_\infty\omega\alpha
  +(\omega^2-a_\infty^2\beta^2)=0.
  \label{eq:kovasznay_acoustic_quadratic}
\end{equation}
The pressure amplitude is normalized as $\widehat{p}^a = 1$.  For each
acoustic root $\alpha_\pm$ of
\eqref{eq:kovasznay_acoustic_quadratic}, with
$\Omega_\pm = \omega - U_\infty \alpha_\pm$, the acoustic mode is
\begin{equation}
  \bm{b}^{\,a}_{\pm}(\bm{x},t;\omega,\beta)
  =
  \begin{bmatrix}
  1\\
  \alpha_\pm/(\rho_\infty\Omega_\pm)\\
  \beta/(\rho_\infty\Omega_\pm)\\
  0
  \end{bmatrix}
  \exp\left\{i\left(\alpha_\pm x_\parallel
  +\beta x_\perp-\omega t\right)\right\}.
  \label{eq:kovasznay_acoustic_mode}
\end{equation}

\paragraph{Projection onto the bow shock.}
The Kov\'asznay modes of
\eqref{eq:kovasznay_entropy_mode}--\eqref{eq:kovasznay_acoustic_mode}
are plane waves of the uniform freestream. The bow shock enters only as a curved locus
of sampling points: each mode is a known function of the
freestream-aligned coordinates $(x_\parallel,x_\perp)$, and it is
evaluated at the freestream-frame positions
$\bigl(x_\parallel(s_k),x_\perp(s_k)\bigr)$ of the discrete shock
stations $s_k$.

For each HYMOR frequency $\omega_l$, the optimized
conservative-variable shock trace $\widehat{\bm q}'_{\infty,l}(s_k)$ is
first converted to the HYMOR/Chu variables
$\widehat{\bm{\psi}}^{\,H}_l(s_k)=[\widehat p,\widehat u,\widehat
v,\widehat S]^T$.  The Kov\'asznay library is built directly from the
mode templates of
\eqref{eq:kovasznay_entropy_mode}--\eqref{eq:kovasznay_acoustic_mode}.
We introduce the shock-evaluation operator $\bm{b}_{k,l,m,r}$,
which samples the freestream plane wave of frequency $\omega_l$,
transverse wavenumber $\beta_m$, and mode type
$r\in\{e,v,a_+,a_-\}$ at the $N_s$ shock stations $s_k$,
\begin{equation}
  \bm{b}_{k,l,m,r}
  =
  \widehat{\bm\psi}_{l,m,r}\,
  \exp\!\left\{i\bigl(\alpha_{l,m,r}\,x_\parallel(s_k)
  +\beta_{m}\,x_\perp(s_k)\bigr)\right\} ,
  \label{eq:kovasznay_eval_operator}
\end{equation}
where $\widehat{\bm\psi}_{l,m,r}$ and $\alpha_{l,m,r}$ are the
eigenvector and freestream-parallel wavenumber carried by the chosen
mode.  The freestream-parallel wavenumber follows without any
externally imposed dispersion assumption: the convected entropy and
vortical modes take $\alpha_c=\omega_l/U_\infty$, while the two
acoustic modes take the roots $\alpha_\pm$ of the dispersion quadratic
\eqref{eq:kovasznay_acoustic_quadratic}.

Sweeping a set of transverse wavenumbers
$\{\beta_m\}_{m=1}^{N_\beta}$ and the mode types $r$, the fitted
shock-trace variables are obtained as the contraction of the
evaluation operator with the Kov\'asznay coefficients $c_{l,m,r}$,
\begin{equation}
  \widehat{\bm{\psi}}^{\,K}_l(s_k)
  =
  \sum_{m,r} \bm{b}_{k,l,m,r}\,c_{l,m,r}.
  \label{eq:kovasznay_einsum}
\end{equation}
The coefficients are
obtained from the weighted least-squares problem
\begin{equation}
  c_{l,m,r}
  =
  \operatorname*{arg\,min}_{c_{l,m,r}}
  \sum_k
  \left|
  \bm{W}_l^{1/2}(s_k)
  \left(\bm{b}_{k,l,m,r}\,c_{l,m,r}
  -\widehat{\bm{\psi}}^{\,H}_l(s_k)\right)
  \right|^2 ,
  \label{eq:kovasznay_lsq}
\end{equation}
where the weight $\bm{W}_l$ is the freestream Chu-energy flux weight
used in the HYMOR input norm.

\paragraph{Reconstruction of the freestream disturbance.}
Once the coefficients $c_{l,m,r}$ have been obtained from
\eqref{eq:kovasznay_lsq}, the same modes can be evaluated at any
upstream point.  Denoting by $\widehat{\bm{b}}_{l,m,r}$ the eigenvector
of the $r$th mode template carried by frequency $\omega_l$ and
transverse wavenumber $\beta_m$ (with parallel wavenumber
$\alpha_{l,m,r}$), the reconstructed freestream disturbance is
\begin{equation}
  \bm{\psi}'_\infty(\bm{x},t)
  =
  \Re\left\{
  \sum_l\sum_m\sum_r
  c_{l,m,r}\,\widehat{\bm{b}}_{l,m,r}\,
  \exp\left[i\left(\alpha_{l,m,r}\,x_\parallel
  +\beta_m\,x_\perp-\omega_l t\right)\right]
  \right\},
  \label{eq:kovasznay_upstream_reconstruction}
\end{equation}
where $r$ runs over the admissible mode types
$\{e,v,a_+,a_-\}$ retained for $\beta_m$.
The quality of the reconstruction is measured by the captured energy
fraction
\begin{equation}
  \eta_{\rm cap}
  =
  1-
  \frac{
  \sum_l
  (\widehat{\bm{\psi}}^{\,H}_l-\widehat{\bm{\psi}}^{\,K}_l)^\dagger
  \bm{W}_l
  (\widehat{\bm{\psi}}^{\,H}_l-\widehat{\bm{\psi}}^{\,K}_l)}
  {
  \sum_l
  (\widehat{\bm{\psi}}^{\,H}_l)^\dagger
  \bm{W}_l
  \widehat{\bm{\psi}}^{\,H}_l},
  \label{eq:kovasznay_capture_fraction}
\end{equation}
where $\widehat{\bm{\psi}}^{\,K}_l$ is the fitted trace assembled from
\eqref{eq:kovasznay_einsum}.
For the optimal disturbances discussed in \S\ref{sec::Optimal_disturbance},
the residual energy is negligible, so the upstream packet shown in
figure~\ref{fig:freestream_dist_first} may be interpreted as the
Kov\'asznay extension of the HYMOR-optimized shock trace.

\paragraph{Practical computation of the freestream projection.}
The projection and reconstruction are performed through the following
steps:
\begin{itemize}
  \item Select the optimal HYMOR freestream-receptivity vector and
  extract the conservative-variable perturbation trace immediately
  upstream of the fitted bow shock for all retained frequencies.

  \item For each frequency $\omega_l$, reshape the extracted trace into
  the shock-station values $\widehat{\bm q}'_{\infty,l}(s_k)$ and
  convert it to the HYMOR/Chu variables
  $\widehat{\bm{\psi}}^{\,H}_l(s_k)=[\widehat p,\widehat u,\widehat
  v,\widehat S]^T$.

  \item Compute the freestream-aligned coordinates
    $x_\parallel(s_k)$ and $x_\perp(s_k)$ of each shock station, which
    set the spatially varying phase of each plane wave along the
    curved shock (see above).  Evaluate also the projected shock-area
    weights entering the Chu-energy flux inner product $\bm W_l$.

  \item For each swept transverse wavenumber $\beta_m$ and mode type
    $r\in\{e,v,a_+,a_-\}$, evaluate the entry
    $\bm{b}_{k,l,m,r}$ of the shock-evaluation operator
    \eqref{eq:kovasznay_eval_operator}: the convected entropy and
    vortical templates $\bm{b}^{\,e},\bm{b}^{\,v}$ use
    $\alpha_c=\omega_l/U_\infty$, while the acoustic templates
    $\bm{b}^{\,a}_{\pm}$ are built from the real roots $\alpha_\pm$ of
    the dispersion quadratic
    \eqref{eq:kovasznay_acoustic_quadratic} using the sound speed
    $a_\infty$.

  \item Solve the weighted least-squares problem
    \eqref{eq:kovasznay_lsq} for the coefficients $c_{l,m,r}$, where
    the fitted trace is the contraction
    $\bm{b}_{k,l,m,r}\,c_{l,m,r}$ over the swept transverse wavenumbers
    $m$ and mode types $r$.

  \item Reconstruct the optimal disturbance over the bow shock using
    $\widehat{\bm{\psi}}^{\,K}_l(s_k)=\bm{b}_{k,l,m,r}\,c_{l,m,r}$ from
    \eqref{eq:kovasznay_einsum}, and evaluate the captured energy
    fraction \eqref{eq:kovasznay_capture_fraction}.

  \item Evaluate the same fitted Kov\'asznay expansion away from the
  shock using \eqref{eq:kovasznay_upstream_reconstruction} to obtain
  the upstream freestream disturbance field implied by the optimal
  shock trace.
\end{itemize}

\end{document}